\documentclass[11pt,a4paper]{article}
\pdfoutput=1
\usepackage{jheppub}

\usepackage{multirow, graphicx,amssymb,url,mathrsfs,amsmath}
\usepackage{wrapfig,boxedminipage,setspace,subfigure,epsfig}
\usepackage{amsxtra,amstext,latexsym,dsfont,amsfonts}
\usepackage{color,eucal}
\usepackage[dvipsnames]{xcolor}
\usepackage{float}
\usepackage{slashed,comment}
\usepackage{kotex}
\usepackage{tikz}
\usetikzlibrary{calc,patterns,angles,quotes}
\usetikzlibrary{decorations.pathreplacing,decorations.markings,snakes}
\usepackage{tabularx, array}
\usepackage{mdframed,mathtools}
\newcolumntype{L}[1]{>{\raggedright\arraybackslash}p{#1}}
\newcolumntype{C}[1]{>{\centering\arraybackslash}p{#1}}
\newcolumntype{R}[1]{>{\raggedleft\arraybackslash}p{#1}}

\usepackage{xcolor}
\definecolor{amethyst}{rgb}{0.54, 0.17, 0.89}
\definecolor{coral}{rgb}{1.0, 0.3, 0.4}

\newcommand{\nn}{\nonumber}

\newcommand{\bra}[1]{\mbox{$\langle #1 |$}}
\newcommand{\ket}[1]{\mbox{$| #1 \rangle$}}

\newcommand{\Tr}{{\rm Tr}\,}

\newcommand{\SL}{\mathrm{SL}}
\newcommand{\SU}{\mathrm{SU}}
\newcommand{\be}{\begin{equation}}
\newcommand{\ee}{\end{equation}}
\newcommand{\bea}{\begin{eqnarray}}
\newcommand{\eea}{\end{eqnarray}}

\newcommand{\mt}[1]{\textrm{\tiny #1}}

\title{Variations on a Theme of Krylov}

\author[a,b,c]{Vijay Balasubramanian,}
\author[d,e,f]{Pawel Caputa,}
\author[g]{Joan Sim\'on}

\emailAdd{vijay@physics.upenn.edu}
\emailAdd{pawel.caputa@fysik.su.se}
\emailAdd{j.simon@ed.ac.uk}

\preprint{YITP-25-171}

\affiliation[a]{David Rittenhouse Laboratory, University of Pennsylvania,
209 S. 33rd Street, Philadelphia, PA 19104, USA}
\affiliation[b]{Santa Fe Institute,
1399 Hyde Park Road, Santa Fe, NM 87501, USA}
\affiliation[c]{Theoretische Natuurkunde, Vrije Universiteit Brussel,
Pleinlaan 2, B-1050 Brussels, Belgium}
\affiliation[d]{
The Oscar Klein Centre and Department of Physics, Stockholm University, AlbaNova, 106 91 Stockholm, Sweden}
\affiliation[e]{Yukawa Institute for Theoretical Physics, Kyoto University, Kitashirakawa Oiwakecho, Sakyo-ku, Kyoto 606-8502, Japan}
\affiliation[f]{Faculty of Physics, University of Warsaw, Pasteura 5, 02-093 Warsaw, Poland}
\affiliation[g]{School of Mathematics and Maxwell Institute for Mathematical Sciences, University of Edinburgh, Edinburgh EH9 3FD, UK}

\abstract{Spread complexity uses the distribution of support of a time-evolving  state in the Krylov basis to quantify dispersal across  accessible dimensions of a Hilbert space. Here, we describe how variations in initial conditions, the Hamiltonian, and the dimension of the Hilbert space affect spread complexity and Krylov basis structure. We introduce {\it Koherence}, the entropy of coherence between perturbed and unperturbed Krylov bases, which can,  e.g., quantify  dynamical amplification of differences in initial conditions in chaos. To illustrate, we show that dynamics on $\mathrm{SL}(2,\mathbb{R})$, SU(2), and Heisenberg-Weyl group manifolds, often used as paradigmatic settings for contrasting chaotic and integrable (semi-)classical behavior, display distinctively different responses to variations of the initial state or Hamiltonian.  We then describe a lattice model that displays linear growth of spread complexity, saturating for bounded lattices  and continuing forever in a thermodynamic limit. The latter example illustrates a breakdown of continuum/classical effective descriptions of complexity growth in bounded quantum systems.}
\begin{document}

\maketitle

\section{Introduction}\label{sec:Introduction}
In classical chaotic systems a typical initial state ergodically explores the configuration space as it evolves in time.  Likewise, small differences in initial conditions lead to large deviations in final states.  It has been difficult to devise analogous crisp characterizations of quantum chaotic  systems.  This is because the Schr\"{o}dinger equation is linear; so small additive perturbations of an initial state propagate separately, and additively, forward in time without the sort of non-linear amplification that is evident classically. 

Recently, the authors of \cite{Balasubramanian:2022tpr} defined a new quantity,  {\it spread complexity}, for characterizing the first of these traits -- how an initial quantum state explores the Hilbert space  \cite{Balasubramanian:2022tpr}.\footnote{A related quantity, the Krylov complexity \cite{Parker:2018yvk}, characterizes the spread of operators.  See \cite{Nandy:2024evd,Baiguera:2025dkc,Rabinovici:2025otw} fo recent reviews.}  Given an initial state and the Hamiltonian governing time evolution, the new quantity  measures how widely the state has spread in the Krylov basis, which minimizes the spread over all possible bases at least for some initial duration \cite{Balasubramanian:2022tpr}.  In this distinguished basis, the Hamiltonian is tri-diagonalized, and its two non-vanishing diagonal bands constitute the Lanczos spectrum of the theory relative to the given initial state.  A conjecture in \cite{Balasubramanian:2023kwd} states that quantum chaotic systems display a Lanczos spectrum well described by Random Matrix Theory (RMT), leading to characteristic long-term linear growth, followed by decline and saturation of  spread complexity, characterizing how generic initial states explore the Hilbert space. This conjecture generalizes spectral characterizations of quantum chaos -- e.g.,
the energy levels have a Wigner-Dyson distribution, show level repulsion, or approach the statistics of an RMT universality class \cite{Dyson:1962es,Bohigas:1983er}.

Here, we propose a new quantity, the  entropy of coherence between Krylov bases, or {\it Koherence}, for characterizing the second trait of chaos -- amplification of differences in initial conditions.  To calculate this quantity we consider the Krylov bases associated to  perturbations of an initial state. Koherence is calculated from the distribution of overlaps between the unperturbed and perturbed bases.  We will see that Koherence, and related measures of late time amplification of initial condition differences that we will define, show important differences between integrable and chaotic systems for some types of initial states. As such, the new measures complement Lyapunov exponent-like early time measures of chaos defined in terms of the rates of growth of Out-Of-Time-Ordered-Correlators (OTOC) \cite{Maldacena:2015waa}, and may provide new ways of separating chaotic and integrable systems.

Another way of thinking about chaos is to ask about the effects of small variations of the Hamiltonian on the dynamics of a  system.  We might expect that two similar chaotic Hamiltonians will produce very different trajectories for the same initial state, although statistical features of these trajectories would presumably be universal and could be uncovered by computing ensemble averages.  Indeed, this is the paradigm of RMT, which gives a canonical description of maximally chaotic systems in terms of Hamiltonians drawn randomly from a fixed distribution.  The standard approach is to study the spectral properties of these Hamilonians, but one could also study the dynamics they generate.  Ensembles of Hamiltonians also appear in the Sachdev–Ye–Kitaev (SYK) model (see the review \cite{Sarosi:2017ykf}), and recent studies also suggest that  the gravitational path integral is actually computing a quantity that is coarse-grained over Hamiltonians (see \cite{Saad:2019lba,Chandra:2022bqq}), or  over the statistics of matrix elements consistently with the Eigenstate Thermalization Hypothesis (ETH)\footnote{For example, see \cite{Sasieta:2022ksu,Balasubramanian:2022gmo} for an explicit application to the path integral construction of a class of black hole microstates and their overlaps.} \cite{Deutsch:1991msp,Srednicki:1994mfb} or perhaps over OPE coefficients \cite{Belin:2020hea}.

Thus we also develop techniques for studying how spread complexity and the Krylov basis vary if we fix the initial state and vary the Hamiltonian.  The variation of spread complexity with the initial state or with the Hamiltonian can both be studied by a similar technique: by asking how the moments of the Hamiltonian vary in the initial state.  This is because the time-varying spread complexity of state can be computed from the list of moments of the Hamiltonian in the initial state.  We will show how all this data can be written in a form that resembles the first law of thermodynamics.  We will also explain how the variations in the moments of the Hamiltonian are related to the distribution of overlaps defining Koherence.  To illustrate our methods we apply them to motion on the $\mathrm{SL}(2,\mathbb{R})$, SU(2), and Heisenberg-Weyl group manifolds, which show strikingly different dynamics.

Dynamics starting from an initial state $\ket{\psi}$ explores a subspace of the Hilbert space controlled by the support of $\ket{\psi}$ on the energy eigenbasis.  Suppose the full Hilbert space has dimension $N$.  Then we know that spread complexity must be bounded because the Krylov basis can have no more than $N$ elements.  Initial states with support on $K<N$ energy eigenstates will have Krylov bases of dimension $\leq K$. The dimension of the span of the Krylov basis also depends on the nature of the spectrum and the dynamics.   If the energy levels are mutually incommensurate, perhaps because the dynamics is chaotic, we expect that the evolving state vector will explore all dimensions of the Hilbert space; if the energy levels have some orderly structure, perhaps because the system has some degree of integrability, we expect the evolving state vector to explore a subspace of the energetically accessible part of the Hilbert space.   Either way, for a bounded system spread complexity will be bounded, and, as we will see, initially grows quadratically, then passes through a phase of linear growth, and then undergoes oscillations around a plateau.    Often we wish to study systems in a thermodynamic, large system limit, in which the dimension of the accessible Hilbert space approaches infinity.  This may happen because the dimension of the Hilbert space diverges ($N\to\infty$), or because the  support of the initial state in an already infinite energy eigenbasis broadens, perhaps because we are considering a high temperature limit.  In these situations we will see that it is possible to get unbounded linear growth of spread complexity.

Large system limits often lead, after coarse-graining to a continuum, to a semiclassical description of a system. The effective descriptions of such limits can erase phenomena that are present at finite size, because the probes that sense these phenomena do not survive the limit, either because they do not approach it smoothly \cite{Longo:2022lod,Chandrasekaran:2022eqq,Liu:2025krl}, or because they become arbitrarily complex.  In that case, an effective continuum model derived from a large system size limit will break down as a description of the underlying discrete theory for some, sufficiently precise, questions.   Precisely this sort of phenomenon seems to be occurring in gravity where the effective field theory around a semiclassical spacetime has  excess  degrees of freedom which are only projected out by considering the  non-perturbative quantum theory (see, e.g., \cite{Marolf:2020xie,Akers:2022qdl}). Understanding the breakdown of the semiclassical limit played a key role in recent progress towards resolving the information paradox \cite{Almheiri:2019psf,Penington:2019kki,Almheiri:2019qdq}, and towards understanding how to count states in gravity \cite{Balasubramanian:2022gmo,Balasubramanian:2025hns}. Likewise the unbounded classical growth of the wormhole behind the horizon of an eternal black hole may be truncated by quantum effects associated associated with the breakdown of the semiclassical effective theory \cite{Lin:2022rbf,Rabinovici:2023yex,Iliesiu:2024cnh,Nandy:2024zcd,Balasubramanian:2024lqk,Miyaji:2025yvm,Miyaji:2025ucp}.

These considerations motivate us to also study how spread complexity and the structure of the Krylov basis vary as we change the dimension of the Hilbert space.  We do so by varying the size of an analytically solvable lattice model.  The large lattice limit, after coarse-graining, gives an effective continuum description of the system, and we examine how this description breaks down, i.e., fails to match the fine-grained lattice model, at late times in the dynamics, and relatedly, in the structure of the higher elements of the Krylov basis.

Four sections follow. In Sec.~\ref{sec:Krylov} we review methods for computing the Krylov basis for a given initial state. 
We focus on the {\it recursion} and {\it moment} methods which are by now conventional, and what we call the {\it Krylov polynomial} method, which exploits properties of a certain infinite family of polynomials in the Hamiltonian which are orthogonal in measures defined by the initial states.  We then define quantities including the Koherence and relative Krylov entropy, that measure how perturbations of the initial state affect the spread of the wavefunction across the Hilbert space and the associated Krylov basis.  These measures also allow us to quantify how the dynamics amplifies or damps differences in  initial conditions. In Sec.~\ref{sec:SolvableEx} we apply these methods to analytically-tractable examples describing motion on the $\mathrm{SL}(2,\mathbb{R})$, SU(2) and Heisenberg-Weyl group manifolds.  In Sec.~\ref{sec:SolvmodLin}  we  discuss a solvable lattice model in which we can construct the Krylov basis as we vary the initial state of the Hamiltonian and the number of sites of the lattice.  We use the polynomial method described in Sec.~\ref{sec:Krylov} to solve this model and find that for the initial states we study an RMT density of states makes a mysterious appearance in the measure under which the associated Krylov polynomials built out of the Hamiltonian are orthogonal.  This model also enables us to study how the  continuum effective field theory describing the large system limit  breaks down in describing the underlying microscopic system. We conclude with a discussion Sec.~\ref{sec:Conclusions}, and expand on details of material in the main text in four appendices. 

\section{Krylov methods}\label{sec:Krylov}
We begin with a review of the construction of the Krylov basis. Consider unitary time evolution of an initial quantum state $\ket{\psi_0}$ by the action of a time-independent Hamiltonian $H$ 
\be
\ket{\psi(t)}=e^{-iHt}\ket{\psi_0}=\sum^{\mathcal{K}-1}_{n=0}\psi_n(t)\ket{K_n}\,.\label{TimEvState}
\ee
In the second equality we expanded the time-evolved state in the $\mathcal{K}$-dimensional Krylov basis. To construct this  basis we use the {\it recursion method} provided by the Lanczos algorithm \cite{Lanczos:1950zz,LanczosBook,pettifor2012recursion}
\be
\ket{A_{n+1}}=(H-a_n)\ket{K_n}-b_{n}\ket{K_{n-1}},\qquad \ket{K_n}=b^{-1}_n\ket{A_n}\,,\label{LanczosAlgGen}
\ee
where $\ket{K_0}=\ket{\psi_0}$ and $b_0=0$, and the Lanczos coefficients are 
\be
a_n=\langle K_n|H|K_n\rangle\,,\qquad b_n=\langle A_n|A_n\rangle^{1/2}\,.\label{LanczosCoeff}
\ee
The algorithm terminates at $n=\mathcal{K}-1$ where $b_{n}=0$. The Krylov basis could be finite or infinite dimensional -- it depends on the dimension of the subspace of the full Hilbert space that is explored by dynamics starting with the given initial state.

By construction \eqref{LanczosAlgGen}, the  Hamiltonian  acts tri-diagonally in the Krylov basis:
\be
H\ket{K_n}=a_n\ket{K_n}+b_n\ket{K_{n-1}}+b_{n+1}\ket{K_{n+1}}\,.\label{LanczosAlg}
\ee
This structure allows us to extract the coefficients of the expansion of a state in the Krylov basis, i.e., the wave functions $\psi_n(t)$ in \eqref{TimEvState}. By differentiating \eqref{TimEvState} and using \eqref{LanczosAlg}, it is easy to show that the $\psi_n(t)$ satisfy a discrete Schr\"{o}dinger equation
\be
i\partial_t\psi_n(t)=a_n\psi_n(t)+b_n\psi_{n-1}(t)+b_{n+1}\psi_{n+1}(t)\,, \label{SEq}
\ee
with initial condition $\psi_n(0)=\delta_{n,0}$. This equation  maps the quantum dynamics \eqref{TimEvState} to a particle hopping on a 1D chain with sites labeled by $n$. The amplitudes for staying on a given site and for jumping to the neighboring ones are  $a_n$ and $b_n$ respectively.

For practical purposes, it is useful to rewrite this equation as
\be
\psi_{n+1}(t)=\frac{(i\partial_t-a_n)\psi_n(t)-b_n\psi_{n-1}(t)}{b_{n+1}}\,,\label{eq:wf-rec}
\ee
which shows that once we know the $n=0$ amplitude and the Lanczos coefficients,  all the higher wave functions can be computed. In fact, $\psi_0(t)$ contains all the information required to determine  dynamics in the Krylov subspace. Indeed, the key ingredient  is the return amplitude 
\be
S(t)\equiv\langle \psi_0|e^{iHt}|\psi_0\rangle=\psi^*_0(t)\,,
\ee
as it contains all the Lanczos coefficients. To see this, we can  perform a moment expansion, i.e., we write the return amplitude in terms of the moments of the  Hamiltonian $H$ in the initial state 
\be
S(t)=\sum^\infty_{k=0}\frac{t^k}{k!}\mu_k\,,\qquad \mu_k=\langle \psi_0|(iH)^k|\psi_0\rangle=\langle K_0|(iH)^k|K_0\rangle\,.\label{MomRetAmpl}
\ee
Since $H$ is tri-diagonal in the Krylov basis, taking its k-th power and equating the k-th moment to the $((iH)^k)_{00}$  matrix element provides a polynomial relation between $\mu_k$ and the  $(a_n,b_n)$ that we can solve for the Lanczos coefficients.\footnote{For the k-th moment, only the  Lanczos coefficients up to $n=k-1$ are involved.}  This is called the {\it moment method}. In particular, two key quantities that will play a central role throughout this work are $a_0$, which corresponds to the expectation value of the Hamiltonian in the initial state, $a_0=\langle \psi_0|H|\psi_0\rangle$, and $b_1$ which represents the energy variance in that state: $b_1^2=\langle \psi_0|H^2|\psi_0\rangle-\langle \psi_0|H|\psi_0\rangle^2$.

The relation between the moments and Lanczos coefficients has an  alternative derivation in terms of the Laplace transform of the wavefunctions:
\be
\Psi_n(z)=\mathcal{L}[\psi_n(t)]\equiv\int^\infty_0 dt e^{-zt}\psi_n(t)\,,\qquad \text{Re}(z)>0\,.
\ee
Applying this transform to \eqref{SEq} we find\footnote{Recall  that the transform of derivative is $\mathcal{L}[f'(t)]=zF(z)-f(0^+)$, and that for us $\psi_n(0^+)=\delta_{n,0}$.}
\be
iz\Psi_n(z)=i\delta_{n,0}+a_n\Psi_n(z)+b_n\Psi_{n-1}(z)+b_{n+1}\Psi_{n+1}(z)\, ,
\ee
which can be written, after distinguishing  $n=0$ and $n\ge 1$, as
\be
\Psi_0(z)=\frac{i}{iz-a_0-R_1(z)}\,,\qquad R_n(z)=\frac{b^2_n}{iz-a_n-R_{n+1}(z)}\,,
\ee
where we defined $R_n(z)\equiv b_n \Psi_n(z)/\Psi_{n-1}(z)$. These equations provide a continued fraction expansion of $\Psi_0(z)=S(z)^*$. It is easy to check that the coefficients of different powers of $z^{-k}$, as $z\to\infty$, are  polynomials of Lanczos coefficients that equal the moments $\mu_k$. The Laplace transform also sheds light on the allowed growth of return amplitudes, since convergence of the transform requires functions that do not grow faster than exponentially $|\psi_0(t)|\lesssim Ce^{\alpha t}$, as $t\to\infty$.

There is yet another way of looking at the recursion method that shows the connection to the spectrum of the  Hamiltonian and the support of the initial state on Hamiltonian's eigenstates. Namely, we can define the Krylov basis vectors $\ket{K_n}$ as polynomials $P_n(H)$, which we will call {\it Krylov polynomials}, of degree n in the Hamiltonian $H$ acting on the initial sate
\be
\ket{K_n}\equiv P_n(H)\ket{\psi_0}\,.
\ee
In terms of these polynomials, \eqref{LanczosAlg} becomes a three-term recursion relation for the Krylov polynomials
\be
HP_n(H)=a_nP_n(H)+b_nP_{n-1}(H)+b_{n+1}P_{n+1}(H)\,,\label{Poly3Term}
\ee
with $P_0(H)=1$ and $b_0=0$.  This approach  opens a path to a  family of analytical solutions for specific choices of $a_n$'s and $b_n$'s for which \eqref{Poly3Term} coincides with known three-term recursion relations for families of classical orthogonal polynomials \cite{LanczosBook,Muck:2022xfc}.  We will call this the {\it polynomial method} (see \cite{Kar:2021nbm,Muck:2022xfc,Muck:2024fpb,Chhetriya:2025ndi,Alishahiha:2024vbf} for recent applications).

Favard's theorem \cite{chihara2011introduction,koornwinder2013orthogonal} guarantees existence of a positive measure on $\mathbb{R}$ with respect to which the polynomials satisfying \eqref{Poly3Term} are orthogonal. Specifically, we  define the measure $\mu(E)$ by
\be
\langle \psi_0|f(H)|\psi_0\rangle\equiv \int d\mu(E)f(E)\,,
\ee
where the right hand side is understood as a Riemann–Stieltjes integral. The orthonormality of the Krylov basis vectors is then equivalent to
\be
\langle K_n|K_m\rangle=\langle \psi_0|P_n(H)P_m(H)|\psi_0\rangle=\int d\mu(E)P_n(E)P_m(E)=\delta_{n,m}\,.\label{PnPm}
\ee
This resembles standard formulae for systems of orthogonal polynomials, but we should remember that the measure here  depends on the initial state
\be
\frac{d\mu(E)}{dE}=\rho(E)=\sum_n\delta(E-E_n)|\langle E_n|\psi_0\rangle|^2\,,\label{MeasureGen}
\ee
where the sum is over the entire spectrum of the Hamiltonian.  In other words, the measure is a product of the density of states with the support on the energy eigenstates on the initial state.  So we are really talking here about infinite families of  polynomials that are orthogonal in different measures defined by the initial states.

Suppose a generic, normalized initial state admits an expansion in the energy basis
\be
\ket{\psi_0}=\sum_{k}c_k\ket{E_k}\,,\qquad \sum_k|c_k|^2=1\,,\label{InStateEn}
\ee
with coefficients $c_k$ that can be zero for some $k$. Then we can rewrite \eqref{PnPm} as
\be
\langle K_n|K_m\rangle=\sum_k |c_k|^2P_n(E_k)P_m(E_k)=\delta_{n,m}\,.\label{OrthoPnPmD}
\ee
In this formulation of the recursion, the return amplitude is again useful. It is written as
\be
S(t)=\langle \psi_0|e^{iHt}|\psi_0\rangle=\int d\mu(E)e^{iEt}\,,
\ee
so that its Laplace transform becomes
\be
S(z)=\int d\mu(E)\frac{1}{z-iE}=\langle\psi_0|\frac{1}{z-iH}|\psi_0\rangle\,.
\ee
This expression is related to the resolvent\footnote{More precisely, to its average in the initial state $\ket{\psi_0}$, hence the notation $G_0$.}
\be
G_0(z)=iS(iz)\,,\qquad \rho(E)=\frac{1}{2\pi i}\lim_{\epsilon\to 0}\left[G_0(E-i\epsilon)-G_0(E+i\epsilon)\right]\,.
\ee
Here, we implicitly assumed that the spectrum of the Hamiltonian is discrete but in general we will have both, discrete and continuous parts of the spectrum (see, e.g., \cite{marino2021advanced}). 

Moreover, we can  write the  Krylov wave functions 
\be
\psi_n(t)=\langle \psi_0|P_n(H)e^{-iHt} |\psi_0\rangle=\int d\mu(E)P_n(E)e^{-iEt}\,,
\ee
with the boundary condition $\psi_n(0)=\delta_{n,0}$ (which also follows from \eqref{PnPm}), and their Laplace transform 
\be
\Psi_n(z)=\int d\mu(E)\frac{P_n(E)}{z+iE}\,.
\ee

Next, we can bridge the algorithmic approach with the Krylov polynomials:  the Lanczos algorithm  provides an explicit form for the $P_n(H)$
\be
P_n(H)=\left(\prod^n_{i=1}b_i\right)^{-1}\det(H I_n-h_n)\,,
\label{eq:DetP}
\ee
where $h_n$ are $n\times n$ tri-diagonal (sub)matrices of Lanczos coefficients representing $H$ in the Krylov basis
\be
h_1=(a_0)\,,\qquad h_2=\left(
\begin{array}{cc}
 a_0 & b_1 \\
 b_1 & a_1 \\
\end{array}
\right)\,,\qquad h_3=\left(
\begin{array}{ccc}
 a_0 & b_1 & 0 \\
 b_1 & a_1 & b_2 \\
 0 & b_2 & a_2 \\
\end{array}
\right)\,,...\,.
\ee
This expression allows us to formally write the Krylov basis vectors using the energy and initial state data. Since we can expand the determinant 
\be
\det(\lambda I_n-A)=\sum^{n}_{k=0}\lambda^{n-k}(-1)^k\chi_{(1^k)}(A)\,,\label{detSch}
\ee
where $\chi_{(1^k)}(A)$ are Schur polynomials labeled by single column Young tableaux with k-boxes.\footnote{Such polynomials compute ``sub-determinants'' -- see \cite{Balasubramanian:2005mg} for an example.}   Using \eqref{InStateEn} we can write
\be
\ket{K_n}=P_n(H)\ket{\psi_0}=\sum_{p}c_pP_n(E_p)\ket{E_p}\equiv \sum_{p}C_{n,p}\ket{E_p}\,,
\label{eq:Kn}
\ee
where
\be
C_{n,p}=c_p\left(\prod^n_{i=1}b_i\right)^{-1}\sum^n_{k=0}(-1)^k\chi_{(1^k)}(h_n)E^{n-k}_p\,.
\label{eq:Kn-coe}
\ee

Finally, given the above formulation, we define  spread complexity as the average position of the hopping particle on the 1D chain in the probability distribution $p_n(t)=|\psi_n(t)|^2$. More precisely
\be
C_K(t)=\langle n\rangle=\sum^{\mathcal{K}-1}_{n=0} n|\psi_n(t)|^2=\langle \psi(t)|K|\psi(t)\rangle\,,\label{SpreadComplexityF}
\ee
where, in the last step, we formally defined a complexity operator
\be
K=\sum^{\mathcal{K}-1}_{n=0}n\ket{K_n}\bra{K_n}\,.
\ee
This definition, and recursion procedure, was first employed to generalize the notion of the operator size, pioneered in \cite{Roberts:2014isa} (and in the SYK model in \cite{Roberts:2018mnp,Qi:2018bje}), to arbitrary quantum many-body systems by \cite{Parker:2018yvk}. The operator size is often simply called Krylov complexity (see the reviews \cite{Nandy:2024evd,Rabinovici:2025otw,Baiguera:2025dkc}). The generalization of the recursion to the spread of states (spread complexity) was carried out in \cite{Balasubramanian:2022tpr}, where the precise connection to complexity measures defined by the minimization of a cost function over choices of basis was also explained.\\ 
The physical connection between these two approaches, and  subtleties in the relationship, can be studied by treating the evolution of density matrices  $\rho(t)=\ket{\psi(t)}\bra{\psi(t)}$ \cite{Caputa:2024vrn} in the operator size framework.

One lesson from  analytical and  numerical studies of spread (and Krylov) complexity in chaotic systems, including studies in random matrix models \cite{Rabinovici:2022beu,Balasubramanian:2022tpr,Balasubramanian:2022dnj,Balasubramanian:2023kwd,Erdmenger:2023wjg} and billiard systems \cite{Hashimoto:2023swv,Camargo:2023eev,Balasubramanian:2024ghv}, is its characteristic behavior under time evolution. Namely, spread complexity starts with quadratic initial growth, followed by a linear ramp up to a peak that is exponential in the entropy, followed by a slope down to a plateau at very late times. If the Hilbert space is infinite-dimensional, the initial quadratic growth is just followed by monotonic growth with time. In the rest of this work, we will explore how these features depend on the initial state,  the  Hamiltonian, and their variations. 
   
For computations, it will be useful to write  spread complexity directly in terms of the Krylov polynomials
\be
C_K(t)=\sum^{\mathcal{K}-1}_{n=0}n\int d\mu(E)d\mu(E')P_n(E)P_n(E')e^{-i(E-E')t}\,,
\ee
which, using \eqref{MeasureGen}, we can also write as
\be
C_K(t)=\sum_{k,l}\sum_nnP_{n}(E_k)P_n(E_l)e^{i(E_k-E_l)t}|c_k|^2|c_l|^2\,.
\label{eq:gen-Ck}
\ee
We will sometimes discuss averages of complexity over  ensembles of theories, denoted $\langle C_K(t)\rangle$.  We will carry out these averages  by integrating over a distribution of energy (or energy differences) or by taking average over many Hamiltonians. In addition, we will write  the time average of spread complexity as
\be
\overline{C_K(t)}\equiv\frac{1}{t}\int^t_0C_K(t')dt'\,.\label{TAComplexity}
\ee
In particular, using the diagonal part of \eqref{eq:gen-Ck} as well as \eqref{eq:DetP}, we can show that the infinite time average satisfies
\begin{eqnarray}
    \overline{C_K(\infty)} &\equiv& \lim_{t\to \infty} \overline{C_K(t)}=\sum_{n=0}^{\mathcal{K}-1} n\, \sum_k |c_k|^4P_n(E_k)^2\nn\\
    &=&\sum_{n=0}^{\mathcal{K}-1} n\, \sum_k |c_k|^4\,\frac{\left(\text{det}(E_kI_n - h_n)\right)^2}{\left(\prod^n_{i=1}b_i\right)^{2}}\,.
\label{eq:T-ave}
\end{eqnarray}
Just as the initial state defines the probability distribution $\tilde{p}_k = |c_k|^2$, we could introduce a second distribution
\begin{equation}
  \tilde{q}_k =\frac{|c_k|^4}{\hat{Z}} = \frac{(\tilde{p}_k)^2}{\hat{Z}}\,, \qquad \text{with} \qquad \sum_ k |c_k|^4 = \hat{Z}\,.
\end{equation}
Since from \eqref{OrthoPnPmD} (for $n=m$) we have the identity $\left(\prod_{i=1}^{n}b_i\right)^2=\sum_k \tilde{p}_k\,\left(\text{det}(E_kI_n - h_n)\right)^2$,  the time-averaged spread complexity can be written as\footnote{This appears to be a generalization of the plateau value found for the TFD (or the spectral form factor $Z(2\beta)/(Z(\beta))^2$ \cite{brezin1997spectral,Guhr:1997ve,Cotler:2016fpe}) to an arbitrary initial pure state.}
\begin{equation}
    \overline{C_K(\infty)}= \overline{\sum_n n |\psi_n(\infty)|^2} = \hat{Z}\,\sum_{n=0}^{\mathcal{K}-1} n\, \frac{\langle \mathcal{H}_{k,n}\rangle_q}{\langle \mathcal{H}_{k,n}\rangle_p}\,, 
\end{equation}
where $\mathcal{H}_{k,n} = \left(\text{det} (E_k I_n -h_n)\right)^2$ and $\langle \mathcal{H}_{k,n}\rangle_p=\sum_k p_k \mathcal{H}_{k,n}$. Appendix~\ref{sec:A0} shows a simple example where the definitions and steps above are worked out explicitly.

Finally, the Krylov approach  maps the quantum dynamics of operators or states into a probability distribution $p_n(t)$ that, in general, contains more information than just the complexity \eqref{SpreadComplexityF}.  Standard information-theoretic probes can  extract this information. A commonly used one is the Shannon entropy, sometimes called K-entropy \cite{Barbon:2019wsy}
\be
S_K(t)=-\sum^{\mathcal{K}-1}_{n=0}p_n(t)\log p_n(t)\,,
\ee
and its R\'enyi generalizations. Before proceeding, we introduce two more tools that will play important roles in our work.
\subsection{Relative Krylov Entropy}\label{sec:RelativeK}
We will want to compare complexity and dynamics over Krylov subspaces after small variations of parameters. To compare different probability distributions over subspaces of the same dimension $\mathcal{K}$, we can define\footnote{See \cite{Caputa:2021sib,Patramanis:2021lkx} for early discussions on the relative entropy in the Krylov complexity context.} the {\it Relative Krylov Entropy (RKE)} as the Kullback-Leibler (KL) divergence in the Krylov basis
\be
S_K(p\Vert q)=\sum^{\mathcal{K}-1}_{n=0}p_n(t)\log\left(\frac{p_n(t)}{q_n(t)}\right)\,,\label{eq:RelKEntr}
\ee
where the probabilities
\be
\sum^{\mathcal{K}-1}_{n=0}p_n(t)=\sum^{\mathcal{K}-1}_{n=0}|\psi_n(t)|^2=1\,,\qquad\sum^{\mathcal{K}-1}_{n=0}q_n(t)=\sum^{\mathcal{K}-1}_{n=0}|\phi_n(t)|^2=1\,,
\ee
are defined for two initial states that explore Krylov subspaces of the same dimension
\be
\ket{\psi(t)}=\sum^{\mathcal{K}-1}_{n=0}\psi_n(t)\ket{K_n}\,,\qquad \ket{\phi(t)}=\sum^{\mathcal{K}-1}_{n=0}\phi_n(t)\ket{\tilde{K}_n}\,.
\ee
Recall here that the Krylov basis is ordered, so that there is an unambiguous correspondence between $\ket{K_n}$ and $\ket{\tilde{K}_n}$. 
Again, in general, $\mathcal{K}$ can be finite or infinite, but to evaluate the relative entropy in this formulation we  need two probability distributions of the same dimension. If the dimensions of the two Krylov subspaces satisfy $\rm{dim}(\{ \ket{K_n} \}) < \rm{dim}(\{ \ket{\tilde{K}_n} \})$ we can still define $S_K(p\Vert q)$ by assigning $p_n(t)=0$ for $\rm{dim}(\{ \ket{K_n} \}) < n \leq \rm{dim}(\{ \ket{\tilde{K}_n} \})$.   However, in this case the relative entropy with the reversed order of arguments,  $S_K(q\Vert p)$, will not be well defined.  Indeed, recall that the relative entropy between distributions is in general not symmetric between its arguments.

To summarize, the RKE measures how much more or less a perturbed state spreads in the Hilbert space as compared to an unperturbed reference state. In our explicit computations in the following sections, we will employ this quantity to compare two probability distributions on the Krylov chains of the same dimension,  obtained by evolving the same initial state with two different Hamiltonians.
 
Note that the Kullback-Leibler divergence between probabilities $p_n(t)$ and $q_n(t)$ on the Krylov chains with sites $n = 0, \ldots, \mathcal{K}-1$, and hence the RKE, is a natural measure of distinguishability between these distributions.
Indeed, in classical information theory KL divergence quantifies the expected log-likelihood ratio between the two distributions and is always non-negative, vanishing if and only if $p_n(t) = q_n(t)$ for all $n$ (and $t$).  By Pinsker’s inequality, the $L^1$ distance between these probabilities is bounded by the KL divergence
\be
\sum_n |p_n(t) - q_n(t)| \le \frac{1}{2}S_K(p \Vert q)\,.
\ee
Hence, a small KL divergence implies that the two distributions are statistically hard to distinguish in any single-shot measurement \cite{cover1999elements}. 

Intuitively, $S_K(p \Vert q)$ measures the information lost when $q_n(t)$ is used to approximate $p_n(t)$, or equivalently, how strongly data drawn from $p_n(t)$ favor the hypothesis that the underlying distribution is $p_n(t)$ rather than $q_n(t)$. The operational meaning of $S_K(p \Vert q)$ becomes clear in the framework of hypothesis testing. Suppose one wishes to decide between two competing hypotheses: $h_0$, that data are drawn from $q_n(t)$, or $h_1$, that they are drawn from $p_n(t)$. The optimal strategy, given by the Neyman-Pearson lemma \cite{neyman1933ix}, is to choose $h_1$ when the likelihood ratio $p_n/q_n$  exceeds one half. In a single trial, there is always a non-zero probability of error, but if the test is repeated independently $m$ times, the probability of mistakenly identifying $p_n$ as $q_n$ decreases asymptotically as $\exp[-m S_K(p \Vert q)]$. So the KL divergence sets the fundamental rate at which evidence accumulates in favor of the correct hypothesis, providing an information-theoretic measure of distinguishability between probability distributions.

\subsection{Koherence: the  entropy of coherence between Krylov bases}\label{sec:OverlapEntropy}
When studying variations in  spread complexity, we compare two different Krylov bases: one involving $\ket{K^{(0)}_n}$, with dimension $\mathcal{K}^0$, and a perturbed one with $\ket{K_m}$ of dimension $\mathcal{K}$. Assume, without loss of generality, that $\mathcal{K} \leq \mathcal{K}^0$ so that any $\ket{K_m}$ can be expanded in the original basis
\be
\ket{K_m}=\sum_n\langle K^{(0)}_n|K_m\rangle\ket{K^{(0)}_n}\,.
\label{eq:b-change}
\ee
By definition, for any fixed $m$, the quantity $\mathrm{P}^m_n\equiv |\langle K^{(0)}_n|K_m\rangle|^2$ defines a probability distribution because
\be
\langle K_m|K_m\rangle=\sum_{n,n'}\overline{\langle K^{(0)}_n|K_m\rangle}\langle K^{(0)}_{n'}|K_m\rangle\langle K^{(0)}_n|K^{(0)}_{n'}\rangle=\sum_n|\langle K^{(0)}_n|K_m\rangle|^2=1.
\ee

One way of describing the change between the two solutions to the Krylov algorithm is to determine how spread each new basis vector is in the old basis.
Indeed, if two given basis vectors match, the probability distribution  $\mathrm{P}^m_n$ will be localized.
At the other end, a Krylov  basis vector of the perturbed system could be uniformly distributed in the old basis. We can quantify the spread for any basis vector $\ket{K_m}$ in terms of the Shannon entropy of the distribution $\mathrm{P}^m_n$,
\be
S^{(m,0)}_K\equiv-\sum_n|\langle K^{(0)}_n|K_m\rangle|^2\log|\langle K^{(0)}_n|K_m\rangle|^2\,.
\label{eq:K-coh}
\ee
To quantify the total divergence between the two finite-dimensional basis, we can  define the average
\be
\overline{S^{(0)}_K}=\frac{1}{\mathcal{K}}\sum^{\mathcal{K}-1}_{m=0}S^{(m,0)}_K=\frac{1}{\mathcal{K}}\sum^{\mathcal{K}-1}_{n,m}|\langle K^{(0)}_n|K_m\rangle|^2\log|\langle K^{(0)}_n|K_m\rangle|^2\,.
\label{eq:av-K-coh}
\ee

In our explicit examples, we will encounter Koherence that grows logarithmically with time (see $\mathrm{SL}(2,\mathbb{R})$ in next section) and, for presentation, it will actually be helpful to plot the exponential of \eqref{eq:K-coh}.  

The measure \eqref{eq:K-coh} can be understood as a quantification of the {\it coherence} of the vector $\ket{K_m}$ in the old basis $\ket{K_n^0}$. In fact, the entropy in \eqref{eq:K-coh} is precisely the {\it relative entropy of coherence} \cite{baumgratz2014quantifying} for the special case of pure states.\footnote{The relative entropy of coherence is a proper monotone, i.e., it is non-negative and vanishes only for  states which are incoherent in a given basis.  The quantity is basis dependent in the sense that coherence is measured related to a reference basis.}  
Specifically, given a density matrix $\rho$ and a  $\mathcal{K}$-dimensional basis $\{B_n\}$, the relative entropy of coherence is defined as \cite{baumgratz2014quantifying}
\be
C_r(\rho)=S(\rho^B_{\text{diag}})-S(\rho)=S(\rho\Vert \rho^{B}_{\text{diag}})\,,
\label{eq:rel-coh}
\ee
where $S(\rho)=-\Tr(\rho\log(\rho))$ and $S(\rho\Vert \rho^{B}_{\text{diag}})$ are the von-Neumann and relative entropies respectively, and
\be
\rho^{B}_{\text{diag}}\equiv\sum^{\mathcal{K}-1}_{n=0}\langle B_n|\rho|B_n\rangle|B_n\rangle\langle B_n|\,.
\ee
In other words, \eqref{eq:rel-coh} measures the difference in von Neumann entropies between the depolarised density matrix in the $\ket{B_n}$ basis and the original $\rho$. Hence, it quantifies the information-theoretic distinguishability between a quantum state and its decohered counterpart (with off-diagonal elements removed), measuring how much information (or quantum uncertainty) is lost when coherence is destroyed.

In our discussion, we identify $\ket{B_n}$ with the reference Krylov basis $\ket{K_n^0}$. For any $m$ we also define a density matrix $\rho =\ket{K_m}\bra{K_m}$. Since the latter describes a pure state, $S(\rho)=0$. It follows that
\be
S^{(m,0)}_K=C_r(\ket{K_m}\bra{K_m}).
\ee
So, we will refer to this  entropy in \eqref{eq:K-coh}  as the ``entropy of coherence between Krylov bases'', or {\it Koherence} for short.  Note that larger Koherence implies greater delocalization of the new Krylov basis in the old one.  Equivalently, the new basis elements require a more quantum coherent description in the old basis -- i.e., the associated density matrices are more distant from their decohered versions.

We are going to consider settings in which the initial state $\ket{K_0}$ is perturbed relative to the reference $\ket{\tilde{K}_0}$.  As we discussed, subsequent elements of the Krylov chain $\ket{K_{n>0}}$ are reached by hopping from the initial state.  Thus, elements further down the chain have increasing support as time passes.   So by measuring the Koherence of $\ket{K_n}$ for larger $n$ relative to the Koherence of $\ket{K_0}$, we can study the degree to which differences in initial conditions are amplified or damped by the dynamics.  The {\it mean Koherence} in \eqref{eq:av-K-coh} computes this divergence on average over the entire perturbed Krylov basis, and thus on average over the dynamics.

To summarize, {\it Koherence} \eqref{eq:K-coh} and  {\it mean Koherence} \eqref{eq:av-K-coh}  quantify the dynamical amplification or damping  of differences in initial conditions by measuring the spread of  perturbed Krylov basis elements in the reference basis. 

\subsection{General results}\label{sec:1stlaw}
To characterize the effect that a change in the initial state has on the Krylov basis and spread complexity, suppose first that the state can be characterized by the expectation values of some  set of operators $\{ \mathcal{O}_i \}$.  Then we could try to express the variation in the complexity as a sum:
\begin{equation}
\begin{aligned}
    \delta C_K(t) \equiv \sum_n n \left(|\tilde\psi_n(t)|^2- |\psi_n(t)|^2\right) &=   \sum_i \nu_i(t) \left(\langle K^\prime_0|\mathcal{O}_i|K_0^\prime\rangle - \langle K_0|\mathcal{O}_i| K_0 \rangle\right)\\
    & 
    \equiv \sum_i \nu_i(t)\,\delta \mathcal{O}_i \,.
\end{aligned}
\end{equation}
Alternatively, if we parametrize a family of initial states by $\lambda_i$, we could also write:
\begin{equation}
  \delta \mathcal{C}_{\mt{K}}(t) = \sum_i \frac{\partial \mathcal{C}_{\mt{K}}}{\partial \lambda_i}(t)\,\delta \lambda_i\,. 
  \label{eq:lambdaExp}
\end{equation}
To map between these formulations we could use quantum state tomography to identify the state in terms of its observable (see, e.g., \cite{DAriano:2003txn,PhysRevA.64.052312}).

In our case, given the special role played by the energy eigenbasis $|E_i\rangle$ in computing  time evolution of states in quantum mechanics, we will expand the initial state  as
\begin{equation}
  |K_0\rangle = \sum_k c_k |E_k\rangle\,.
\end{equation}
This expansion determines the Krylov subspace and a set of probabilities $\tilde{p}_i\equiv|c_i|^2$, which we identify with the set  $\{\lambda_i\}$ in \eqref{eq:lambdaExp}.

Recall that  spread complexity is entirely determined by the moments of the Hamiltonian in the initial state. Consequently, any variation in the spread complexity can be decomposed in terms of the  variations of these moments
\begin{equation}
    \delta H^n \equiv \langle K_0^\prime |H^n|K_0^\prime \rangle - \langle K_0 |H^n| K_0\rangle\,,
\end{equation}
where $\ket{K_0^\prime} = \sum_i c_i^\prime |E_i\rangle$ with $\tilde{p}_i^\prime = \tilde{p}_i + \delta \tilde{p}_i$. Alternatively, using quantum state tomography, it should be possible to reconstruct the change in the probability amplitudes  in terms of  measurements of the Hamiltonian moments. Either way, 
\begin{equation}
    \delta \mathcal{C}_{\mt{K}}(t) = \sum_n \nu_n(t)\,\delta H^n\,,
\label{eq:uni-1law}
\end{equation}
This expression defines a set of dimensionful, time-dependent ``chemical potentials'' $\nu_n(t)$, that repackage the energy gaps within the Krylov subspace. We will refer to this expression as a ``first law'' of spread complexity.\footnote{This sort of first law relating variations in a macroscopic quantity to variations in the underlying parameters at the leading order has been considered for entanglement \cite{Bhattacharya:2012mi,Blanco:2013joa},  circuit complexity \cite{Bernamonti:2019zyy}, and general quantum resources \cite{Sparaciari:2020xfm}.}
We will next use  general features of  spread complexity at short and long times, to infer some features of the $\nu_n(t)$.

\paragraph{Short time considerations.} Spread complexity is an even function of time. Indeed, rewriting \eqref{eq:gen-Ck} as
\begin{equation}
\begin{aligned}
    C_K(t) &= \sum_{n=0}^{\mathcal{K}-1} n\,\sum_k |c_k|^4\,\left(P_n(E_k)\right)^2 \\
    & + 2 \sum_{k<j} \sum_{n=0}^{\mathcal{K}-1} n\,|c_k|^2|c_j|^2\,P_n(E_k)\,P_n(E_j) \cos ((E_k-E_j)t)\,,
\end{aligned}
\label{eq:even-Ck}
\end{equation}
shows that the first diagonal term matches the infinite time average $\overline{C}_K(\infty)$ (see \eqref{eq:T-ave}), and that the temporal oscillations are even functions of the  set of energy gaps $\Delta_{kj}\equiv E_k - E_j$ within the Krylov subspace.

At short times, such that $\Delta_{kj}\,t \ll 1\,,$ $\forall\,k\neq j$, we can Taylor expand \eqref{eq:even-Ck} as
\begin{equation}
    C_K(t) = \sum_{r=1}^\infty (-1)^r\frac{t^{2r}}{(2r)!} \sum_n n \sum_{i=0}^{2r} \binom{2r}{i} \langle K_n|H^i|K_0\rangle \langle K_0|H^{2r-i}|K_n\rangle\,,
\end{equation}
where we used the  definition of the $P_n(E_k)$ polynomials to rewrite rewrite the expression in terms of transition amplitudes of different powers of the Hamiltonian of the system. The tri-diagonal action of the Hamiltonian in the Krylov basis implies that the short-time expansion depends on the energy moments, as argued before.

Working up to $\mathcal{O}(t^6)$, we get
\begin{equation}
    C_K(t) = b_1^2 t^2 + \frac{t^4}{12}\left(2\tilde{\mu}_4 - 3\frac{\tilde{\mu}_3^2}{b_1^2}-6b_1^4 \right) + \mathcal{O}(t^6)\,,
\label{eq:short-t}
\end{equation}
where
\begin{equation}
    \tilde{\mu}_k \equiv \langle K_0 |\left(H - a_0\right)^k |K_0\rangle\,,
\end{equation}
gives the higher-order centered energy moments in the initial state $\ket{K_0} = \ket{\psi_0}$. Also, from the Lanczos algorithm \eqref{LanczosAlgGen}, $b^2_1$ equals the variance of the Hamiltonian in $\ket{K_0}=\ket{\psi_0}$: 
\begin{equation}
b_1^2=\langle \psi_0|(H-a_0)^2|\psi_0\rangle \, .
\end{equation}
Thus, as time evolves, spread complexity becomes more sensitive to finer-grained data on the initial energy distribution.

The short-time expansion \eqref{eq:short-t} is compatible with the Mandelstam-Tamm bound \cite{mandelstam1945uncertainty} for pure states, i.e., the minimal time $\tau_\perp$ for a quantum state to evolve into an orthogonal state\footnote{See \cite{Hornedal:2022pkc} for applications to Krylov complexity and speed limits for operator growth.}
\begin{equation}
  \tau_\perp \geq \frac{\pi\,\hbar}{2b_1}\,.
\label{eq:MT-bound}  
\end{equation}
As time evolves, our state may  be in any of the other orthogonal states $\ket{K_n}$. Due to the tri-diagonal form of H in the Krylov basis, the variance of the energy in these states equals
\begin{equation}
  \langle K_n | \left(H - \langle K_n|H|K_n\rangle\right)^2 | K_n\rangle = b_n^2 + b_{n+1}^2\,, \qquad n\geq 1\,.
\end{equation}
These variances involve higher energy moments in the initial state $\ket{K_0}$. 
They correspond to  time scales that probe the fine-grained structure of the energy distribution in $\ket{K_0}$ and govern the higher-order terms in the short-time expansion of the spread complexity~\eqref{eq:short-t}.

Since \eqref{eq:short-t} is universal, we can also derive a universal first law of spread complexity at early times
\begin{equation}
    \delta C_K(t) = t^2\left[1 + \frac{t^2}{4}\left(\frac{\tilde{\mu}_3^2}{\tilde{\mu}_2^2}-4 \tilde{\mu}_2\right) \right]\,\delta \tilde{\mu}_2 + \frac{t^4}{6}\left(\delta \tilde{\mu}_4 - 3\frac{\tilde{\mu}_3}{\tilde{\mu}_2}\,\delta\tilde{\mu}_3 \right) + \mathcal{O}(t^6,\tilde{\mu}_k)\,.
\end{equation}
This equation captures two main features. First, at any new order, there is a new independent higher order moment entering the first law. Second, such higher order contributions also involve combinations of the lower moments. Thus, the chemical potentials $\nu_n(t)$ are polynomials in $t^2$ (in the short time expansion) with non-trivial dependence on the $\tilde{\mu}_k$ evaluated in the original state $\ket{K_0}$\footnote{The constructive nature of the Krylov algorithm guarantees that we can work out these specific polynomials to any order $t^{2k}$.}.

\paragraph{Late times.} The expression in  \eqref{eq:even-Ck} is a linear combination of  functions with periodicities set by the energy gaps $\Delta_{kj}$ in the Krylov subspace. These oscillations have amplitudes determined by the Krylov polynomials $P_n(E_k)$ and the probability distribution $\tilde{p}_i=|c_i|^2$ coming from the initial state. Formula \eqref{eq:even-Ck} shows that these oscillations fluctuate around the time-average of the spread complexity 
\begin{equation}
   \overline{C_K(\infty)} = \sum_{n=0}^{\mathcal{K}-1} n\,\sum_k |c_k|^4\,\left(P_n(E_k)\right)^2  \,.
\end{equation}
Since this depends on all polynomials $P_n(E_k)$ within the Krylov subspace, it depends on all energy moments evaluated on $\ket{K_0}$. It follows that the methods above would allow us to compute the change in the plateau value with the change in the original state. Unfortunately, the dependence on the state is convoluted and it is hard to extract  universal information. So below we will instead discuss instructive examples to understand how variations in the state and Hamiltonian affect the late time spread complexity.

\paragraph{Relation to Koherence.} The relation~\eqref{eq:uni-1law} may at first seem unrelated to Koherence which we defined in Sec.~\ref{sec:OverlapEntropy}. However, the two quantities are in fact related, as we explain below.

First, let us compare the wave functions determining the spread complexities. Consider two different initial states
\begin{equation}
    |K_0\rangle = \sum_k c_k |E_k\rangle\,, \qquad \qquad |K_0^\prime\rangle = \sum_\alpha c^\prime_\alpha |E_\alpha \rangle\,.
\label{eq:start0}
\end{equation}
By construction, the time evolution $|\psi^\prime(t)\rangle$ of the perturbed state $|K^\prime_0\rangle$ is
\begin{equation}
    |\psi^\prime(t)\rangle = \sum_n \psi^\prime_n(t) | K^\prime_n\rangle\,.
\end{equation}
Assuming the perturbed Krylov subspace is within the original one, the perturbed wave functions can be written as
\begin{equation}
    \psi^\prime_n(t) = \sum_{s,r} \langle K_n^\prime|K^0_s\rangle \langle K_r^0|K^\prime_0\rangle\,\langle K_s^0|e^{-itH}|K_r^0\rangle\,.
\label{eq:wave-basis-change}
\end{equation}
Thus, the set of $|\psi^\prime_n(t)|^2$ encoding the spread complexity of the perturbed state is determined by the matrix elements \eqref{eq:b-change} defining the probability distribution controlling Koherence \eqref{eq:K-coh}, together with the transition amplitudes $\langle K_s^0|e^{-itH}|K_r^0\rangle$ within the original Krylov subspace. Hence, while Koherence carries information about the perturbation, the transition amplitudes will be determined by the set of unperturbed Krylov coefficients given the tri-diagonal action of the Hamiltonian in the Krylov basis. This information is repackaged in the chemical potentials $\nu_n(t)$.

Second, let us compute some explicit low-order Krylov vector overlaps to explicitly see how they encode  information about the variation of the Hamiltonian moments, as expected from the first law \eqref{eq:uni-1law}. Consider the two initial states \eqref{eq:start0}.  
Split the energy label $\alpha$ of the perturbed state into a set $k$ that labels levels that also appear in $\ket{K_0}$ and a set $r$ that do not.  Then we can write
$c^\prime_k = c_k + \delta c_k$ and $c^\prime_r = \delta c_r$, and find that 
\begin{equation}
    \langle K^\prime_0 |K_0\rangle = \sum_k (c^\prime_k)^*\,c_k = 1 + \sum_k (\delta c_k)^*\,c_k\,.
\end{equation}
Geometrically, this is the projection of the perturbed state into the original one in the space of quantum states. 

Next, we can use the first step of the Lanczos algorithm
\begin{equation}
    |K_1 \rangle = \frac{1}{b_1}\sum_k c_k (E_k - a_0)|E_k\rangle\,, \qquad \qquad |K_1^\prime \rangle = \frac{1}{b^\prime_1}\sum_\alpha c_\alpha (E_\alpha - a^\prime_0)|E_\alpha\rangle\,,
\end{equation}
to compute the overlap 
\begin{equation}
    \langle K_1^\prime|K_1\rangle = \frac{1}{b_1b_1^\prime} \sum_k (c_k^\prime)^* c_k (E_k-a_0^\prime)(E_k-a_0)\,.
\end{equation}
To unpack the physical significance, let us compute the Krylov coefficients $a_0^\prime$ and $b_1^\prime$ appearing above, and relate them to $a_0$ and $b_1^2$. For example,
\begin{equation}
    a_0^\prime = a_0 + \sum_k \delta \tilde{p}_k\,E_k + \sum_r \delta \tilde{p}_r\,E_r \equiv a_0 + \delta a_0\,.
    \label{eq:a0prime}
\end{equation}
where $\tilde{p}_\alpha = |c_\alpha|^2$.
Working at first order in the perturbation, we find
\begin{equation}
  b_1^\prime \approx b_1\left(1 + \frac{1}{2}\delta b_1^2\right)\,,
\end{equation}
with
\begin{equation}
    \delta b_1^2 =\frac{1}{b_1^2} \left(\sum_k \delta \tilde{p}_k (E_k-a_0)^2 + \sum_r \delta \tilde{p}_r (E_r-a_0)^2\right)\,.
\end{equation}
It follows that
\begin{equation}
    \langle K_1^\prime|K_1\rangle \approx 1 + \frac{1}{b_1^2}\sum_k (\delta c_k)^* c_k\,(E_k-a_0)^2 - \delta a_0 - \frac{1}{2}b_1^2\,.
\end{equation}
Recalling the $b_1^2 = \tilde{\mu}_2$ and \eqref{eq:a0prime}, this short calculation confirms that the set of matrix elements $\langle K^\prime_n |K_m\rangle$  knows about the variations of the  energy moments $\delta H^n$, as they appear in \eqref{eq:uni-1law}, together with the phases turned on in $\delta c_k$ and $\delta c_r$.

\paragraph{Monotonicity of spread complexity.} 
We can also ask if there is a ``2nd law for spread complexity'', i.e., whether the first derivative is positive, at least in some circumstances. Within a finite-dimensional Krylov space and assuming no degeneracies in the spectrum, the spread complexity \eqref{eq:even-Ck} has oscillating behavior at long times. Hence, it does not  have a positive first derivative. However, if the Krylov subspace is infinite-dimensional the spread complexity can grow monotonically -- we will see an explicit example in section \ref{sec:SL2}. This suggests that we should study the variation in spread complexity with the dimension $\mathcal{K}$ of the Krylov subspace, as well as the $\mathcal{K}\to \infty$ limit. Note that even in classical thermodynamics the second law if strictly speaking emergent in a coarse-grained, large system limit.  Finite systems can show oscillations, recurrences and other non-monotonic patters in the entropy. We will discuss this in Sec.~\ref{sec:SolvmodLin}.

We can also consider the entropy of the distribution of the time evolving state over the Krylov basis.  Since this is simply Shannon's entropy evaluated for the Krylov probability distribution $p_n(t) = |\psi_n(t)|^2$,
\begin{equation}
  \frac{dS_{\mt{K}}}{dt} = -\sum_n \dot{p}_n(t)\,\log p_n(t)
\end{equation}
where we used $\sum_n \dot{p}_n(t) = 0$.  If these time derivatives satisfy an analog of  Fermi's golden rule (suppressing explicit time dependence in $p_n(t)$)
\begin{equation}
  \dot{p}_n = \sum_m \nu_{nm}\left(p_m - p_n\right)\,, \qquad  \dot{p}_m = \sum_n \nu_{nm}\left(p_n - p_m\right)\,,
\label{eq:fermi-gr}
\end{equation}
then
\begin{equation}
  \frac{dS_{\mt{K}}}{dt} = \frac{1}{2} \sum_{\alpha,\beta} \nu_{\alpha\beta} \left(\log p_\beta - \log p_\alpha\right)\left(p_\beta - p_\alpha\right) \, .
\end{equation}
Then if $p_\beta < p_\alpha$, it follows that $\log p_\beta < \log p_\alpha$, so that the product $\left(\log p_\beta - \log p_\alpha\right)\left(p_\beta - p_\alpha\right)$ will be positive and $dS_K/dt > 0$. However, the derivatives $\dot{p}_n$ within the Krylov subspace satisfy the exact relation
\begin{equation}
  \dot{p}_n = ip_n\left(b_{n+1}\, \frac{\psi^*_{n+1}}{\psi^*_n} + b_n\,\frac{\psi^*_{n-1}}{\psi^*_n}\right) -ip_n \left(b_{n+1}\, \frac{\psi_{n+1}}{\psi_n} + b_n\,\frac{\psi_{n-1}}{\psi_n}\right)\,.
\end{equation}
These are neither of the form \eqref{eq:fermi-gr} nor one can simply prove the positivity of $\frac{dS_{\mt{K}}}{dt}$ in general. That said, there may be specific dynamics for which the necessary relations hold.

\section{Solvable examples: motion on group manifolds}\label{sec:SolvableEx}
Next, we proceed with exactly-solvable examples where the Krylov chain dynamics is governed by symmetries. In these examples Hamiltonian evolution from the initial state, once mapped onto the Krylov chain, can be described by geodesic motion on the associated Lie group manifold. In these cases, the Lanczos coefficients, extracted from the moments of the return amplitude, exhibit a structured pattern, leading to solvable recursion relations in the Krylov basis that  can be solved explicitly using coherent states \cite{Caputa:2021sib}, or equivalently Toda systems \cite{Dymarsky:2019elm} (see also \cite{Takahashi:2025iol}), or orthonormal polynomials \cite{Muck:2022xfc}. We will consider three examples: the $\SL(2,\mathbb{R})$, the $\SU(2)$, and the Heisenberg-Weyl groups for which the Krylov basis  is related to the Lie algebra basis. A more detailed discussion of these systems appears in \cite{Caputa:2021sib,Balasubramanian:2022tpr}.
\subsection{$\mathrm{SL}(2,\mathbb{R})$}
\label{sec:SL2}
We start with the SL(2,$\mathbb{R}$) algebra defined by commutators
\be
[L_{0},L_{\pm 1}]=\mp L_{\pm 1},\qquad [L_1,L_{-1}]=2L_0\,,
\ee
where $L_{+1}$ and $L_{-1}$ play the role of raising and lowering operators. 
Then we consider unitary time evolution
\be
\ket{\psi(t)}=e^{-iHt}\ket{\psi_0}\equiv e^{-iHt}\ket{z,h}\,,\label{InState}
\ee
by a Hamiltonian defined by a linear combination of the algebra generators
\be
H=\gamma L_0+\alpha (L_1+L_{-1})\, .\label{HamSL2R}
\ee
By varying $\gamma$ and $\alpha$ we obtain a parametrized family of Hamiltonians. This symmetry algebra and Hamiltonian can be represented on many different physical systems, and our analysis here applies to any realization.  We take the initial state to be $\ket{\psi_0}=\ket{z,h}$ defined as a generalized coherent state
of the SL(2,$\mathbb{R}$) algebra \cite{Perelomov:1971bd}. The coherent state is obtained by acting with a displacement operator on a highest weight state $\ket{h}$ (defined by $L_0\ket{h}=h\ket{h}$ and $L_1\ket{h}=0$) as
\be
\ket{z,h}=D(\xi)\ket{h}\equiv e^{\xi L_{-1}-\bar{\xi}L_1}\ket{h}\,,\label{CSSU11}
\ee
and can be parametrized by a complex number $\xi=\rho/2 e^{i\phi}$.  We can write $\rho$ and $\phi$ as a complex coordinate on the  Poincar\'{e} disc by instead defining 
\be
z=\frac{\xi}{|\xi|}\tanh(|\xi|)=\tanh\left(\frac{\rho}{2}\right)e^{i\phi},\qquad |z|<1\,.
\label{eq:poincareCoords}
\ee  
Below we will study the spread complexity of \eqref{InState}, and regard $\rho$ and $\phi$ as parameters of the initial state that can be varied. This will allow us to analytically derive the variations of the Lanczos coefficients, the Krylov basis, and spread complexity with respect to variations of these parameters.  

As explained above, the key object for the computation of the Lanczos coefficients is the return amplitude.  For motion on a group manifold, we can calculate this  amplitude explicitly by applying the Baker–Campbell–Hausdorff (BCH) formula. For the SL(2,$\mathbb{R}$) algebra, a short calculation leads to  (see Appendix~\ref{appena})
\be
S(t)=\bra{z,h}e^{iHt}\ket{z,h}=\left(\cosh\left(\frac{\mathcal{D}}{2}t\right)-\frac{if(z)}{\mathcal{D}}\sinh\left(\frac{\mathcal{D}}{2}t\right)\right)^{-2h},
\ee
in terms of the functions
\be
\mathcal{D}=\sqrt{4\alpha^2-\gamma^2},\qquad f(z)=\frac{2\alpha (z+\bar{z})+\gamma(1+|z|^2)}{1-|z|^2}\,.\label{Dandf}
\ee
Information about the initial state enters through $h$ and $f(z)$, whereas $\mathcal{D}$ only depends on parameters of the evolving Hamiltonian.  

After following the standard procedure to compute Lanczos coefficients from the moments, we find the infinite set of Lanczos coefficients
\be
a_n=f(z)(h+n)\,,\qquad b_n=\frac{\sqrt{\mathcal{D}^2+f(z)^2}}{2}\sqrt{n(n+2h-1)}\,.
\ee
The dependence on $n$ and $h$ is the same as reported in \cite{Balasubramanian:2022tpr}, but the overall coefficients are different, capturing the physical information about the family of initial states and  Hamiltonians that we are studying here.

\paragraph{Variations of Lanczos coefficients.}
Below we will denote Lanczos coefficients for $\rho=0$ (or equivalently $z=0$) as
\be
a^{(0)}_n=\gamma(h+n)\,,\qquad b^{(0)}_n=\alpha\sqrt{n(n+2h-1)}\,.\label{Lanczos0th}
\ee
Clearly, the variation of Lanczos coefficients with respect to $\rho$ or $\phi$ enters through the overall proportionality coefficients, but the algebraic structure, i.e., the dependence on $n$ is unaffected. For example, expanding to the second order in $\rho$ we have
\bea
f(z)&\simeq&\gamma+2\alpha\cos(\phi)\rho+\frac{\gamma}{2}\rho^2 \,,\\
\frac{\sqrt{\mathcal{D}^2+f(z)^2}}{2}&\simeq& \alpha+\frac{\gamma}{2}\cos(\phi)\rho+\frac{4\alpha^2+\gamma^2+(4\alpha^2-\gamma^2)\cos(2\phi)}{16\alpha}\rho^2\,.
\eea
We see that the original coefficients $(\gamma,\alpha)$ get mixed for non-zero $\rho$. Also, the direction along $\phi=\pi/2$ is singled out, namely the first-order variation vanishes for this angle and we only get contributions at $O(\rho^2)$. 

Since we have the exact form of the Lanczos coefficients for this family of Hamiltonians and states, we can also vary them explicitly with respect to the highest weight $h$ that we used to define the initial state. To first order around some initial value $h=h_0$ we have
\be
\delta a_n\sim\gamma(h-h_0)\,,\qquad \delta b_n\sim \frac{\alpha n}{\sqrt{n(n+2h-1)}}(h-h_0)\,.
\ee
When embedded in a conformally invariant system, the parameter $h$ may be related to the conformal dimension $\Delta$ of the operator that creates the state~\cite{Dymarsky:2021bjq}, or, for local quenches, to the central charge~\cite{Caputa:2025dep} $c$ of the 2D CFT. However,  this $h$ should not be confused with the chiral conformal weight of the CFT operator, for which the total conformal dimension is $\Delta = h + \bar{h}$. In the context of Krylov dynamics, we are considering  a single $\mathrm{SL}(2,\mathbb{R})$ algebra governing  evolution on the Krylov chain for suitable Hamiltonians and initial states, whereas in two-dimensional CFTs there are two copies of $\mathrm{SL}(2,\mathbb{R})$. In some dynamical settings, these two copies may effectively combine into a single $\mathrm{SL}(2,\mathbb{R})$ structure governing  the Krylov chain picture \cite{Caputa:2023vyr,Caputa:2024sux}.

More generally, observe that, in this example governed by the SL(2,$\mathbb{R}$)  algebra, we only need to fix three parameters to completely specify Lanczos coefficients. To see this more explicitly, recall that the moments of the return amplitude are defined as expectation values of  powers of the Hamiltonian in the initial state $\ket{K_0}=\ket{\psi_0}$ \eqref{MomRetAmpl}. Then, the first three relations between them and the Lanczos coefficients are
\be
\langle \psi_0|H|\psi_0\rangle=a_0,\qquad \langle \psi_0|H^2|\psi_0\rangle=a^2_0+b^2_1,\qquad \langle\psi_0|H^3|\psi_0\rangle=a^3_0+2a_0b^2_1+a_1b^2_1\,,
\ee
and they can be solved for Lanczos coefficients as follows
\be
a_0=\langle \psi_0|H|\psi_0\rangle,\quad b^2_1=\langle \psi_0|H^2|\psi_0\rangle-\langle \psi_0|H|\psi_0\rangle^2,\quad a_1=\frac{\langle \psi_0|H^3|\psi_0\rangle-a^3_0}{b^2_1}-2a_0\,.
\label{eq:sl2rLanczos}
\ee
Clearly, $a_0$ is just the average energy in the initial state, $b^2_1$ is the variance, and $a_3$ contains non-trivial information about the third moment of $H$, and hence  the skewness.  We also have the following relation between our parameters
\be
\frac{a_1-a_0}{b_1}=\langle \psi_0|\left(\frac{H-a_0}{b_1}\right)^3|\psi_0\rangle\,.
\ee
To recap, SL(2,$\mathbb{R}$) Lanczos coefficients are determined by three free parameters $(h,f(z),\mathcal{D})$. Therefore,  knowledge of the first three moments: the mean, the average and the skewness of the Hamiltonian $H$ in the initial state $\ket{\psi_0}$, allows us to fix the Lanczos coefficients completely. Consequently, variations of Lanczos coefficients can be physically interpreted as variations of the mean, the variance and the skewness of the energy spectrum describing a given quantum dynamics on the Krylov chain.  This simple interpretation applies in the case of motion on SL(2,$\mathbb{R}$), and also in other cases studied in this section that are highly constrained by symmetry, but not more generally.

\paragraph{Variations of spread complexity.}
The spread complexity  can be derived analytically and follows the general form for SL(2,$\mathbb{R}$) \eqref{SComplSL2R}. For us it becomes
\be
C_K(t)=2h\left(1+\frac{f(z)^2}{\mathcal{D}^2}\right)\sinh^2\left(\frac{\mathcal{D}}{2}t\right)\,.
\label{eq:Ck-SL2}
\ee
A short calculation shows that  we can rewrite this expression as\footnote{Actually, at early times one can show more generally that Krylov and spread complexity should grow quadratically as $b^2_1t^2$.} 
\be
C_K(t)=\frac{4b^2_1}{\mathcal{D}^2}\sinh^2\left(\frac{\mathcal{D}}{2}t\right)\,.\label{ComplGenForm}
\ee
In other words, all the  information about the initial state that makes its way into the spread complexity is encoded in $b_1^2$, which in turn equals the variance of the Hamiltonian in the initial state according to \eqref{eq:sl2rLanczos}.  On the other hand, the late time evolution depends only on the details of the Hamiltonian through $\mathcal{D}$. Thus, for SL(2,$\mathbb{R}$), we get the exponential, periodic, or quadratic growth  $C_K(t)$ for real $\mathcal{D}>0$, imaginary $\mathcal{D}$, or $\mathcal{D}\to 0$ respectively.

In particular for real, positive $\mathcal{D}$ and $t\ge \frac{2}{\mathcal{D}}$ spread complexity grows  exponentially as 
\be
C_K(t)\sim e^{\mathcal{D}(t-t_s)}\,,
\ee
where we defined the scrambling time 
\be
t_s=\frac{2}{\mathcal{D}}\log\left(\frac{\mathcal{D}}{b_1}\right)\,.
\ee
Hence, after the scrambling time, the information about initial state is ``washed away" and we transition to a universal regime of  exponential growth with Lyapunov exponent $\mathcal{D}$ that depends only on the parameters of the Hamiltonian.

The above result also implies that the {\it variation} of spread complexity with respect to the initial state depends on the change in  the variance
\be
\delta C_K(t)=\frac{4(\delta b^2_1)}{\mathcal{D}^2}\sinh^2\left(\frac{\mathcal{D}}{2}t\right)\,,
\label{eq:SL2-var}
\ee
and so changes in the initial state do not significantly affect the functional form of the time dependence (although the scrambling time can change). This is consistent with the expectation that a Hamiltonian $\SL(2,\mathbb{R})$ symmetry and positive $\mathcal{D}$ represents a ``universality class" of fast scramblers of information. 

The change \eqref{eq:SL2-var} is consistent with the first law formulation in \eqref{eq:uni-1law}, since higher order Hamiltonian moments for the coherent states under consideration are not independent. Notice that the non-vanishing chemical potentials interpolate between the universal short time $(\mathcal{D} t \ll 1)$ behavior in \eqref{eq:short-t} and exponential growth at long times $(\mathcal{D} t \gg 1)$. We may infer that exponential time dependence in some of the chemical potentials appearing in the first law may be signaling chaotic behavior.

Finally, note that for generic Hamiltonians \eqref{HamSL2R} with $\mathcal{D}>0$, the 
$\SL(2,\mathbb{R})$ class of models provides examples of infinite-dimensional Krylov subspaces where the spread complexity increases monotonically
\begin{equation}
    \frac{dC_K(t)}{dt}= \frac{2b_1^2}{\mathcal{D}}\,\sinh (\mathcal{D} t) >0\,, \qquad \qquad \forall\,t > 0\,.
\end{equation}
Nevertheless, we can still tune the parameters such that $4\alpha^2-\gamma^2<0$, and $\mathcal{D}$ is purely imaginary such that complexity oscillates in time violating the second law. Similar conclusions can be reached for the K-entropy.
\paragraph{Variations of the Krylov basis.}
Above, we obtained the Lanczos coefficients and  spread complexity directly from the return amplitude, without explicitly constructing the Krylov basis vectors. Nevertheless, the use of coherent states enables us to build these vectors explicitly and to examine their response to variations in the parameters of the initial state. 

To derive Krylov basis for non-trivial $z$, it is straightforward to use the algorithm \eqref{LanczosAlg} (see App.~\ref{appena}) and we find the general answer
\be
\ket{K_n}=\left(\frac{\alpha(z)}{\bar{\alpha}(z)}\right)^{n/2}D(\xi)\ket{h,n}\equiv e^{\frac{in}{2}\Phi(\rho,\phi)}D(\xi)\ket{K^{(0)}_n}\, ,\label{KBSU2RF}
\ee
where we defined the unperturbed basis $\ket{K^{(0)}_n}=\ket{h,n}$, as the Krylov basis for the initial state $\ket{0,h}$ in the notation of \eqref{CSSU11}. The overall coefficient, a pure phase, is expressed in terms of the $(\rho,\phi)$ coordinates for SL(2,$\mathbb{R}$) in \eqref{eq:poincareCoords} as  
\be
\frac{\alpha(z)}{\bar{\alpha}(z)}\equiv  e^{i\Phi(\rho,\phi)}=e^{2i\phi}\frac{2\alpha\cos\phi\cosh\rho-2i\alpha \sin\phi+\gamma\sinh\rho}{2\alpha\cos\phi\cosh\rho+2i\alpha \sin\phi+\gamma\sinh\rho}\,.\label{ComplPhaseSL2R}
\ee
This  result allows us to address a key question:  What is the relation between Krylov bases for two, different initial states?

Before deriving the general answer, note that, to  first order in the perturbation $\rho$, we have
\be
\ket{K_n}=\ket{h,n}+\frac{\rho}{2}\left[\frac{in\gamma\sin(\phi)}{\alpha}+e^{i\phi}L_{-1}-e^{-i\phi}L_1\right]\ket{h,n}+O(\rho^2)\,,
\ee
and using \eqref{ActionLns} we can express this as 
\be
\ket{K_n}\simeq\ket{K^{(0)}_n}+\frac{\rho}{2\alpha}\left[i(a^{(0)}_n-h\gamma)\sin(\phi)\ket{K^{(0)}_n}+e^{i\phi}b^{(0)}_{n+1}\ket{K^{(0)}_{n+1}}-e^{-i\phi}b^{(0)}_{n}\ket{K^{(0)}_{n-1}}\right]\,,
\ee
where $a_n^{(0)}$ and $b_n^{(0)}$ are given by \eqref{Lanczos0th}.
Then, to first order, the overlap between the two states is
\be
\langle K^{(0)}_m|K_n\rangle=\delta_{n,m}\left(1+i\,n\frac{\rho \gamma\sin(\phi)}{2\alpha}\right)+\frac{\rho}{2\alpha}\left(e^{i\phi}b^{(0)}_{n+1}\delta_{m,n+1}-e^{-i\phi}b^{(0)}_n\delta_{m,n-1}\right)\,.\label{OverlapsInf}
\ee
This shows that infinitesimal variations of the initial state tri-diagonally  ``spread in the old basis" to first order in $\rho$.

In fact,  we can do better and evaluate general overlaps between the new and old Krylov bases using coherent states and standard techniques from quantum optics \cite{gerry2023introductory}. For the case at hand, using the BCH identity we obtain
\bea
e^{-i\frac{n}{2}\Phi(\rho,\phi)}\langle K^{(0)}_m|K_n\rangle
&=&\frac{z^{m-n}(1-z\bar{z})^{h+n}}{\mathcal{N}_{m}\mathcal{N}_{n}}\sum^n_{l=0}\frac{\mathcal{N}^2_{n-l}}{l!(m-n+l)!}\left(-\frac{z\bar{z}}{1-z\bar{z}}\right)^l\,,
\eea
where we denote $\mathcal{N}^2_n=\Gamma(2h)/(n!\Gamma(2h+n))$. This can be elegantly written in terms of Jacobi polynomials\footnote{with generating function 
\be
\sum^\infty_{n=0}P^{(\alpha,\beta)}_n(x)t^n=\frac{2^{\alpha+\beta}}{R(1-t+R)^{\alpha}(1+t+R)^\beta}\,,\nn
\ee
where $R=R(x,t)=(1-2xt+t^2)^{1/2}$ and branch of the square root is chosen such that $R(x,0)=1$.}
\be
\langle K^{(0)}_m|K_n\rangle=e^{i\frac{n}{2}\Phi(\rho,\phi)}(1-|z|^2)^{h}\sqrt{\frac{n!\Gamma(2h+m)}{m!\Gamma(2h+n)}}z^{m-n}P^{(m-n,2h-1)}_n(1-2|z|^2)\,.\label{OverlapsMatrixSL2R}
\ee
Note that the information about the parameters $\gamma$ and $\alpha$ of the Hamiltonian \eqref{HamSL2R} only enters the phase $\Phi(\rho,\phi)$ \eqref{ComplPhaseSL2R} (which disappears in the absolute value).  Thus, surprisingly, for the family of Hamiltonians, and initial coherent states that we are considering, the Krylov bases for different Hamiltonians are the same up to an initial condition dependent phase.  This will not the be case for general Hamiltonians and for general initial states even for motion on a group manifold.

\begin{figure}[t!]
\centering
\includegraphics[width=.49\textwidth]{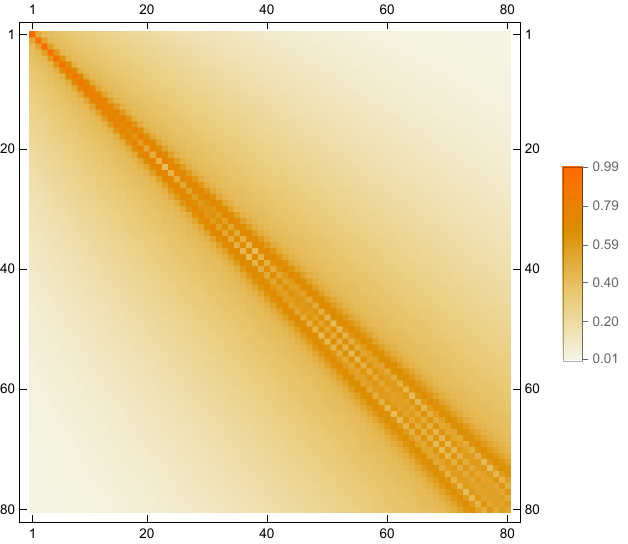}
\includegraphics[width=.49\textwidth]{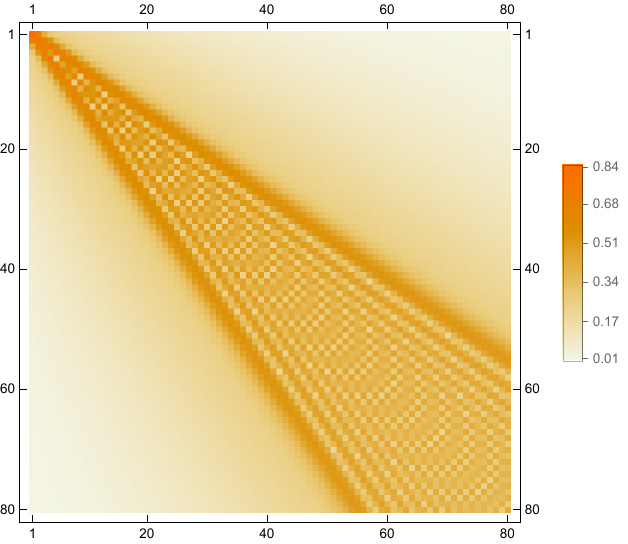}
\caption{Absolute values of the matrix of overlaps \eqref{OverlapsMatrixSL2R} between perturbed and unperturbed Krylov basis vectors for motion on  SL(2,$\mathbb{R}$).  Results shown for $n,m \leq 80$ and $h=1$. Left panel: for $\rho=0.1$,
right panel: for $\rho=0.4$.}
\label{fig:MatrOvermn}
\end{figure}

We plot the absolute value of these overlaps as a matrix in Fig.~\ref{fig:MatrOvermn} for a truncated range of $m,n\in[0,80]$, considering different values of the initial-state parameters. For small $\rho$, the overlaps are predominantly localized near the (tri-)diagonal band, in agreement with \eqref{OverlapsInf}. However, as 
$\rho$ increases, the overlaps develop a distinct “ballistic” pattern that broadens with $n$.   Since motion on SL(2,$\mathbb{R}$) is paradigmatic of chaotic dynamics, our results suggest that{\it } ballistic broadening of the support of the perturbed Krylov vectors in the unperturbed basis is a characteristic of quantum chaos, including systems exhibiting maximal chaos~\cite{Maldacena:2015waa} such as those described by RMT.  

Below we will compare the results above with motion on SU(2), a group which typically appears in integrable settings, and on the Heisenberg-Weyl group which occupies a status intermediate  between the chaotic  SL(2,$\mathbb{R}$) and the integrable SU(2) cases.

\subsection{SU(2)}\label{sec:SU2}
An analysis very similar to that for  $\SL(2,\mathbb{R})$ can be performed for motion on  $\SU(2)$ and so we will be brief. The algebra is defined by the commutation relations
\be
[J_0,J_{\pm}]=\pm J_\pm,\qquad [J_+,J_-]=2J_0\,.\label{SU2alg}
\ee
Now, consider a Hamiltonian given by a combination of the generators
\be
H=\gamma J_0+\alpha(J_++J_-)\, .
\ee
This leads to a time evolution 
\be
\ket{\psi(t)}=e^{-iHt}\ket{z,j}\,,
\ee
where we choose the initial state $\ket{z,j}$ to be a $\SU(2)$ generalized coherent state \cite{Perelomov:1971bd} labeled by a point on the sphere 
\be
\ket{z,j}=D(\xi)\ket{j,-j},\qquad D(\xi)=e^{\xi J_+-\bar{\xi}J_-},\qquad z=\frac{\xi}{|\xi|}\tan(|\xi|)=\tan\left(\frac{\theta}{2}\right)e^{i\phi}\,,
\label{eq:SU(2)Coords}
\ee
where $\xi=\theta/2e^{i\phi}$. Here the index $j$ indicates that we are working with $2j+1$ dimensional representation of $\SU(2)$, and $\ket{j,-j}$ indicates a state in the $j$ representation with polarization $-j$.

Again, the BCH formula for this Lie algebra allows us to evaluate the return amplitude
\be
S(t)=\left(\cos\left(\frac{\mathcal{D}}{2}t\right)-\frac{if(z)}{\mathcal{D}}\sin\left(\frac{\mathcal{D}}{2}t\right)\right)^{2j}\,,
\ee
this time parametrized by the following two functions
\be
\mathcal{D}=\sqrt{4\alpha^2+\gamma^2},\qquad f(z)=\frac{\gamma(1-z\bar{z})-2\alpha(z+\bar{z})}{1+z\bar{z}}\,.
\ee
As for $SL(2,\mathbb{R})$, the algorithm to extract Lanczos coefficients from the moments also requires three steps (fixing the mean, the variance and the skewness) and we derive the following $2j$ coefficients
\be
a_n=f(z)(-j+n),\qquad b_n=\frac{\sqrt{\mathcal{D}^2-f(z)^2}}{2}\sqrt{n(2j-n+1)}\,,
\ee 
fixed by the algebraic data \cite{Caputa:2021sib}, up to  physical prefactors that encode the data of the initial state and the Hamiltonian. This leads to  spread complexity with a form analogous to \eqref{ComplGenForm}, proportional to $b^2_1$, but now oscillating in time
\be
C_K(t)=2j\left(1-\frac{f(z)^2}{\mathcal{D}^2}\right)\sin^2\left(\frac{\mathcal{D}}{2}t\right)=\frac{4b^2_1}{\mathcal{D}^2}\sin^2\left(\frac{\mathcal{D}}{2}t\right)\,.
\label{SpreadGenSU2}
\ee
Its variation with respect to the initial state is equivalent to the variation of $b^2_1$ giving a balance equation consistent with the first law \eqref{eq:uni-1law}
\begin{equation}
    \delta C_K(t) = 4\sin^2\left(\frac{\mathcal{D} t}{2}\right)\,\frac{\delta b_1^2}{\mathcal{D}^2}\,.
\end{equation}
Thus $C_K$ does not increase monotonically in time because of the oscillatory behavior in \eqref{SpreadGenSU2}, in agreement with our general arguments about finite-dimensional Krylov spaces. Finally, variations due to the change in the Hamiltonian are encoded in both $\delta b_1^2$ and $\delta \mathcal{D}$.
\paragraph{Variations of the Krylov basis.}
The Krylov basis for $\xi=0$ was studied for this scenario in \cite{Caputa:2021sib} and contains $2j+1$ vectors, from $\ket{j,-j}$ to $\ket{j,j}$ denoted by $\ket{K^{(0)}_n}=\ket{j,-j+n}$, $n=0,..,2j$. 
Fortunately, the framework of coherent states allows us to derive it even for the general initial coherent state. Following the Lanczos algorithm, and using \eqref{DHDSU2}, we find the new Krylov basis that has the same dimension $2j+1$ but a non-trivial dependence on the SU(2) coordinates $\theta$ and $\phi$ (see \eqref{eq:SU(2)Coords})
\be
\ket{K_n}=\left(\frac{\alpha(z)}{\bar{\alpha}(z)}\right)^{n/2}D(\xi)\ket{j,-j+n}=e^{i\frac{n}{2}\Phi(\theta,\phi)}D(\xi)\ket{K^{(0)}_n} \,,\label{KBSU2}
\ee
where the complex phase in now expressed as
\be
\frac{\alpha(z)}{\bar{\alpha}(z)}\equiv e^{i\Phi(\theta,\phi)}=e^{2i\phi}\frac{2\alpha\cos\theta\cos\phi-2i\alpha \sin\phi+\gamma\sin\theta}{2\alpha\cos\theta\cos\phi+2i\alpha \sin\phi+\gamma\sin\theta}\,.
\ee
As before, we ask how small perturbations of the initial state alter the basis. This time, the overlap between the new and old Krylov basis vectors is
\begin{figure}[t!]
\centering
\includegraphics[width=.49\textwidth]{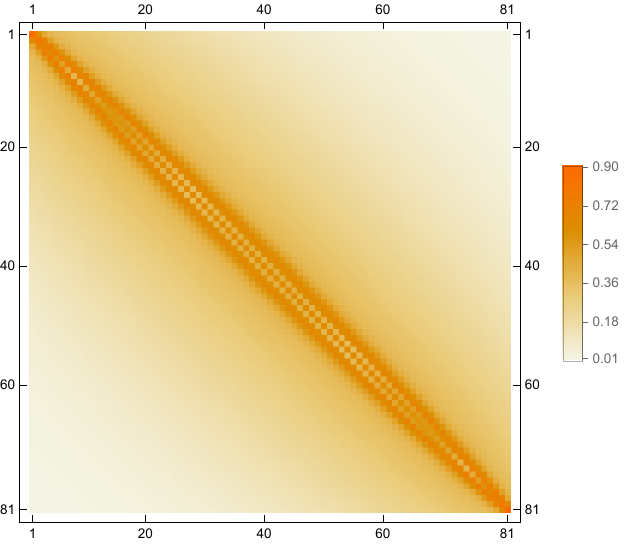}
\includegraphics[width=.49\textwidth]{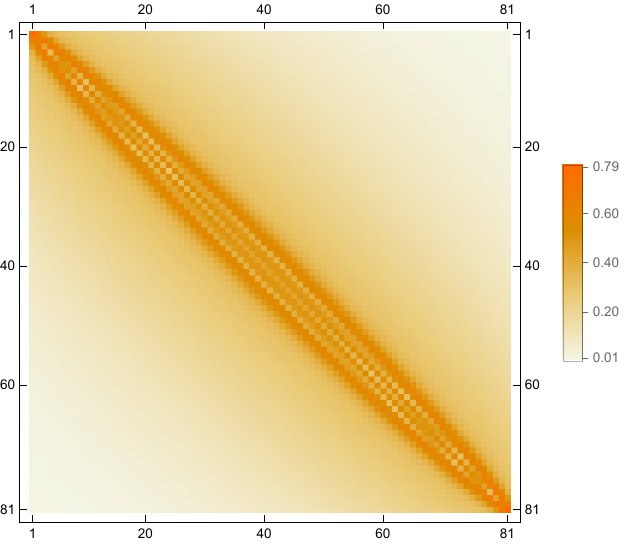}
\caption{Absolute value of matrix overlaps \eqref{MOHOLagSU} between perturbed and unperturbed Krylov basis vectors for motion on SU(2). Results shown  for $j=40$,  $n,m\in[0,80]$. Left panel: for $\theta=0.1$,
right panel: for $\theta=0.15$, both for $\phi=\pi/3$.
}
\label{fig:MatrOverSUmn}
\end{figure}
\be
e^{-i\frac{m}{2}\Phi(r,\theta)}\langle K^{(0)}_n|K_m\rangle=\sum^n_{l=0}\sum^{2j-n+l}_{k=0}\frac{z^k(-\bar{z})^l(1+|z|^2)^{-j+n-l}\tilde{\mathcal{N}}^2_{n-l}}{l!k!\tilde{\mathcal{N}}_n\tilde{\mathcal{N}}_{n-l+k}}\delta_{m,n-l+k}\,,\label{MOHOLagSU}
\ee
where $\tilde{\mathcal{N}}^2_n=\Gamma(2j-n+1)/(n!\Gamma(2j+1))$. 
We plot their absolute value on Fig.~\ref{fig:MatrOverSUmn} for a two of small values of $\theta$. The overlaps are in striking contrast with their SL(2,$\mathbb{R}$) counterparts, and are now supported mostly near the tri-diagonal band. There is very little broadening of the new Krylov vectors in the old Krylov basis i.e., perturbing initial state can be expressed by only a few nearby vectors.  

One may worry that there is a confound here because the SU(2) Hilbert space for representation $j$ is $2j+1$ dimensional,  constraining how broad the support of the perturbed Krylov vectors can be in the old basis.  We will mitigate this doubt below by considering a large $j$ limit in which we can compare more directly with the SL(2,$\mathbb{R}$) case.

\subsection{Heisenberg-Weyl}
Finally, we analyze Krylov dynamics governed by the Heisenberg–Weyl (HW) algebra, which typically lies at the boundary between the two cases discussed above: the Krylov basis is infinite-dimensional, yet the spread complexity does not grow exponentially \cite{Caputa:2021sib}. In this setting, we can  exploit ordinary coherent states to obtain analytical results.

The HW algebra, with the creation $a^\dagger$, the annihilation $a$, and the number $n=a^\dagger a$ operators is simply
\be
[a,a^\dagger]=1\,,\qquad [n,a]=-a\,,\qquad [n,a^\dagger]=a^\dagger\,.\label{HWalgebra}
\ee
Consider a Hamiltonian of the form
\be
H=\gamma a^\dagger a+\alpha(a^\dagger+a)\,,
\ee
and  time evolution of a coherent state labeled by a complex number $z$
\be
\ket{\psi(t)}=e^{-iHt}\ket{z},\qquad \ket{z}=D(z)\ket{0}\equiv e^{za^\dagger-\bar{z}a}\ket{0}\,,\qquad z=re^{i\theta}\,.
\label{eq:HWcoords}
\ee
The return amplitude can be simply computed using the BCH formula and reads
\be
S(t)=\bra{z}e^{iHt}\ket{z}=\exp\left(-i\frac{\alpha^2}{\gamma}t-\left|z+\frac{\alpha}{\gamma}\right|^2\left(1-e^{i\gamma t}\right)\right)\,.
\ee
Following the standard procedure, we derive infinite sequences of Lanczos coefficients  
\be
a_n=\gamma (n+|z|^2)+\alpha(z+\bar{z})\,,\qquad b_n=
|\alpha+\gamma z|\sqrt{n}\,.
\ee
This way, to the first order in $r$ we have
\bea
a_n&\simeq&\gamma n+2r\alpha\cos\theta=a^{(0)}_n+2r\alpha\cos\theta\,,\nn\\
b_n&\simeq&\alpha\sqrt{n}+r\gamma\cos\theta\sqrt{n}=b^{(0)}_n+r\gamma\cos\theta\sqrt{n}\,,
\eea
where the superscript $0$ denotes Lanczos coefficients for $r=0$ derived in \cite{Balasubramanian:2022tpr}.

Although the structure of Lanczos coefficients is now more involved,  we can still evaluate the spread complexity analytically
\be
C_K(t)=\frac{4|\alpha+\gamma z|^2}{\gamma^2}\sin^2\left(\frac{\gamma}{2}t\right)=\frac{4b^2_1}{\mathcal{D}^2}\sin^2\left(\frac{\mathcal{D}}{2}t\right)\,,
\label{eq:C-Weyl}
\ee
with $\mathcal{D}=\gamma$. Its variation with the initial state 
\begin{equation}
    \delta C_K(t) = 4\sin^2\left(\frac{\gamma}{2}t\right)\,\frac{\delta \tilde{\mu}_2^2}{\mathcal{D}^2}\,,
\end{equation}
recalling again that $b_1^2 = \tilde{\mu}_2$ is the variance of the Hamiltonian in the initial state.
This is consistent with \eqref{eq:uni-1law}. Interestingly,
despite the infinite-dimensionality of the Krylov space, the oscillatory behavior in \eqref{eq:C-Weyl} for generic $\gamma$  mean that spread complexity does not grow monotonically. However, if $\gamma\to0$, complexity grows quadratically $C_K(t)=\alpha^2t^2$ and a ``second law'' of monotonic growth holds. Finally, variations due to the change in the Hamiltonian are encoded in both $\delta b_1^2$ and $\delta \mathcal{D}$.
\paragraph{Variations of the Krylov basis.}
To find the new Krylov basis, we use the action of the displacement operator on the algebra generators \cite{gerry2023introductory}
\be
D^\dagger(z)aD(z)=a+z,\qquad D^\dagger(z)a^\dagger D(z)=a^\dagger+\bar{z}\,,
\ee
and consequently on the Hamiltonian
\be
D^\dagger(z)HD(z)=\gamma a^\dagger a+(\alpha+\gamma z)a^\dagger +(\alpha+\gamma \bar{z})a+\gamma z\bar{z}+\alpha(z+\bar{z})\,.
\ee
This way, after following the Lanczos algorithm, we find the new Krylov basis
\be
\ket{K_n}=\left(\frac{\alpha+\gamma z}{\alpha+\gamma\bar{z}}\right)^{n/2}D(z)\ket{n}\equiv e^{i\frac{n}{2}\Phi(r,\theta)}D(z)\ket{K^{(0)}_n}\,,
\ee
where by $\ket{K^{(0)}_n}$ we denoted the unperturbed Krylov basis for $z=0$, i.e.,  $\ket{K^{(0)}_n}=\ket{n}$. Both Krylov bases are infinite-dimensional.

To quantify how perturbations spread and modify the Krylov vectors, we again compute the overlap between the old and new basis vectors. In this case, the overlaps can be written in terms of matrix elements of the displacement operator that are well known in quantum optics
\be
\langle K^{(0)}_n|K_m\rangle=e^{i\frac{m}{2}\Phi(r,\theta)}\langle n|D(z)\ket{m}\,.
\ee
These quantities can be expressed in terms of the Laguerre polynomials\footnote{With generating function
\be
\sum^\infty_{n=0} L^{(\alpha)}_n(x)t^n=\frac{e^{-\frac{tx}{1-t}}}{(1-t)^{\alpha+1}}\,.\nn
\ee} as follows \cite{gerry2023introductory}
\be
\langle K^{(0)}_n|K_m\rangle=e^{i\frac{m}{2}\Phi(r,\theta)}\left\{
\begin{array}{c}
 \sqrt{\frac{n!}{m!}}e^{-\frac{1}{2}|z|^2}(-\bar{z})^{m-n}L^{(m-n)}_n(|z|^2)\,,\qquad m\ge n\, \\
 \sqrt{\frac{m!}{n!}}e^{-\frac{1}{2}|z|^2}(z)^{n-m}L^{(n-m)}_m(|z|^2)\,,\qquad n\ge m\, \\
\end{array}
\right.\,.\label{MOHOLag}
\ee
\begin{figure}[h!]
\centering
\includegraphics[width=.49\textwidth]{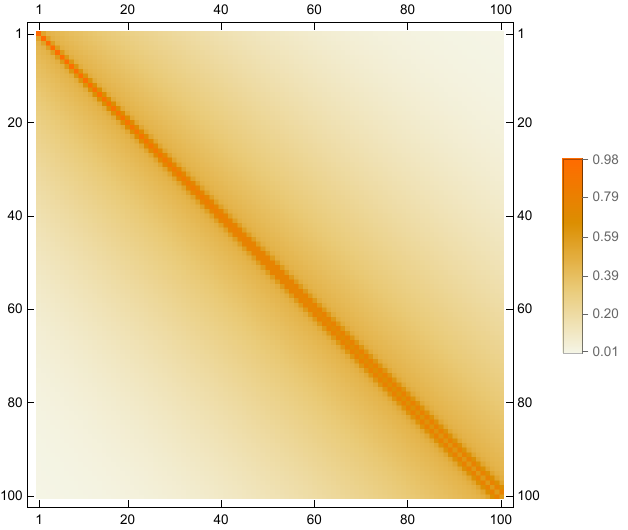}
\includegraphics[width=.49\textwidth]{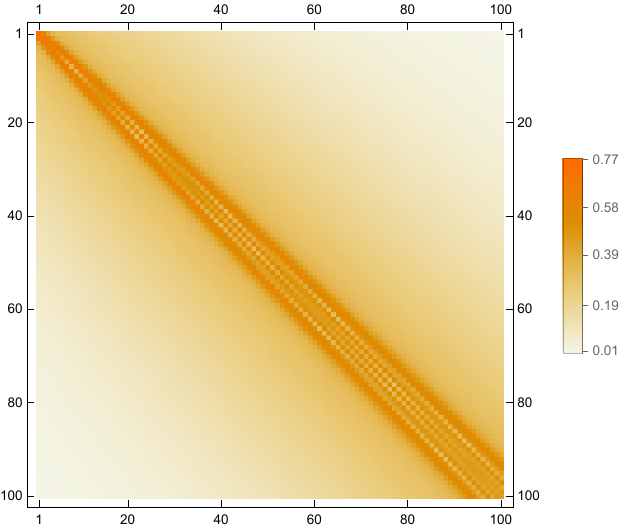}
\caption{Absolute value of matrix overlaps \eqref{MOHOLag} between perturbed and unperturbed Krylov vectors for motion on the Heisenberg-Weyl group manifold.  Results shown for $n,m$ up to 100. Left panel: for $r=0.1$, right panel: for $r=0.4$.}
\label{OverlaspHO}
\end{figure}
We plot the absolute values of the overlaps for small and large values of $r$, i.e., the displacement of the perturbed state, on Fig.~\ref{OverlaspHO}. For small perturbations, the overlaps remain localized along the (tri-)diagonal, whereas for larger $r$  they spread significantly over the original basis. Nevertheless, this spreading is much slower and less pronounced than in the SL(2,$\mathbb{R}$) case shown in Fig.~\ref{fig:MatrOvermn}, supporting the conjecture that effective dynamics governed by SL(2,$\mathbb{R}$) Hamiltonians correspond to “fast scrambling” models, and that the Heisenberg-Weyl case is intermediate between SL(2,$\mathbb{R})$ and the SU(2) group which usually appears in integrable settings

\subsection{Koherence and Relative Krylov Entropy }

\subsubsection{Koherence}
\begin{figure}[h!]
\centering
\includegraphics[width=.49\textwidth]{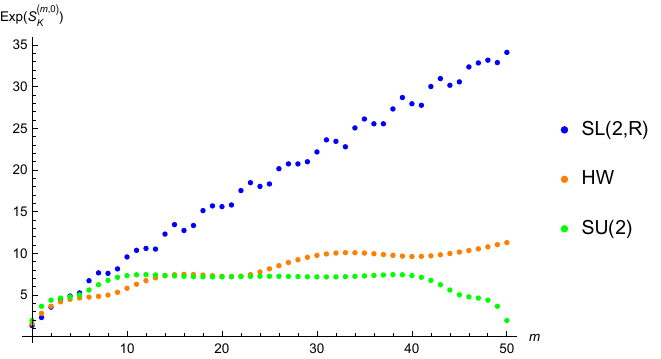}
\includegraphics[width=.49\textwidth]{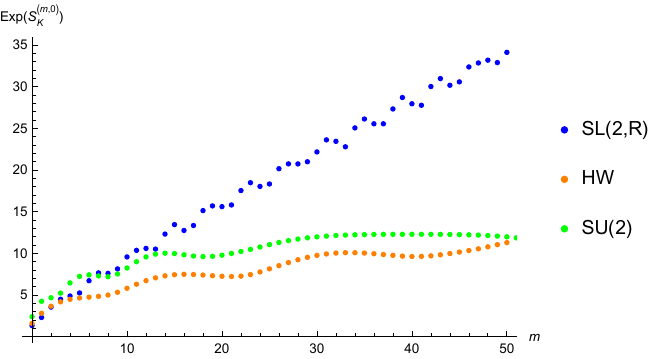}
\caption{Exponential of Koherence computed from  overlaps in the Krylov bases of SL(2,$\mathbb{R}$), SU(2) and HW coherent states. Two different initial states, the highest-weight state, and coherent state obtained by displacing the highest-weight state by a displacement operator, were evolved by the same Hamiltonian. Left: Figure for SU(2) with $j=25$ (Hilbert space dimension 51) and Right: SU(2) with $j=40$ (Hilbert space dimension 81).}
\label{fig:SKmLie}
\end{figure}

Fig.~\ref{fig:SKmLie} shows  the exponent of Koherence, defined in \eqref{eq:K-coh}, for $\SL(2,\mathbb{R})$ \eqref{OverlapsMatrixSL2R}, for $\SU(2)$ \eqref{MOHOLagSU} and for the Heisenberg-Weyl group \eqref{MOHOLag}.  To better compare the two infinite dimensional bases for $\SL(2,\mathbb{R})$ and Heisenberg-Weyl with the $2j+1$-dimensional $\SU(2)$ group, we show plots for $j=25$ (left) and $j=40$ (right).  Fig.~\ref{fig:SKmLie} shows a dramatic difference between $\SL(2,\mathbb{R})$ and the other two cases.  We already saw in Fig.~\ref{fig:MatrOvermn} that the Krylov basis elements for the perturbed $\SL(2,\mathbb{R})$ state appear to have support in the unperturbed basis that grows linearly with the Krylov index.  Here we see that the exponential of the Koherence, i.e., the entropy of the distribution of that support,  increases linearly with the Krylov index.  This is in sharp contrast to $\SU(2)$, for which the exponential of Koherence plateaus at about half the size of the Hilbert space. Meanwhile, for  the Heisenberg-Weyl case, which also has an infinite dimensional Hilbert space, does not display a plateau, but instead shows slow growth of Koherence.    In these examples, motion on $\SL(2,\mathbb{R})$ is paradigmatic of the diverging trajectories expected in chaotic systems.  So Fig.~\ref{fig:SKmLie}  may be displaying a characteristic difference in Koherence between systems showing different degrees of chaos, at least for appropriately chosen initial states.  We leave a systematic treatment to the future. 

\subsubsection{Relative Krylov Entropy}
Let us finally discuss the relative K-entropy and compare evolutions of the same initial state with two different Hamiltonians. We start with the $\mathrm{SL}(2,\mathbb{R})$ symmetry with the Hamiltonian expressed as linear combination of the generators
\be
H_{\text{SL}(2,\mathbb{R})}=s_0L_0+s_1 L_1+s_{-1}L_{-1}\,,
\ee
and the evolution of the initial state represented by the highest weight vector $\ket{h}$\footnote{In this section we use slightly more general Hamiltonian than in \eqref{HamSL2R} but a simpler initial state.}. Varying with respect to parameters $s_i$ corresponds to different choices of the evolving Hamiltonian.  
Then the BCH relation implies (see App.~\ref{appena})
\be
\ket{\psi(t)}=e^{-iH_{\text{SL}(2,\mathbb{R})}t}\ket{h}=\mathcal{N}e^{z(s_i,t) L_{-1}}\ket{h}\,,\qquad A_i=z\bar{z}\,,
\ee
where $A_i$, $i=1,2$, will depend on the two, particular choices of parameters $\{s_{-1},s_0,s_1\}$ as
\be
A_i=\frac{4s_1s_{-1}}{s^2_0+\mathcal{D}^2\coth^2\left(\frac{\mathcal{D}t}{2}\right)}\,,\qquad \mathcal{D}=\sqrt{4s_1s_{-1}-s^2_0}\,.
\ee
These two choices yield two probability distributions on the infinite-dimensional Krylov chain with the same initial state
\be
p_n=(1-A_1)^{2h}A^n_1\frac{\Gamma(2h+n)}{n!\Gamma(2h)}\,,\qquad q_n=(1-A_2)^{2h}A^n_2\frac{\Gamma(2h+n)}{n!\Gamma(2h)}\,.\label{probsRKESL2R}
\ee
Their RKE \eqref{eq:RelKEntr} can be summed to the following expression
\be
S_K(p\Vert q)=\sum^\infty_{n=0}p_n\log\left(\frac{p_n}{q_n}\right)=2h\left[\log\left(\frac{1-A_1}{1-A_2}\right)+\frac{A_1}{1-A_1}\log\left(\frac{A_1}{A_2}\right)\right].\label{eq:RelEntrSL2R}
\ee
To get some feeling about its time dependence, consider first the simplest scenario where in both cases $s_0=0$ and $s_1=s_{-1}=\alpha_i$ (implying $\mathcal{D}=2\alpha_i$) \footnote{studied e.g. in \cite{Parker:2018yvk,Caputa:2021sib}.} and
\be
A_i=\tanh^2\left(\alpha_i t\right)\,.
\ee
The corresponding probabilities \eqref{probsRKESL2R} yield exponentially growing Krylov complexities with Lyapunov exponents $\lambda_i=2\alpha_i$. Geometrically, this quantum dynamics can be mapped to a particle on the hyperbolic disc, starting at the origin at $t=0$ and, as time progresses, moving radially towards the boundary with velocity $\alpha_i$.

The time evolution of RKE for these parameters evaluates to
\be
S_K(p\Vert q)=4h\left[\log\left(\frac{\cosh(\alpha_2 t)}{\cosh(\alpha_1 t)}\right)+\sinh^2(\alpha_1 t)\log\left(\frac{\tanh(\alpha_1 t)}{\tanh(\alpha_2 t)}\right)\right]\,,\label{RKEOG}
\ee
and (without loss of generality) for $\alpha_2>\alpha_1>0$ RKE grows linearly at late times with the coefficient equal to the difference between the two velocities 
\be
S_K(p_1|p_2)\sim 4h\Delta \alpha\, t\,,\qquad \Delta\alpha=\alpha_2-\alpha_1\,.
\ee

Similar analysis can be done for the $\SU(2)$ and HW Hamiltonians so we just briefly summarize the steps. For $SU(2)$, we consider Hamiltonians expressed in terms of generators \eqref{SU2alg} as
\be
H_{\text{SU(2)}}=s_0J_0+s_1 J_-+s_{-1}J_{+}\,.
\ee
The time evolution of the initial highest weight state $\ket{j,-j}$ can be parametrized by
\be
\ket{\psi(t)}=e^{-iH_{\text{SU(2)}}t}\ket{j,-j}=\mathcal{N}e^{z(s_i,t) J_{+}}\ket{j,-j},\qquad A_i=z\bar{z}\,,
\ee
where, using BCH formula for SU(2), we now get
\be
A_i=\frac{4s_1s_{-1}}{s^2_0+\mathcal{D}^2\cot^2\left(\frac{\mathcal{D}t}{2}\right)}\,,\qquad \mathcal{D}=\sqrt{4s_1s_{-1}+s^2_0}\,.
\ee
As before, we focus on the evolution of the same initial state with two different Hamiltonians parametrized by distinct set of coefficients $s_i$, $i\in(-1,0,1)$. The (2j+1)-dimensional probability distributions with $n=0,1,...,2j$ are now
\be
p_n=(1+A_1)^{-2j}A^n_1\frac{\Gamma(2j+1)}{n!\Gamma(2j-n+1)}\,\qquad q_n=(1+A_2)^{-2j}A^n_2\frac{\Gamma(2j+1)}{n!\Gamma(2j-n+1)}\,,
\ee
and the RKE can be written as 
\be
S_K(p\Vert q)=\sum^{2j}_{n=0}p_n\log\left(\frac{p_n}{q_n}\right)=2j\left[\log\left(\frac{1+A_2}{1+A_1}\right)+\frac{A_1}{1+A_1}\log\left(\frac{A_1}{A_2}\right)\right].\label{eq:RelEntrSU2}
\ee
To get some intuition on its evolution, similarly to the $\mathrm{SL}(2,\mathbb{R})$, we can fist take a simpler range of parameters with $s_0=0$ and $s_1=s_{-1}=\alpha_i$ that yields
\be
S_K(p\Vert q)=4j\left[\log\left(\frac{\cos(\alpha_2 t)}{\cos(\alpha_1 t)}\right)+\sin^2(\alpha_1 t)\log\left(\frac{\tan(\alpha_1 t)}{\tan(\alpha_2 t)}\right)\right]\,.\label{RKESU21}
\ee
As we can see, as the time approaches $t\to \frac{\pi}{2\alpha_i}$, the formula diverges implying that we can perfectly distinguish the two probability distributions\footnote{From Stein’s or Neyman–Pearson lemma, the error in distinguishing them over m iid samples scales as $\exp\left(-m S_K(p\Vert q)\right)$.} coming from the two evolutions with different Hamiltonians on the Krylov chains.

Analogously, for the Heisenberg-Weyl algebra we consider two Hamiltonians parametrized as
\be
H_{\text{HW}}=s_0a^\dagger a+s_1 a+s_{-1}a^\dagger\,,
\ee
and evolve the initial state $\ket{0}$ by
\be
\ket{\psi(t)}=e^{-iH_{\text{HW}}t}\ket{0}=\mathcal{N}e^{z(s_i,t) a^\dagger}\ket{0}\,,\qquad A_i=z\bar{z}\,,
\ee
where the BCH yields
\be
A_i=\frac{4s_1s_{-1}}{s^2_0}\sin^2\left(\frac{s_0t}{2}\right)\,.
\ee
The infinite-dimensional probability distributions  are now
\be
p_n=e^{-A_1}\frac{A^n_1}{n!}\,,\qquad q_n=e^{-A_2}\frac{A^n_2}{n!}\,,
\ee
with RKE
\be
S_K(p\Vert q)=\sum^{\infty}_{n=0}p_n\log\left(\frac{p_n}{q_n}\right)=A_2-A_1+A_1\log\left(\frac{A_1}{A_2}\right).\label{eq:RelEntrHW}
\ee
In the similar, simple choice of parameters as before, $s_0=0$ and $s_1=s_{-1}=\alpha_i$, we now observe a quadratic growth towards distinguishability
\be
S_K(p\Vert q)=t^2\left(\alpha^2_2-\alpha^2_1+2\alpha^2_1\log\left(\frac{\alpha_1}{\alpha_2}\right)\right).\label{eq:RelEntSimplHW}
\ee
\begin{figure}[h!]
\centering
\includegraphics[width=.6\textwidth]{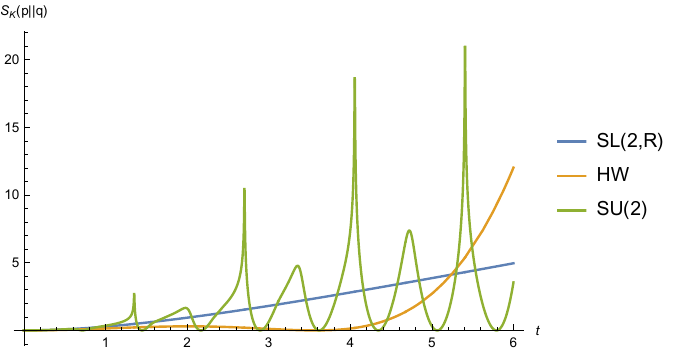}
\caption{Early time evolution of RKE entropies for SL(2,$\mathbb{R}$), SU(2) and HW with choices of parameters for one Hamiltonian $\{s_0,s_1,s_{-1}\}=\{0.5,4,1\}$, and $\{s_0,s_1,s_{-1}\}=\{0.7,4.1,1.2\}$ for the other, both corresponding the positive $\mathcal{D}$ for SL(2,$\mathbb{R}$). Plot for $h=j=1$.}
\label{fig:RKEAll}
\end{figure}
To give a comparison between these three different RKE with more generic choices of parameters, we choose the same two sets of  $s_i$ and plot RKE for early times on Fig.~\ref{fig:RKEAll}. It is clear that for $\SU(2)$ the RKE diverges rapidly, indicating perfect distinguishability. In contrast, $\SL(2,\mathbb{R})$ exhibits a faster, linear initial growth, but is eventually surpassed by the HW case, which grows quadratically. This behavior is also natural from the scrambling perspective: distinguishing between the two probability distributions remains harder for $\SL(2,\mathbb{R})$, even at later times.  We leave a more systematic analysis  of different classes of models (e.g. integrable vs chaotic) as an interesting future problem.  However, we note here that motion on $\SL(2,\mathbb{R})$ models chaotic dynamics, and the RKE results in Fig.~\ref{fig:RKEAll} suggest that the way in which an initial state spreads across the Krylov chain will be more universal for chaotic systems.

\section{Lattice model: varying the Hilbert space dimension}\label{sec:SolvmodLin}
As discussed in Sec.~\ref{sec:Introduction}, we would like study how the Krylov basis and spread complexity vary as the dimension of the Hilbert space changes, especially with a view to understanding how effective descriptions of the large system limit differ from an underlying finite dimensional theory.  To this end, 
we will study in this section a tight-binding lattice  Hamiltonian with constant, positive onsite energy ($a$) and constant, positive nearest-neighbor hopping ($b$), and open boundary conditions
\be
H=\sum^{N-1}_{n=0}a\ket{n}\bra{n}+\sum^{N-2}_{n=0}\left(b\ket{n+1}\bra{n}+b\ket{n}\bra{n+1}\right)\,.\label{HamTB}
\ee
This model can also be regarded as a quantum random walk quantum  \cite{Venegas-Andraca:2012zkr}, or as an instance of Anderson's model of localization \cite{anderson1958absence} (see also \cite{Peacock:2025mwl}). 
We can express this Hamiltonian in terms of discrete shift operators
\be
H=aI+b(T_-+T_+)\,,\qquad T_\pm\ket{n}=\ket{n\pm 1}\,,\qquad I\ket{n}=\ket{n}\,.
\ee
Similar models have  appeared in  complexity-related contexts, including the double-scaled SYK model (DSSYK) \cite{Berkooz:2018qkz,Berkooz:2024lgq,Miyaji:2025ucp,Basu:2025mmm}, Krylov complexity \cite{Barbon:2019wsy}, and the correspondence between the spread complexity of the TFD state and the geodesic length in JT gravity \cite{Lin:2022rbf,Rabinovici:2023yex,Iliesiu:2024cnh,Nandy:2024zcd,Balasubramanian:2024lqk,Miyaji:2025yvm,Miyaji:2025ucp}. However, our discussion and application to variations of spread complexity as well us finite vs. infinite $N$ is novel.

The model has a simple form in momentum space: its energy eigenstates can be written as 
\be
\ket{E_j}=\sum^{N-1}_{n=0}\phi_n(j)\ket{n}\,,\qquad \langle E_i|E_j\rangle=\delta_{i,j}\,,\qquad\sum^{N}_{j=1}\ket{E_j}\bra{E_j}=1\,,
\label{eq:finite-spectrum}
\ee
where the $\phi_n(j)$ are a set of $N$ orthonormal functions, known as the discrete sine transform
\be
\phi_{n}(j)=\sqrt{\frac{2}{N+1}}\sin\left(\frac{\pi(n+1)j}{N+1}\right)\,,\qquad j=1,...,N,\quad n=0,...,N-1\,,\label{SineBasis}
\ee
satisfying
\be
\langle \phi_n|\phi_m\rangle\equiv\sum^N_{j=1}\phi_n(j)\phi_m(j)=\delta_{n,m}\,.\label{OrhonormDisc}
\ee
We will later be interested in the continuum, ``thermodynamic" limit when we take $N\to\infty$, while the discrete variable $j$ (``momentum") is replaced by continuous variable $\theta$
\be
\frac{\pi j}{N+1}\to \theta\,,\qquad \theta\in[0,\pi]\,,\qquad n=0,...,\infty\,.
\label{eq:theta-def}
\ee
In this limit, the wave functions become
\be
\phi_n(j)\to \phi_n(\theta)=\sqrt{\frac{2}{\pi}}\sin\left((n+1)\theta\right)\,,\qquad \int^\pi_0\phi_n(\theta)\phi_m(\theta)d\theta=\delta_{n,m}\,.\label{OrthnmCont}
\ee

In this basis  the Hamiltonian acts as
\be
H\ket{E_j}=E_j\ket{E_j}\,,\qquad E_j=a+2b\cos\left(\frac{\pi j}{N+1}\right)\,,\qquad j=1,...,N\,,\label{DiscEnergyCL}
\ee
so it has $N$ eigenvalues between $[a-2b,a+2b]$. Notice that the position vectors themselves can also be written in the energy basis as
\be
\ket{k}=\sum^{N}_{j=1}\ket{E_j}\bra{E_j}k\rangle=\sum^{N}_{j=1}\phi_k(j)\ket{E_j}\,.\label{kinEnBas}
\ee
In the thermodynamic, continuum limit we  have
\be
\ket{E(\theta)}=\sum^{\infty}_{n=0}\phi_n(\theta)\ket{n}\,,\qquad \langle E(\theta)|E(\theta')\rangle=\delta(\theta-\theta')\,,\qquad \int^\pi_0d\theta \ket{E(\theta)}\bra{E(\theta)}=1\,, 
\ee
and the energies are continuously supported on an interval $[a-2b,a+2b]$
\be
E(\theta)=a+2b\cos(\theta)\,.\label{EnSpCont}
\ee
Analogously, we can express the discrete position vectors
\be
\ket{k}=\int^\pi_0d\theta \ket{E(\theta)}\bra{E(\theta)}k\rangle=\int^\pi_0d\theta \phi_k(\theta)\ket{E(\theta)}\,.
\ee
Below, we analyze the spread complexity of different initial states evolved by the finite $N$ and continuum Hamiltonians described above.

\subsection{Localized initial state}\label{sec:ToyIS0}
We will start by computing spread complexity of the evolution of the initial state that is localized on the first site, namely $\ket{0}$.   As explained in Sec.~\ref{sec:Krylov},  we start from the return amplitude 
\bea
S_0(t)&=&\langle 0|e^{iHt}|0\rangle=\sum^N_{i,j=1}\langle 0|E_i\rangle\langle E_i|e^{iHt}|E_j\rangle\langle E_j|0\rangle=\sum^N_{j=1}\phi_0(j)^2e^{iE_jt}\nn\\
&=&e^{iat}\frac{2}{N+1}\sum^N_{j=1}\sin^2\left(\frac{\pi j}{N+1}\right)e^{i2b t\cos\left(\frac{\pi j}{N+1}\right)}\,.
\eea
The Lanczos coefficients derived from this return amplitude are 
\be
a_n=a\,,\  \text{for}\   n=0,1,...,N-1; \qquad b_n=b, \   \text{for}\   n=1,...,N-1\,.
\ee
This implies that the Krylov basis for the evolution of the initial state $\ket{0}$ coincides with $\ket{K_n}=\ket{n}$ in which $H$ is  by definition \eqref{HamTB} tri-diagonal with precisely constant Lanczos coefficients $a$ and $b$. Note that this is an initial state dependent fact and will not hold for other choices of initial state (see, e.g., Appendix~\ref{sce:InStatek} for $\ket{\psi_0}=\ket{k}$).

Using \eqref{kinEnBas} and \eqref{SineBasis}, the time-dependent wave functions $\psi_n(t)$, with $n=0,...,N-1$, are equal to
\bea
\psi_n(t)&\equiv&\langle K_n|e^{-iHt}|0\rangle=\langle n|e^{-iHt}|0\rangle=\sum^N_{j=1}\phi_n(j)\phi_0(j)e^{-iE_jt}\nn\\
&=&e^{-iat}\frac{2}{N+1}\sum^N_{j=1}\sin\left(\frac{\pi(n+1)j}{N+1}\right)\sin\left(\frac{\pi j}{N+1}\right)e^{-i2b t\cos\left(\frac{\pi j}{N+1}\right)}\,,\label{WFnDiscr}
\eea
and allow us to compute spread complexity for various finite $N$. Notice that the parameter $a$ only enters via overall phase and will not affect the complexity. 

\begin{figure}[h!]
\centering
\includegraphics[width=.49\textwidth]{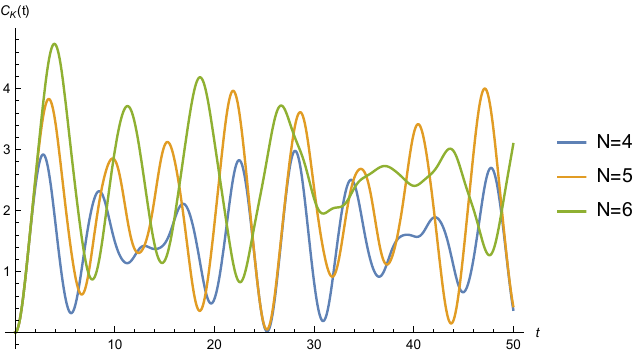}
\includegraphics[width=.49\textwidth]{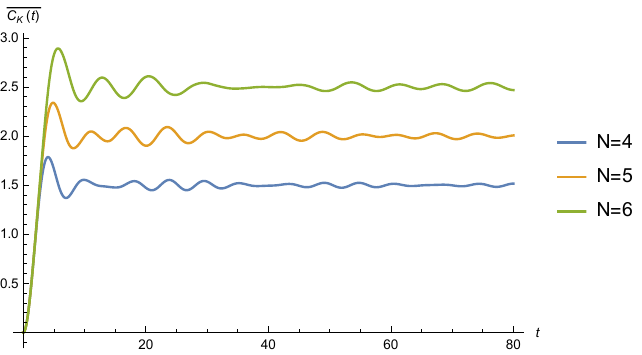}
\caption{Spread complexity (left) and its time average (right)  from \eqref{WFnDiscr} for the discrete lattice model. Late-time, constant plateau is equal $(N-1)/2$. Plot for $b=1$.}
\label{fig:DiscreteAvC}
\end{figure}

In agreement with considerations reviewed in section \ref{sec:1stlaw}, the spread complexity grows as $b^2t^2$ at early times for all values of $N$, but oscillates around different plateau values that are equal $(N-1)/2$ (Fig.~\ref{fig:DiscreteAvC}), as we expect if the late time wavefunction is spread approximately uniformly across the entire Hilbert space.  The time-averaged spread complexity shows a ramp-peak-slope-plateau structure similar to Random Matrix Theory and chaotic systems in general \cite{Balasubramanian:2022tpr,Balasubramanian:2022dnj,Balasubramanian:2023kwd}, although we will see below that the model studied here does not display other finer-grained markers of chaos.\footnote{ Time-averaging similarly smooths out erratic behavior in the spectral form factor of an instance of the SYK theory leading to a smooth ramp and plateau structure \cite{Cotler:2016fpe}.  In fact, time averaging can reveal such structures even in integrable theories with large oscillations as  we are seeing here for the spread complexity -- see, e.g., \cite{Balasubramanian:2016ids} for an example taken from the orbifold limit of the CFT dual to AdS$_3$ black holes.}

Next, let us consider the large-$N$ limit defined in \eqref{eq:theta-def} and giving rise to the wave functions \eqref{OrthnmCont}. The resulting return amplitude, with the same initial state $\ket{0}$, is \cite{Barbon:2019wsy}
\be
S_0(t)=\int^\pi_0d\theta\phi_0(\theta)^2 e^{iE(\theta)t}=e^{iat}\frac{1}{bt}J_1(2bt)\,.
\ee
The infinite set of constant Lanczos coefficients is again
\be
a_n=a\,,\qquad b_n=b\,,
\ee
and the infinite-dimensional Krylov basis is simply $\ket{K_n}=\ket{n}$. Consequently, the Krylov wave functions can be extracted using  overlaps between the energy eigenstates with $\ket{n}$, and are written in terms of Bessel functions \cite{Barbon:2019wsy}
\be
\psi_n(t)=\langle n|\psi(t)\rangle=\int^\pi_0d\theta \phi_n(\theta)\phi_0(\theta)e^{-iE(\theta)t}=-\frac{i^n(n+1)}{bt}J_{n+1}(-2bt)e^{-iat}\,.\label{BJKampl}
\ee
The spread complexity becomes
\bea
C_K(t)=\frac{1}{(bt)^2}\sum^\infty_{n=0}n(n+1)^2J^2_{n+1}(-2bt)\,,
\eea
and can be re-summed analytically \cite{Rabinovici:2023yex} to 
\be
C_K(t)=\frac{16b^2t^2+1}{3}J_1(2bt)^2-\frac{8bt}{3}J_0(2bt)J_1(2bt)+\frac{16b^2t^2+3}{3}J_0(2bt)^2-1\,.\label{SCContk0}
\ee
This again has the same quadratic initial growth as the discrete example above $b^2t^2$, but it never saturates and grows linearly for late times
\be
C_K(t)\simeq\frac{16}{3\pi}bt-1+...\,.
\ee
This spread complexity is plotted in Fig.~\ref{fig:EvolCont}. In Appendix~\ref{sce:InStatek} we compare discrete and continuum model evolution of initial states $\ket{k}$, $k\ge1$, and observe similar evolution,  with some interesting differences, e.g., non-constant $b_n$'s.
\begin{figure}[t!]
\centering
\includegraphics[width=.49\textwidth]{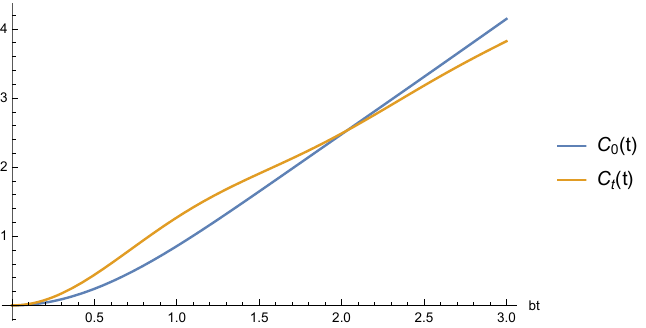}
\includegraphics[width=.49\textwidth]{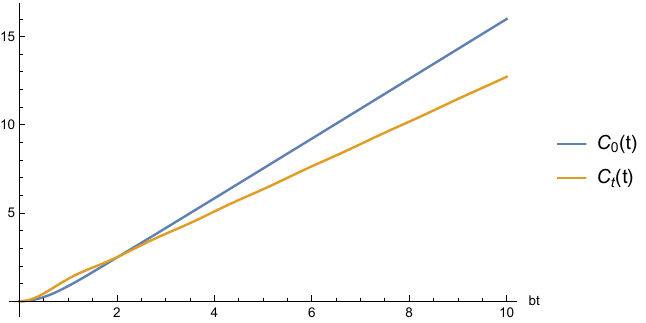}
\caption{Time evolution of spread complexities $C_0(t)$ and $C_t(t)$ in the continuum model for the choice of initial state $\ket{0}$ and \eqref{eq:InTypCont} respectively. Left figure for the early and right for the late times.}
\label{fig:EvolCont}
\end{figure}

Thus, we see that in the large system limit there is an emergent monotonicity in the spread complexity, but at any finite $N$ the spread complexity eventually fails to be monotonic.  In view of this, we could say that there is an emergent second law of monotonic increase in the large system limit, but treated as an effective theory of an underlying finite lattice, the large $N$ theory fails at a sufficiently large timescale.  This is entirely parallel to the standard emergence of the laws of thermodynamics from statistical physics.

\subsubsection{Krylov polynomial approach}\label{subsec:KPolynomi}
We can obtain further insights from  the Krylov polynomials that govern the Krylov basis and their orthogonality properties. The goal is to solve \eqref{Poly3Term} with constant Lanczos coefficients and determine the measure $\mu(E)$, that depends on the initial state $\ket{\psi_0}$ and renders these polynomials orthonormal on the support of the initial state over the energy spectrum.

We will start from finite lattice theory. From the expression for the energy \eqref{DiscEnergyCL}, we first rewrite the momenta in terms of energies as
\be
\frac{\pi j}{N+1}=\arccos\left(\frac{E_j-a}{2b}\right)\,.
\ee
Then, we can check that the solution of \eqref{Poly3Term} with constant Lanczos coefficients $a_n=a$ and $b_n=b$ is given by
\be
P_n(E_j)=\frac{\phi_n(j)}{\phi_0(j)}
=\frac{\sin\left((n+1)\arccos\left(\frac{E_j-a}{2b}\right)\right)}{\sqrt{1-\frac{(E_j-a)^2}{4b^2}}}\,.\label{PnDiscrCL}
\ee
The orthonormality is a consequence of \eqref{OrhonormDisc}, and we  have
\be
\sum^N_{j=1}P_n(E_j)P_m(E_j)w_j=\delta_{n,m}\,,\qquad w_j=|\phi_0(j)|^2=\frac{2}{N+1}\left(1-\frac{(E_j-a)^2}{4b^2}\right)\,,
\ee
fixing the measure to
\be
\frac{d\mu(E)}{dE}=\sum^N_{j=1}\delta(E-E_j)|\langle E_j|\psi_0\rangle|^2=\sum^N_{j=1}\delta(E-E_j)w_j\,.
\ee
This is consistent with \eqref{MeasureGen}, since $\ket{\psi_0}=\ket{0}$ and from \eqref{kinEnBas}
\be
\langle E_j|\psi_0\rangle=\sqrt{\frac{2}{N+1}}\sin\left(\frac{\pi j}{N+1}\right)=\sqrt{\frac{2}{N+1}}\sqrt{1-\frac{(E_j-a)^2}{4b^2}}\,.
\ee
We can also confirm that the Krylov basis vectors coincide with the position vectors \eqref{kinEnBas}
\be
\ket{K_n}=P_n(H)\ket{\psi_0}=\sum^N_{j=1}\phi_0(j)P_n(E_j)\ket{E_j}=\sum^N_{j=1}\phi_n(j)\ket{E_j}=\ket{n}\,.
\ee
Having determined the Krylov polynomials, we can check the late time value of the plateau for spread complexity. Using the first line in \eqref{eq:T-ave} yields
\be
\overline{C_K(\infty)}=\sum^{N}_{j=1}\sum^{N-1}_{n=0}nP_n(E_j)^2|c_j|^4=\sum^{N}_{j=1}\sum^{N-1}_{n=0}n\phi_n(j)^2\phi_0(j)^2=\frac{N-1}{2}\,,
\label{eq:finN-plateau}
\ee
matching the late time values presented in Fig.~\ref{fig:DiscreteAvC}.

The same steps can be repeated in the continuum  $N\to\infty$ limit, where we have
\be
\theta=\arccos\left(\frac{E-a}{2b}\right)\,. \label{eq:thetaandEn}
\ee
The Krylov polynomials $P_n(E)$ have the same form as \eqref{PnDiscrCL} expressed in terms of the continuous energy $E$
\be
P_n(E)=\frac{\phi_n(\theta)}{\phi_0(\theta)}=\frac{\sin\left((n+1)\arccos\left(\frac{E-a}{2b}\right)\right)}{\sin\left(\arccos\left(\frac{E-a}{2b}\right)\right)}\,,
\ee
giving
\be
\ket{K_n}=\int^\pi_0 d\theta\phi_0(\theta)P_n(H)\ket{E(\theta)}=\int^\pi_0 d\theta\phi_n(\theta)\ket{E(\theta)}=\ket{n}\,.\label{eq:Knfor0Cont}
\ee
These Krylov basis vectors  are orthonormal with respect to the measure 
\be
\frac{d\mu(E)}{dE}=\rho(E)=\frac{\sqrt{4b^2-(E-a)^2}}{2\pi b^2}\,,\label{Wignerket0Cont}
\ee
which is supported on the energy interval $E\in[a-2b,a+2b]$ as a consequence of \eqref{OrthnmCont}. 

Notice that this density matches the Wigner's (generally the Marchenko-Pastur \cite{Pastur:1967zca}\footnote{Up to this point, we have not assumed anything about the values of constants $a$ and $b$ in \eqref{HamTB}, other than positivity. There is however a famous subtlety when we parametrize them as $a=1+c$ and $b=\sqrt{c}$ with $c>1$, i.e., the energy lies in the interval $\left[(1-\sqrt{c})^2,(1+\sqrt{c})^2\right]$, and we have to add to the density \eqref{Wignerket0Cont} the so-called ``atom" contribution at $E=0$: $(1-1/c)_{+}\,\delta(E)$, where $(x)_+\equiv \text{max}(0,x)$. Orthogonal polynomials with such a modified measure are also modified, by a shift expressed in terms of the Christoffel–Darboux kernel (to our knowledge, attributed to Uvarov \cite{uvarov1969connection}). We will not explore this further in our work but some discussion about the Marchenko-Pastur distribution and spread complexity can be found in, e.g., \cite{Muck:2024fpb}.}) semi-circle distribution centered at $a$ with variance $\sigma^2=b^2$. This is a surprise.  Our model contains no explicit randomness, and yet the limiting density $\rho(E)$ \eqref{Wignerket0Cont} describing the continuous support of the state $\ket{0}$ precisely matches the characteristic structure of an RMT universality class.  A hint as to why this is happening may come from the Double Scaled SYK model, which has a representation in which the Hamiltonian has a tight-binding form similarly to our lattice model \cite{Berkooz:2018qkz,Berkooz:2024lgq}. In this chord diagram representation of the DSSYK model, the state $\ket{0}$ is  interpreted as the analog of the Thermofield Double of the dual gravity theory state after ensemble averaging \cite{Berkooz:2018qkz,Lin:2022rbf,Rabinovici:2023yex}.  It would be very interesting to better understand this connection and its possible bearing on the surprising emergence that we are seeing of an RMT density of states \eqref{Wignerket0Cont} from the dynamics of an apparently ordered initial state evolving under the action a non-random Hamiltonian.

For both finite and infinite $N$, the Krylov polynomials can be written as Chebyshev polynomials of the second kind defined as
\be
U_n(\cos(\theta))\equiv\frac{\sin((n+1)\theta)}{\sin(\theta)}\,,\qquad \sum^\infty_{n=0}U_n(x)t^n=\frac{1}{1-2tx+t^2}\,,
\ee
and we have the precise relation
\be
P_n(E)=U_n\left(\frac{E-a}{2b}\right)\,.
\ee
We can repeat the computation of the late time value in the continuum limit. We now have
\be
\overline{C_K(\infty)}=\sum^\infty_{n=0}\int^\pi_0d\theta \,nP_n(E(\theta))^2|\langle E(\theta)|0\rangle|^4= \sum^\infty_{n=0}\int^\pi_0d\theta\,n\phi_n(\theta)^2\phi_0(\theta)^2\,.
\ee
This time, we can first perform the integral over $\theta$ to get
\be
\overline{C_K(\infty)}=\sum^\infty_{n=0}\left(\frac{n}{\pi}+\frac{1}{2\pi^2}\frac{\sin(2\pi n)}{n^2+3n+2}\right)\,,
\ee
but the sum does not converge as we can see from writing it as
\be
\overline{C_K(\infty)}=\lim_{N\to\infty}\sum^{N-1}_{n=0}\frac{n}{\pi}=\lim_{N\to\infty}\frac{N(N-1)}{2\pi}\to\infty\,.
\ee
This  confirms that the value of the plateau depends on the dimension of the Krylov subspace, diverging in the continuum limit. The computation also illustrates how variations in the Krylov basis dimension influence the complexity data, including the late-time plateau. 

\subsection{Typical initial states}
In Sec.~\ref{sec:ToyIS0} we considered initial states that were localized on our lattice model.  Below we will consider initial states that are maximally diffuse across the energy eigenstates. First, consider the finite lattice  model with spectrum \eqref{DiscEnergyCL} and initial state
\be
\ket{\psi_0}=\frac{1}{\sqrt{N}}\sum^N_{j=1}\ket{E_j}\,.\label{eqn:TypPS}
\ee
This state is spread uniformly across all the energy eigenstates, and is the pure state analog of an infinite temperature state.  
The typical random state in the Hilbert state  will also have approximately uniform support on all of the energy eigenstates along with random phases. It will therefore have the form (\ref{eqn:TypPS})
if we absorb the phases into the definition of the eigenstates.
Thus, below we will refer to (\ref{eqn:TypPS}) as a  ``typical pure state''.

Unitary evolution gives
\be
\ket{\psi(t)}=e^{-iHt}\ket{\psi_0}=\frac{e^{-iat}}{\sqrt{N}}\sum^N_{j=1}e^{-2ibt\cos\left(\frac{\pi j}{N+1}\right)}\ket{E_j}\,.
\ee
The return amplitude becomes
\be
S(t)=\bra{\psi_0}e^{iHt}\ket{\psi_0}=\frac{1}{N}\sum^N_{j=1}e^{iE_jt}=e^{iat}\frac{1}{N}\sum^N_{j=1}e^{i2bt\cos\left(\frac{\pi j}{N+1}\right)}\,.
\ee
We can find the Lanczos coefficients analytically -- they are
\be
a_n=a,\qquad b_1=\sqrt{2\frac{N-1}{N}}b\,,\qquad b_{n\ge2}=\sqrt{\frac{(N-n)(N-n+3)}{(N-n+1)(N-n+2)}}b\,.
\ee
The sequence of $b_n$'s terminates at $n=N-1$ with $b_{N-1}=\sqrt{2/3}$. At large $N$, $b_1\to\sqrt{2}b$ while $b_{n\ge2}\to b$, and we will confirm this below working directly with the return amplitude in the continuum limit.

Obtaining the wave functions analytically for arbitrary $N$ is more challenging. However,  we can  solve the Lanczos algorithm numerically for fixed $N$ and generate the relevant plots. For concreteness, we pick $N=10$, find the 9 non-trivial Lanczos coefficients
\be
b^2_n=\left\{\frac{9}{5},\frac{44}{45},\frac{35}{36},\frac{27}{28},\frac{20}{21},\frac{14}{15},\frac{9}{10},\frac{5}{6},\frac{2}{3}\right\}b^2\,,
\ee
and the wave functions using \eqref{eq:wf-rec}.

\begin{figure}[h!]
\centering
\includegraphics[width=.49\textwidth]{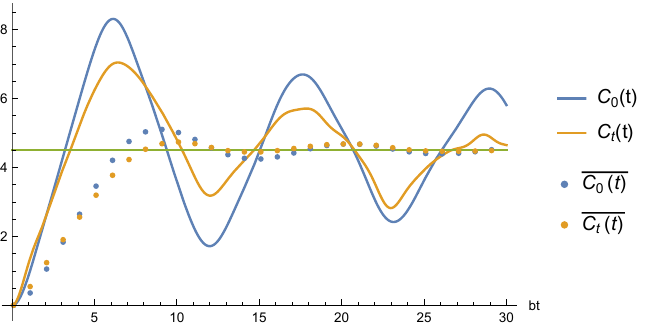}
\includegraphics[width=.49\textwidth]{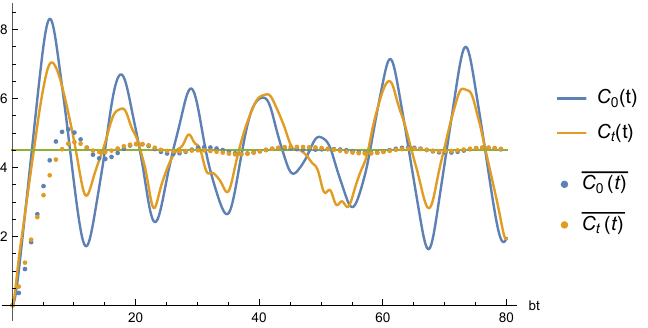}
\caption{Plot of spread complexity for N=10 for the choice of $\ket{0}$ and \eqref{eqn:TypPS} as initial states, for early (left) and late (right) times. Dotted lines denote their time average.  In these plots $C_0(t)$ and $C_t(t)$ are the spread complexities for the localized and typical initial states respectively.   Over-bars denote the time-averaged quantities (see the definition of time-averaging in Sec.~\ref{sec:Krylov}).}
\label{fig:Evol}
\end{figure}

To see the difference that this choice of the initial state makes, in Fig.~\ref{fig:Evol}, we plot spread complexities, denoted as $C_0(t)$ for the localised  initial state $\ket{0}$ and $C_t(t)$ for the typical pure state \eqref{eqn:TypPS}, along with their time averages. The initial growth for the typical pure state is faster, since $b_1^2 > b^2$, but the dynamics for  the localized initial state $\ket{0}$ overtakes and increases to higher peak both with and without time averaging. That said, the general character of the dynamics of spread complexity is the same for both initial states, as is the late time plateau,  reflecting their broad initial support in the energy eigenbasis.

The large-$N$ limit \eqref{eq:theta-def} and \eqref{OrthnmCont} leads to a more analytically-tractable scenario. As before, we pick a normalized initial state 
\be
\ket{\psi_0}=\frac{1}{\sqrt{\pi}}\int^\pi_0d\theta\ket{E(\theta)}\,,\label{eq:InTypCont}
\ee
with the energy spectrum $E(\theta)$ in \eqref{EnSpCont}. 
More generally, we could have started from a ``regulated" state
\be
\ket{\psi^\epsilon_0}=\int^\pi_0d\theta e^{-\frac{\epsilon}{2}H}\ket{E(\theta)}\,,
\ee
which can be seen as the TFD state with $\beta=\epsilon$, but using
\be
I_n(z)=\frac{1}{\pi}\int^\pi_0e^{z\cos(\theta)}\cos(n\theta)d\theta\,,
\ee
we find that the normalization
\be
\langle \psi^\epsilon_0|\psi^\epsilon_0\rangle=\int^\pi_0e^{-\epsilon (a+2b\cos(\theta))}d\theta=\pi e^{-\epsilon a}I_0(2b\epsilon)\,,
\ee
is finite when $\epsilon\to 0$. For this reason, we just directly work with $\ket{\psi_0}$\footnote{One could insist and consider the time evolution
\be
\ket{\psi^\epsilon(t)}=e^{-iHt}\ket{\psi^\epsilon_0}=\frac{e^{-iat}}{\sqrt{\pi I_0(2b\epsilon)}}\int^\pi_0d\theta e^{-(\epsilon+2it)b\cos\theta}\ket{E(\theta)}\,,
\ee
that leads to return amplitude
\be
S(t)=e^{iat}\frac{I_0(2b(\epsilon-it))}{I_0(2b\epsilon)}\,.
\ee
However, the Lanczos coefficients are quite complicated, so we proceed directly with $\epsilon\to0$.}.

The return amplitude is then
\be
S(t)=e^{iat}I_0(2ibt)=e^{iat}J_0(2bt)\,,
\ee
and following our standard procedure, we can find the Lanczos coefficients 
\be
a_n=a\,,\qquad b_1=\sqrt{2}b\,,\qquad b_{n\ge2}=b\,.\label{eqn:LanczoscContTyp}
\ee
In this way, we solve the Schr\"{o}dinger equation to determine the wave functions
\bea
\psi_0(t)&=&e^{-iat}I_0(-2ibt)=e^{-iat}J_0(-2bt)\,,\nn\\
\psi_{n\ge 1}(t)&=&\sqrt{2}e^{-iat}I_n(-2ibt)=\sqrt{2}i^ne^{-iat}J_n(-2bt)\,.\label{eq:weavefunTypCont}
\eea
These expressions are similar to the Bessel functions in \eqref{BJKampl}, but are nevertheless slightly different. Still, using summation rules for Bessel functions, it is easy to check these wave functions provide a normalized  probability distribution over the the Krylov basis. We can then compute the spread complexity 
\be
C_t(t)=2\sum^{\infty}_{n=1}n|J_n(-2bt)|^2=4b^2t^2\left(J_0(2bt)^2+J_1(2bt)^2\right)-2btJ_0(2bt)J_1(2bt)\,.
\ee
At early times, $C(t)$ grows quadratically, consistently with $b_1=\sqrt{2}b$, namely
\be
C_t(t)\sim 2b^2t^2\,,
\ee
as expected from the general considerations in Sec.~\ref{sec:1stlaw}.
However, at late times, it grows linearly 
\be
C_t(t)\sim \frac{4b}{\pi}t-\frac{2\sin(4bt)+1}{8b\pi t}+...\,,
\ee
but with a smaller slope than for the $\ket{0}$ initial state \eqref{SCContk0} (see Fig.~\ref{fig:EvolCont}). 
Note that in the $N\to\infty$ continuum limit the spread complexity increases monotonically , while at finite $N$ it oscillates at late times (Fig.~\ref{fig:Evol}), consistently with the general considerations in Sec.~\ref{sec:1stlaw}.

\begin{figure}[b!]
\centering
\includegraphics[width=.5\textwidth]{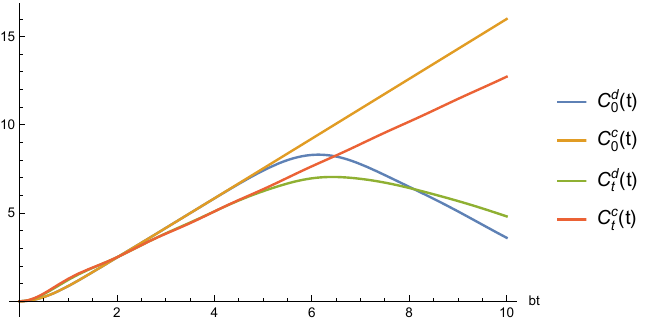}
\includegraphics[width=.4\textwidth]{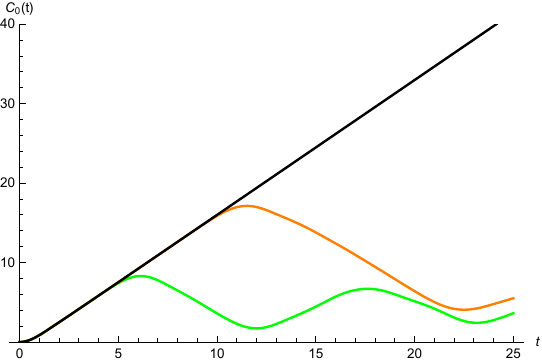}
\caption{Left: comparison of spread complexities in the discrete model (for $N=10$) for initial state $\ket{0}$, denoted by $C^d_0(t)$, and typical pure state denoted by $C^d_t(t)$, with their continuous counterparts $C^c_0(t)$ and $C^c_t(t)$ respectively. Right: spread complexity for the initial state $\ket{0}$ in the continuum limit (black) and discrete models with $N=10$ (green) and $N=20$ (orange). Plot for $b=1$.} 
\label{DiscvsCont}
\end{figure}

We can compare spread complexities for  $\ket{0}$ and the typical initial state at finite and large $N$. The plots in Fig.~\ref{DiscvsCont} show that the discrete and continuous models begin to diverge at a timescale $bt \sim e^S \sim N$, i.e., $t \sim N/b$. To understand this timescale, note first that the energy-time uncertainty relation $\Delta E \, \Delta t \geq \hbar$ lower bounds the amount of time $\Delta t$ that is required to resolve energy gaps of size $\Delta E$.  In the model we are studying here, the dimension of the Hilbert space also controls the gap $|E_k - E_j| \sim \frac{b}{N}$ for $k-j \sim O(1)$ (see \eqref{eq:finite-spectrum}).  So, to resolve these gaps we must have time of order $\Delta t \sim \frac{1}{|E_j - E_k|} \sim \frac{N}{b}$, precisely reproducing the timescale at which the discrete lattice and effective continuum descriptions diverge.  This breakdown occurs at late times at which small differences in energy levels begin to matter.   This is precisely the sort of scenario described in \cite{Balasubramanian:2006iw} where, given enough time, an observer can resolve mass differences between black hole microstates and identify them, signaling a breakdown in the effective semiclassical description in terms of featureless microstates hidden behind a horizon that is valid at early times. Similar considerations are at play in the accounts of the late time, quantum mechanical saturation of wormhole length in \cite{Balasubramanian:2024lqk,Miyaji:2025yvm,Miyaji:2025ucp}.
For a different discussion of the breakdown of effective descriptions of the Krylov chain see \cite{Demulder:2025uda}.

\subsubsection{Krylov polynomial approach}
Next, we derive explicit Krylov basis vectors and corresponding Krylov orthonormal polynomials. For simplicity  we work directly in the continuum limit.

First, using the Lanczos coefficients \eqref{eqn:LanczoscContTyp} in the general expression \eqref{eq:DetP}, we derive $P_0(H)=1$ as well as
\be
P_n(H)=\sqrt{2}\cos(n\theta)=\sqrt{2}\cos\left(n\arccos\left(\frac{H-a}{2b}\right)\right)\,,\qquad n\ge1\,.
\ee
These are the Chebyshev polynomials of the first kind defined as
\be
T_n(\cos(\theta))=\cos(n\theta)\,,
\ee
and the precise relation with our Krylov polynomials is
\be
P_n(E)=\left\{\begin{array}{c}
 T_0\left(\frac{E-a}{2b}\right)=1\,\qquad n=0\,, \\
 \sqrt{2}T_n\left(\frac{E-a}{2b}\right)\,,\qquad n\ge1\,. \\
\end{array}\right.
\ee
We can derive their support on the energy spectrum and the measure  as we did for the initial state $\ket{0}$. Namely, using their orthonormality
\be
\frac{1}{\pi}\int^\pi_0P_n(\theta)P_m(\theta)d\theta=\delta_{n,m}\,,
\ee
and the relation between $\theta$ and the energy $E$ \eqref{eq:thetaandEn}, we have 
\be
\int^{a+2b}_{a-2b}\rho(E)P_n(E)P_m(E)\,dE=\delta_{n,m}\,,
\ee
where the density is now
\be
\rho(E)=\frac{d\mu(E)}{dE}=\frac{1}{\pi}\frac{1}{\sqrt{4b^2-(E-a)^2}}\,.
\ee
Note that it is proportional to the inverse of  Wigner's semi-circle density that appeared for the initial state $\ket{0}$ in \eqref{Wignerket0Cont}, although the significance of this is not clear.
Finally, we can write the Krylov basis vectors for $n\ge1$ as
\be
\ket{K_n}=P_n(H)\ket{\psi_0}=\sqrt{\frac{2}{\pi}}\int^\pi_0d\theta\cos(n\theta)\ket{E(\theta)}\,,\label{eqn:TypKn}
\ee
and it is easy to check that projecting the time-evolved state \eqref{eq:InTypCont} on this basis reproduces the wave functions \eqref{eq:weavefunTypCont}.
\subsection{Koherence}
\begin{figure}[t!]
\centering
\includegraphics[width=.45\textwidth]{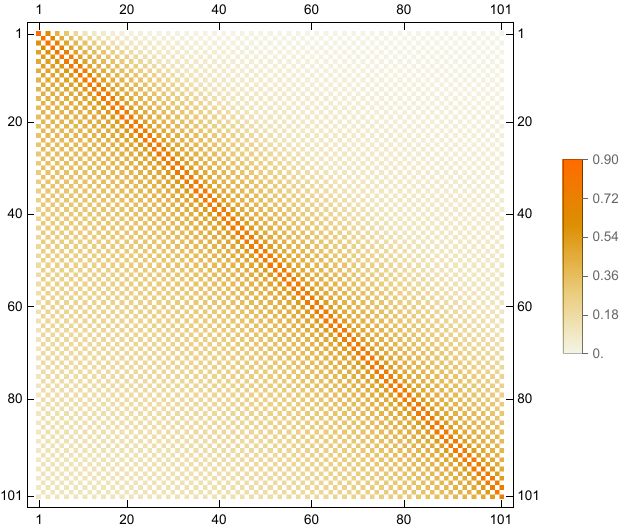}
\includegraphics[width=.52\textwidth]{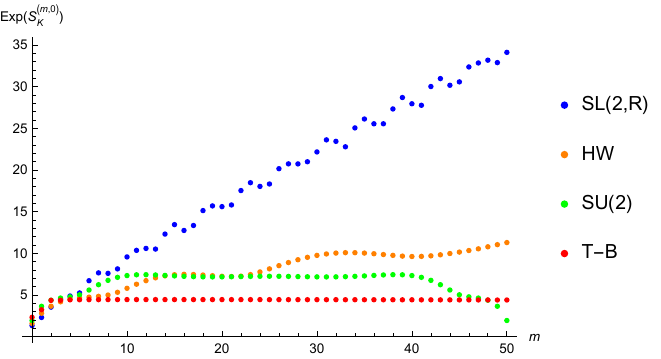}
\caption{Plot of matrix elements \eqref{OverlapToy0Typ} for $n,m\in\{0,...,100\}$, and associated exponential of Koherence (for $n,m$ up to 50) for our tight-binding lattice model. The two Krylov bases whose overlaps are shown in these plots  are constructed with the same Hamiltonian but with the initial states $\ket{0}$ and the typical pure state \eqref{eq:InTypCont}. For comparison, we present  Koherence for this model together with the Lie algebra examples.}
\label{fig:OverlapsToy0Typ}
\end{figure}

Finally, we can compute the overlaps between the Krylov basis vectors associated to the typical initial state \eqref{eqn:TypPS} and the localized initial state $\ket{0}$. Denote the basis \eqref{eq:Knfor0Cont} by $\ket{K^{(0)}_n}=\ket{n}$ and keep the notation of \eqref{eqn:TypKn} for $\ket{K_n}$. Then the overlap with the 0-th vector is
\be
\langle K^{(0)}_n|K_0\rangle=
\frac{\sqrt{2}}{\pi}\int^\pi_0\sin\left((n+1)\theta\right)d\theta=\frac{2\sqrt{2}}{\pi(n+1)}\cos^2\left(\frac{n\pi}{2}\right)\,.
\ee
This vanishes for odd n, whereas it equals $2\sqrt{2}/(\pi(n+1))$ for even $n$. For $m\ge1$, the overlap is
\bea
\langle K^{(0)}_n|K_m\rangle&=&\frac{2}{\pi}\int^\pi_0\sin\left((n+1)\theta\right)\cos(m\theta)d\theta
=\frac{2(n+1)(1+(-1)^{n+m})}{\pi((n+1)^2-m^2)}\,.
\eea
These are non-trivial unless $n+m$ is odd or $m=n+1$\footnote{One should be careful since, for these values, the denominator of our general formula vanishes but the numerator is zero so the limit should be extracted appropriately (explicit computation of these values does not cause a problem).}. This way we can write all of the overlaps analytically
\be
\langle K^{(0)}_n|K_m\rangle=\left\{\begin{array}{c}
\frac{2\sqrt{2}}{\pi(n+1)}\cos^2\left(\frac{n\pi}{2}\right)\,\qquad m=0\,, \\
 \frac{2(n+1)(1+(-1)^{n+m})}{\pi((n+1)^2-m^2)} \,\qquad m\neq n+1\,,\\
 0 \,\qquad\qquad\qquad\qquad m=n+1\,.\\
\end{array}\right.\label{OverlapToy0Typ}
\ee
We plot these matrix elements in Fig.~\ref{fig:OverlapsToy0Typ} (left). It is clear that the overlap between the two bases is centered along the diagonal and does not grow as for our examples in Sec.\,\ref{sec:SolvableEx}. 

The overlaps can be used to compute the Koherence   (Fig.~\ref{fig:OverlapsToy0Typ}, right), which we can compared to results for the group manifold examples in Sec.~\ref{sec:SolvableEx}.  We can clearly see that Koherence in our lattice model quickly saturates to a plateau, similar to the SU(2) example. This signals the integrable structure of our simple model, even though its spread complexity  grows linearly as expected for chaotic models. This suggests  that Koherence is a fine-grained tool that can distinguish chaotic and integrable models which may both have  fast scrambling properties. 

\section{Conclusions and Outlook}\label{sec:Conclusions}
In this work, we studied  how variations in the evolving state and Hamiltonian  shape the dynamics of spread complexity, the Lanczos coefficients, and the Krylov basis.  We introduced several new tools to study variations in complexity, including  an orthogonal polynomial formulation of these variations, a measure of overlaps between Krylov bases corresponding to nearby Hamiltonians or states, their associated overlap Shannon entropy dubbed {\it Koherence}, and the relative entropy between the distributions over different Krylov bases,  

We also established a direct link between Koherence and the relative entropy of coherence familiar from quantum resource theory \cite{baumgratz2014quantifying}. It would be interesting to develop this connection, as it could lay the foundation for a resource-theoretic formulation of quantum spread complexity. Such a framework would offer a new paradigm for probing black hole interiors, possibly with local infalling operators \cite{Caputa:2024sux}, and could place recent discussions of “un-complexity” \cite{Brown:2017jil} on a more rigorous footing. We demonstrated the effectiveness of these methods in a range of analytically tractable examples, such as systems governed by Lie algebras and a tight-binding lattice model with constant hopping amplitudes. Extending this framework to chaotic and disordered systems,  notably the SYK model and random matrix ensembles, will be an important next step.

From our explicit examples, we learned that the behavior of spread complexity captures how the energy distribution of the initial state is encoded within the sequence of Krylov states. At short times, the inability to distinguish non-orthogonal quantum states motivates the expansion into an orthogonal Krylov basis, where the minimal orthogonalization time is set by the energy moments of the initial state. These moments govern the early-time expansion of spread complexity: accessing higher-order moments requires longer evolution times. At late times, the dynamics exhibit oscillations around a plateau whose value is determined by the underlying energy gaps, while the infinite-time average encodes, in a nonlinear fashion, the full energy distribution of the initial state. Since this encoding appears through a hierarchy of orthogonal polynomials whose recursion coefficients depend on spectral statistics, studying the variation of spread complexity with respect to the choice of initial state provides a concrete route to uncovering how microscopic spectral data give rise to emergent, thermodynamic-like relations in quantum complexity. In this sense, our analysis generalizes the “first-law” discussions of Nielsen’s geometric complexity \cite{Bernamonti:2019zyy} into a constructive framework within the Krylov approach.

In  our solvable lattice  model with a tunable Hilbert space dimension we investigated the effects of a thermodynamic limit on spread complexity and the Krylov basis.  In this limit the Hilbert space dimension $N$ becomes infinite with an accompanying transition between discrete and continuous spectra.  At large $N$, the spread complexity exhibits long-term linear growth, 
while for finite N it saturates and undergoes oscillations around a plateau, signaling a breakdown of the limiting effective field theory description. The monotonic linear growth of complexity at large $N$ is consistent qualitatively with Lloyd's bound  \cite{lloyd2000ultimate}. It would be interesting to understand what general properties the underlying density of states must have to enable this linear growth.

Finally the orthogonal-polynomial approach to spread complexity that we described renders the interplay between the Hamiltonian spectrum, the support of the initial state in the energy basis, and the evolution of spread complexity particularly transparent. This mathematical framework, which can exploit tools developed in other contexts \cite{deift2000orthogonal,johansson2006random},
offers a powerful analytic handle on the mechanisms behind linear growth and invites further adaptation of mathematical techniques to uncover and classify universal features of quantum complexity.

Our results concerning the late-time breakdown of large-system, continuous  effective descriptions, offer an alternative perspective on a tension between three standard components of conventional models of the world: (a) the assumption that observers only perform low-complexity operators, (b) the growth of the computational complexity of states with time, and (c) the possibility of making measurements over exponentially long times.  The tension is that an observer with limited computational power, restricted to simple operations, might nonetheless overcome these limitations if granted another resource such as {\it time}.   This is familiar in circuit complexity: low-depth gates applied for sufficiently long times can generate highly complex, effectively deep circuits. In holography, this intuition is mirrored in the code-subspace picture. The code subspace formalizes the set of observables for which a bulk Effective Field Theory description is valid: it essentially contains the  algebra of single-trace operators whose products remain small (do \textit{not} scale with $N$). This algebra does not strictly close, i.e., multiplying operators or evolving them for long times eventually pushes us outside the EFT regime. Dynamics, therefore, generically threaten the consistency of the EFT approximation. 

Black holes, with their classically growing interiors, offer a natural arena in which to test these ideas. The event horizon itself is an emergent feature of the large-$N$ approximation where a finite-dimensional Hilbert space appears effectively infinite. Simple operators cannot resolve fine-grained deviations and thus yield predictions well approximated by large-$N$ EFT in a semiclassical black-hole geometry. However, infalling observers should be able to probe phenomena that would require asymptotic boundary observers to wait parametrically long times. Attempts to construct boundary operators that reproduce interior creation and annihilation operators with standard commutation relations consistently run into obstructions, manifesting as state dependence \cite{Papadodimas:2013jku} or requirements of exponentially large boundary complexity \cite{Harlow:2013tf,Akers:2024wre}. These difficulties may be concrete signatures of long-time, high-complexity regimes in which effective descriptions of the kind we are used to necessarily fail. The spread complexity framework we described is one avenue for investigating these ideas.

\acknowledgments
We would like to thank Brian Creed, Giuseppe Di Giulio, Javier Mag\'an, Onkar Parrikar, Dimitrios Patramanis, Bo Sundborg, Evita Verheijden for discussions. VB was supported in part by the DOE through DE-SC0013528 and QuantISED grant DE-SC0020360, and in part by the Eastman Professorship at Balliol College, University of Oxford. PC is supported by NCN Sonata Bis 9 2019/34/E/ST2/00123 grant and the ERC Consolidator grant (number: 101125449/acronym: QComplexity). Views and opinions expressed are however those of the authors only and do not necessarily reflect those of the European Union or the European Research Council. Neither the European Union nor the granting authority can be held responsible for them. JS is supported by the Science and Technology Facilities Council [grant number ST/X000494/1]. VB is grateful to the organizers of the YITP workshop "Extreme Universe 2025" YITP-T-25-01 held at YITP, Kyoto University for the stimulating environment in which this work was completed.
\appendix
\section{Two-state system} \label{sec:A0}
To illustrate our definitions from Sec.~\ref{sec:Krylov}, we consider arguably the simplest example of a two-level system with initial state and the Hamiltonian in the energy basis
\be
\ket{\psi_0}=\cos\theta\ket{E_1}+\sin\theta e^{i\phi}\ket{E_2}\equiv\sum^2_{k=1}c_k\ket{E_k},\qquad H=\left(
\begin{array}{cc}
 E_1 & 0 \\
 0 &E_2 \\
\end{array}
\right).
\label{eq:N2-ket}
\ee
The time evolution of this state is
\be
\ket{\psi(t)}=e^{-iHt}\ket{\psi_0}=\cos\theta e^{-iE_1 t}\ket{E_1}+\sin\theta e^{i\phi}e^{-iE_2 t}\ket{E_2}\,,
\ee
and the return amplitude is given by
\be
S(t)=\bra{\psi_0}e^{iHt}|\psi_0\rangle=e^{itE_1}\cos^2\theta+e^{itE_2}\sin^2\theta\,.
\ee
Even in this simple example we already see that, for non-trivial complexity, it is crucial that $E_1\neq E_2$ or $\theta\neq 0$ (for which the amplitude is a pure phase and leads to a trivial dynamics on the Krylov chain).\\
From this amplitude we can extract only three non-zero Lanczos coefficients
\be
a_0=E_1\cos^2\theta+E_2\sin^2\theta\,,\qquad a_1=\frac{E_1+E_2}{2}+\frac{\Delta E}{2}\cos(2\theta)\,,
\ee
and
\be
b_1=\frac{E_2-E_1}{2}\sin(2\theta)=\frac{\Delta E}{2}\sin(2\theta)\,,\label{b1Toy2Lev}
\ee
where, without loss of generality, we denoted $\Delta E=E_2-E_1>0$.\\
Next, the Krylov basis vectors are 
\be
\ket{K_0}=\ket{\psi_0}\,,\qquad \ket{K_1}=P_1(H)\ket{\psi_0}=-\sin\theta\ket{E_1}+\cos\theta e^{i\phi}\ket{E_2}\,,
\ee
where we used the polynomials \eqref{eq:DetP}, i.e. $P_0(H)=1$ and $P_1(H)=b^{-1}_1(H-a_0)$, which for our setup become
\be
P_1(E_1)=-\tan\theta\,,\qquad P_1(E_2)=\cot\theta\,.
\ee
Then indeed the non-trivial Krylov vector is written as
\be
\ket{K_1}=\cos\theta P_1(E_1)\ket{E_1}+\sin\theta e^{i\phi}P_1(E_2)\ket{E_2}\,.
\ee
These Krylov polynomials are orthonormal with respect to the measure \eqref{MeasureGen} corresponding to the density 
\be
\rho(E)=\frac{d\mu(E)}{dE}=\sum^2_{k=1}\delta(E-E_k)|c_k|^2=\delta(E-E_1)\cos^2\theta+\delta(E-E_2)\sin^2\theta\,.
\ee
The two amplitudes that solve the Schr\"{o}dinger equation \eqref{SEq} are then 
\bea
\psi_0(t)&=&S(t)^*=e^{-itE_1}\cos^2\theta+e^{-itE_2}\sin^2\theta\,,\nn\\ \psi_1(t)&=&\frac{1}{2}\sin(2\theta)\left(e^{-itE_2}-e^{-iE_1t}\right)\,,
\eea
and the spread complexity is expressed by the energies and data of the initial state as
\bea
C_K(t)&=&\sin^2(2\theta)\sin^2\left(\frac{\Delta E}{2}t\right)=\frac{4b^2_1}{\Delta E^2}\sin^2\left(\frac{\Delta E}{2}t\right)\,.\label{Compl2Levb1}
\eea
It follows the same pattern as observed for the $SU(2)$ algebra \eqref{SpreadGenSU2} with $\mathcal{D}=\Delta E$ and for $j=1/2$ where the Krylov chain has only two sites. Note that, in all the steps above, the dependence on the data of the Hamiltonian (the two energies) as well as the initial state is explicit and can be easily varied. Nevertheless, again the variation of $b^2_1$ controls the initial state and changing the Hamiltonian is equivalent to tuning $\Delta E$. More precisely, due to a change $\delta \theta$, or $\delta b_1^2$, in the original quantum state, complexity satisfies
\begin{equation}
    \delta C_K(t) = \frac{4}{\Delta E^2} \sin^2\left(\frac{\Delta E}{2} t\right)\,\delta b_1^2\,.
\label{eq:1lawN2}
\end{equation}
This is consistent with our universal short time expansion \eqref{eq:short-t} since there are no higher order independent time scales for such simple system. Alternatively, such expansion could be re-summed and it would reproduce the sine function.

Notice that using the identity $\delta b_1^2 = \delta \langle H^2\rangle - 2\langle H \rangle \delta \langle H\rangle$, we could easily rewrite \eqref{eq:1lawN2} in the general framework used in \eqref{eq:uni-1law}. This would allow us to identify
\begin{equation}
    \nu_1(t) = -8 \frac{\langle H \rangle}{\delta E^2} \sin^2\left(\frac{\Delta E}{2} t\right)\,, \qquad \nu_2(t) = \frac{4}{\Delta E^2} \sin^2\left(\frac{\Delta E}{2} t\right)\,.
\end{equation}

It is also instructive to express this result in terms of the Krylov polynomials according to \eqref{eq:gen-Ck}. For us the formula yields
\bea
C_K(t)&=&\sum^2_{k,l=1}P_{1}(E_k)P_1(E_l)e^{i(E_k-E_l)t}|c_k|^2|c_l|^2\nn\\
&=&|c_1|^4P_1(E_1)^2+|c_2|^4P_1(E_2)^2+2|c_1|^2|c_2|^2P_1(E_1)P_1(E_2)\cos(\Delta E t)\,.\label{CAPc}
\eea
Next, we consider the two types of averaged complexity introduced in Sec.~\ref{sec:Krylov}. The first one will be over an ensemble of theories with different energies. To model this, we simply integrate over the energy differences $\Delta E$ with the GUE measure (see e.g. details in \cite{Caputa:2024vrn,Caputa:2025ozd})
\be
\langle C_K(t)\rangle \equiv\sqrt{\frac{2}{\pi}}\int^\infty_0 e^{-\frac{\Delta E^2}{2}}\Delta E^2\,C_K(t)\, d(\Delta E)\,,
\ee
which for \eqref{Compl2Levb1} yields
\be
\langle C_K(t)\rangle =\frac{\sin^2(2\theta)}{2}\left(1+e^{-t^2/2}(t^2-1)\right).
\ee
We can compare it with the time-averaged spread complexity, defined in \eqref{TAComplexity}, which becomes
\be
\overline{C_K(t)}=\frac{\sin^2(2\theta)}{2}\left(1-\frac{\sin(\Delta E\, t)}{\Delta E\, t}\right)\,.
\ee
It is interesting to point that the time derivatives of the exact answer \eqref{Compl2Levb1}, as well as time derivatives the two averages above, do not have definite sign. However, the time average of the first derivative is positive i.e., satisfies a 2nd law for spread complexity
\begin{equation}
    \overline{\dot{C}_K(t)} = \frac{2b_1^2}{t(\Delta E)^2}\left(1-\cos \Delta E\, t\right) \geq 0\,.
\end{equation}

\begin{figure}[h!]
\centering
\includegraphics[width=.6\textwidth]{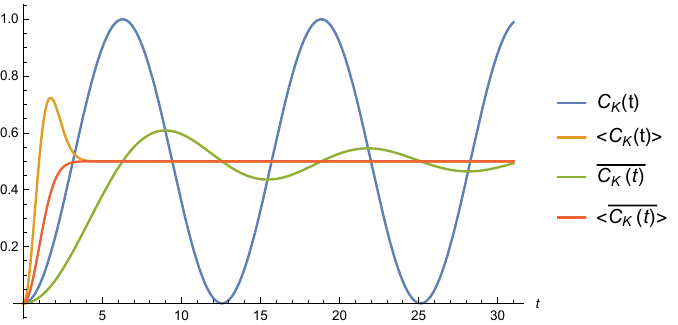}
\caption{Spread complexity and its averages for the two-state system. Plot for $\Delta E=1/2$ and $\theta=\pi/4.$}
\label{fig:Compl2Leve}
\end{figure}

Finally, we can combine the two averages in arbitrary order (procedures commute in this simple model) to derive
\be
\overline{\langle C_K(t)\rangle}=\langle \overline{C_K(t)}\rangle=\frac{\sin^2(2\theta)}{2}\left(1-e^{-t^2/2}\right)\,.
\ee
We present the spread complexity \eqref{Compl2Levb1} and its averages above on Fig. \ref{fig:Compl2Leve}.

Several comments are in order at this point. Firstly, observe that the initial, universal quadratic growth of complexity is altered by the averaging procedures
\bea
\left.
\begin{array}{c}
 C_K(t) \\
 \langle C_K(t)\rangle \\
 \overline{C_K(t)} \\
 \langle \overline{C_K(t)}\rangle \\
\end{array}
\right\}\simeq b^2_1t^2\times\left\{\begin{array}{c}
 1 \\
 3/\Delta E^2 \\
 1/3 \\
 1/\Delta E^2 \\
\end{array}\right.\,,
\eea
where $b_1$ is the Lanczos coefficient \eqref{b1Toy2Lev}. Secondly, averaging over $\Delta E$ (orange) provides a toy example of the RMT evolution (for $N=2$ random matrices) with ramp-peak-slope-plateau curve for spread complexity \cite{Balasubramanian:2022tpr}. Thirdly, applying both averages removes correlations between energies and kills the peak (red). Moreover, at late times, both ways of averaging asymptote to the same plateau expressed as
\be
\lim_{t\to\infty}\langle C_K(t)\rangle=\lim_{t\to\infty}\overline{C_K(t)}=\lim_{t\to\infty}\langle\overline{C_K(t)}\rangle=\frac{1}{2}\sin^2(2\theta)=\frac{2b^2_1}{\Delta E^2}\,,
\ee
where the last equality comes from \eqref{Compl2Levb1}, but the explicit dependence on $\Delta E$ obviously cancels in the ratio. We can also explicitly compute the variation of the plateau value
\begin{equation}
    \delta \overline{C_K(\infty)}=\sin 2\theta \cos2\theta\,\delta \theta = 2\,\frac{\delta b_1^2}{(\Delta E)^2}\,.
\end{equation}

This late-time plateau of averaged complexities can be equivalently expressed using the data of the initial state and Krylov polynomials $P_n(E)$. Indeed, using \eqref{CAPc} we can write it as
\be
\overline{C_K(\infty)}=\langle C_K(\infty)\rangle=|c_1|^4P_1(E_1)^2+|c_2|^4P_1(E_2)^2\,,
\ee
which is just the time-independent (diagonal) part of \eqref{CAPc}.

Fourthly, the value of the peak for $\langle C_K(t)\rangle$, which happens at $t=\sqrt{3}$ in this model, is given by
\be
\langle C_K(t)\rangle_{max}=\frac{1}{2}\left(1+\frac{2}{e^{3/2}}\right)\sin^2(2\theta)\,.
\ee
It is not immediately obvious how this time depends on the data of the system but, since the Krylov subspace is 2-dimensional, the complexity is directly related to the return amplitude $1-|S(t)|^2$ (spectral form factor if we interpret $\ket{\psi_0}$ as the TFD state for this model). This way, we can connect this time to the the Thouless time \cite{thouless1974electrons} for the $2\times 2$ random matrix. The value of the peak is clearly governed by $\theta$ of the initial state. \\
Last but not least, careful reader should notice that none of the Krylov quantities was sensitive to the phase angle $\phi$ of the initial state \eqref{eq:N2-ket}. This suggests that perhaps more fine-grained tools, beyond Krylov methods, are be required to extract such information.

\section{Details of exactly-solvable examples}\label{appena}
We start with the SL(2,$\mathbb{R}$) Lie algebra defined by generators
\be
[L_{0},L_{\pm 1}]=\mp L_{\pm 1},\qquad [L_1,L_{-1}]=2L_0\,,
\ee
where $L_{+1}$ and $L_{-1}$ play the role of lowering and rising ladder operators. The orthonormal basis for the algebra is obtained in a standard way by acting with the rising operators
\begin{equation}
   \vert h,n\rangle\equiv\sqrt{\frac{\Gamma(2h)}{n!\Gamma(2h+n)}} L^n_{-1}\vert h\rangle \,,
    \qquad
   \langle h,n \vert h,m\rangle=\delta_{nm}\,,\label{AlgBasisSL2R}
\end{equation}
and the action of the algebra generators on it is
\begin{eqnarray}
  L_{0} \vert h,n\rangle&=&(h+n)\vert h,n\rangle \,,\nn
   \\
  L_{1} \vert h,n\rangle&=&\sqrt{n(n+2h-1)}\vert h,n-1\rangle \,,\nn
   \\
   L_{-1}\vert h,n\rangle&=&\sqrt{(n+1)(n+2h)}\vert h,n+1\rangle \,.\label{ActionLns}
\end{eqnarray}
In the main text, we used the BCH formulas for evolving with a general Hamiltonian of the form
\be
H=a_0L_0+a_1 L_1+a_{-1}L_{-1},
\ee
such that
\bea
e^{-it H}&=&e^{AL_{-1}}e^{BL_0}e^{CL_{1}},
\eea
where\footnote{This can be checked for any representation of SL(2,$\mathbb{R}$), including e.g. the two-dimensional, non-unitary representation: $L_0=\frac{1}{2}\left(
\begin{array}{cc}
 1 & 0 \\
 0 & -1 \\
\end{array}
\right)$, $L_1=\left(
\begin{array}{cc}
 0 & 0 \\
 -1 & 0 \\
\end{array}
\right)$, $L_{-1}=\left(
\begin{array}{cc}
 0 & 1 \\
 0 & 0 \\
\end{array}
\right)$.}
\be
B=-2\log\left[\cosh\left(\frac{\mathcal{D}t}{2}\right)+\frac{ia_0}{\mathcal{D}}\sinh\left(\frac{\mathcal{D}t}{2}\right)\right],
\ee
and
\be
A=-\frac{2a_{-1}}{a_0-i\mathcal{D}\coth\left(\frac{\mathcal{D}t}{2}\right)},\qquad C=-\frac{2a_{1}}{a_0-i\mathcal{D}\coth\left(\frac{\mathcal{D}t}{2}\right)}
\ee
as well as
\be
\mathcal{D}=\sqrt{4a_1a_{-1}-a^2_0}.
\ee
Given general Lanczos coefficients that are directly related to the above structure
\be
a_n=G(h+n),\qquad b_n=A\sqrt{n(n+2h-1)},
\ee
the spread complexity becomes \cite{Balasubramanian:2022tpr}
\be
C_K(t)=\frac{2h}{1-\frac{G^2}{4A^2}}\sinh^2\left(At\sqrt{1-\frac{G^2}{4A^2}}\right)\,.\label{SComplSL2R}
\ee
To find the Krylov basis, let us  first notice that with the BCH we can derive
\be
D^\dagger(\xi) H D(\xi)=f(z)L_0+\bar{\alpha}(z)L_1+\alpha(z)L_{-1}\,,
\ee
where the SL(2,$\mathbb{R}$) displacement operator $D(\xi)$ was defined in \eqref{CSSU11}, $f(z)$ was introduced in \eqref{Dandf}, and for the moment we will only need
\be
\sqrt{\alpha(z)\bar{\alpha}(z)}=\frac{\sqrt{\mathcal{D}^2+f(z)^2}}{2}\,.
\ee
Then we follow the Lanczos algorithm \eqref{LanczosAlg} step by step from $n=0$. The first two steps are 
\be
\ket{K_0}=D(\xi)\ket{h},\qquad \ket{A_1}=(H-a_0)\ket{K_0}\,,
\ee
and we get
\be
a_0=\bra{K_0}H\ket{K_0}=f(z)h\,.
\ee
Then, we can write our vector as
\be
\ket{A_1}=D(\xi)\left(f(z)L_0+\alpha(z)L_{-1}\right)\ket{h}-f(z)h D(\xi)\ket{h}=\sqrt{2h}\alpha(z)D(\xi)\ket{h,1}\,,
\ee
where we employed the algebra basis \eqref{AlgBasisSL2R} and action \eqref{ActionLns} of the generators on it.\\
Finally, after normalising, we fix the Lanczos coefficient and the first Krylov vector
\be
b_1=\sqrt{\alpha(z)\bar{\alpha}(z)}\sqrt{2h},\qquad \ket{K_1}=\sqrt{\frac{\alpha(z)}{\bar{\alpha}(z)}}D(\xi)\ket{h,1}\,.
\ee
Following these steps yields \eqref{KBSU2RF}.

Similar analysis can be done for the SU(2) algebra defined by
\be
[J_0,J_{\pm}]=\pm J_\pm,\qquad [J_+,J_-]=2J_0\,.
\ee
The basis is labeled by spin $j$ and has $2j+1$ elements 
\be
\ket{j,-j+n}=\sqrt{\frac{\Gamma(2j+1-n)}{n!\Gamma(2j+1)}}J^n_{+}\ket{j,-j}\,,
\ee
on which the generators act as
\bea
J_0\ket{j,-j+n}&=&(-j+n)\ket{j,-j+n}\,,\nn\\
J_+\ket{j,-j+n}&=&\sqrt{(n+1)(2j-n)}\ket{j,-j+n+1}\,,\nn\\
J_-\ket{j,-j+n}&=&\sqrt{n(2j-n+1)}\ket{j,-j+n-1}\,.\label{Jmrepn}
\eea
For general Lanczos coefficients in this class
\be
a_n=G(-j+n)\,,\qquad b_n=A\sqrt{n(2j-n+1)}\,,
\ee
the spread complexity becomes \cite{Balasubramanian:2022tpr}
\be
C_K(t)=\frac{2j}{1+\frac{G^2}{4A^2}}\sin^2\left(At\sqrt{1+\frac{G^2}{4A^2}}\right)\,.
\ee
To derive the Krylov basis for non-trivial $\xi$, analogously to SL(2,$\mathbb{R}$), we first compute
\be
D^\dagger(z)HD(z)=f(z)J_0+\alpha(z)J_++\bar{\alpha}(z)J_-,\qquad \sqrt{\alpha\bar{\alpha}}=\frac{\sqrt{\mathcal{D}^2-f(z)^2}}{2}\,,\label{DHDSU2}
\ee
and, after applying Lanczos algorithm, \eqref{KBSU2} follows. 
\section{More on the linear complexity  growth: initial states $\ket{k}$}\label{sce:InStatek}
In this appendix, we investigate the sensitivity of the results in Sec.~\ref{sec:ToyIS0} to changes in the initial conditions. To this end, we consider the time evolution under the same Hamiltonian~\eqref{HamTB}, but starting from different position eigenstates $\ket{k}$. Equivalently, this setup can be viewed as the evolution of the Krylov basis vectors $\ket{K_n} = \ket{n}$ generated from the initial state $\ket{0}$ under the same Hamiltonian $H$. This perspective provides an interesting protocol for probing scrambling dynamics and for examining how the evolution progressively “forgets” (or, as we will see, remembers) information about its initial configuration. In particular, we will focus on how these features are reflected in the structure of the Lanczos coefficients and in the time evolution of the spread complexity.

\begin{figure}[t!]
\centering
\includegraphics[width=.49\textwidth]{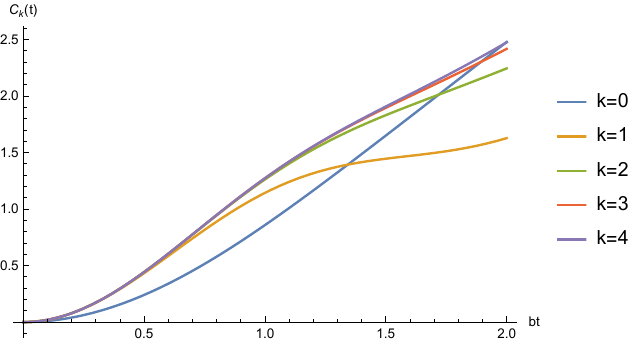}
\includegraphics[width=.49\textwidth]{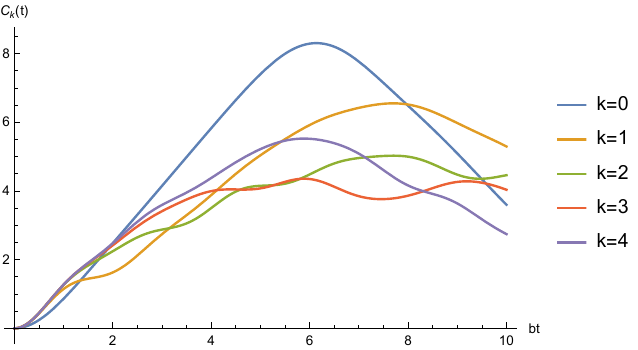}
\caption{Spread complexities for $N=10$ (as functions of $bt$) at early times (left) and later times (right), for several different initial states $\ket{k}$.}
\label{fig:N10ET}
\end{figure}

\begin{figure}[b!]
\centering
\includegraphics[width=.50\textwidth]{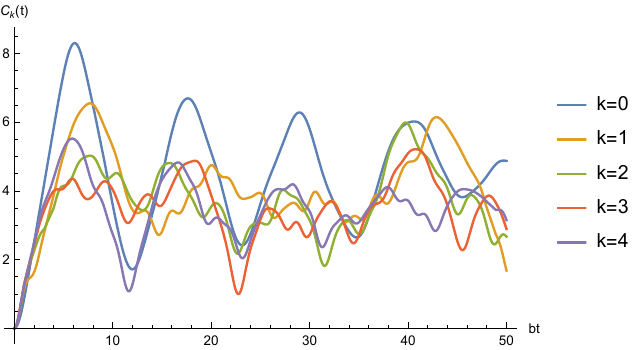}
\includegraphics[width=.48\textwidth]{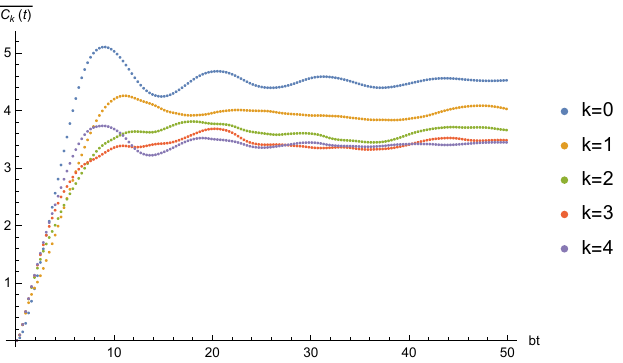}
\caption{Spread complexities for $N=10$ (left) and their time averages (right) for late times, for several different initial states $\ket{k}$.}
\label{fig:N10LT}
\end{figure}

We begin from the discrete model, and compute return amplitudes for $\ket{k}$ as follows
\bea
S_k(t)=\sum^N_{j=1}\phi_k(j)^2e^{iE_jt}=e^{iat}\frac{2}{N+1}\sum^N_{j=1}\sin^2\left(\frac{\pi(k+1)j}{N+1}\right)e^{i2b t\cos\left(\frac{\pi j}{N+1}\right)}\,.
\eea
In general, we see that they have the symmetry $S_k=S_{N-k-1}$. This way, we only get $\lceil N/2\rceil$ non-trivial cases with different $\ket{k}$. Since $a$ only enters via complex phase, we find that for all $\ket{k}$, $a_n=a$. In each case we find $N-1$ $b_n$'s from $b_1$ to $b_{N-1}$. Firstly, for all $k\ge1$,  we have the same Lanczos coefficient
\be
b_1=\sqrt{2}b\,,
\ee
while for $k=0$ we had simply $b_1=b_n=b$. However, higher Lanczos coefficients depend on $k$ more non-trivially. For concreteness, lets consider the case of $N=10$ where we have $k\in\{0,...,4\}$ different possibilities. For $k=1$ Lanczos coefficients split into odd and even
\be
b_{2n-1}=\sqrt{\frac{n+1}{n}}b\,,\qquad b_{2n}=\sqrt{\frac{n}{n+1}}b\,,\label{OddEvenbDiscr}
\ee
and go up to $b_{8}$, finished by the last coefficients $b_{9}=\frac{1}{\sqrt{5}}b$. We can write them explicitly
\be
b_n=\left\{\sqrt{2},\frac{1}{\sqrt{2}},\sqrt{\frac{3}{2}},\sqrt{\frac{2}{3}},\frac{2}{\sqrt{3}},\frac{\sqrt{3}}{2},\frac{\sqrt{5}}{2},\frac{2}{\sqrt{5}},\frac{1}{\sqrt{5}}
   \right\}b\,,\qquad \text{for k=1}.
\ee
For higher $k$ this pattern repeats but these odd and even coefficients for $k=1$ are now separated from $b_1$ by $k-1$ coefficients equal to $b$. If we increase $N$, then $k-1$ values of $b_n=b$ appears after each non-trivial pair and also final coefficients have more non-trivial dependence on $N$ and $k$ as we approach final $n=N-1$. We can see this explicitly 
\bea
b_n&=&\left\{\sqrt{2},1,\frac{1}{\sqrt{2}},\sqrt{\frac{3}{2}},1,\sqrt{\frac{2}{3}},\frac{2}{\sqrt{3}},\frac{1}{2},\frac{\sqrt{3}}{2}\right\}b\,,\qquad \text{for k=2}\,\nn\\
b_n&=&\left\{\sqrt{2},1,1,\frac{1}{\sqrt{2}},\sqrt{\frac{3}{2}},1,\frac{1}{\sqrt{3}},\sqrt{\frac{2}{3}},1\right\}b\,,\qquad \text{for k=3}\,\nn\\
b_n&=&\left\{\sqrt{2},1,1,1,\frac{1}{\sqrt{2}},\frac{1}{\sqrt{2}},1,1,1\right\}b\,,\qquad \text{for k=4}\,.
\eea
In any case, it is relatively straightforward to fix a given $N$, find all $b_n$'s, the wave functions using \eqref{eq:wf-rec}, and plot the spread complexity. In Fig.~\ref{fig:N10ET} and Fig.~\ref{fig:N10LT} we show the example of total 5 possible spread complexities for $N=10$ at early and late times as well as their time averages.  

As a consequence of equal $b_1$'s for all $k\ge1$'s we see that the early time growth $C_K(t)\sim b^2_1t^2$ is the same for all these initial states and is faster than for $k=0$ (left plot on Fig.~\ref{fig:N10ET}). However, after this initial period, spread complexity for $k=0$ overtakes and evolves to a higher peak (right plot). As time progresses, all complexities show erratic oscillations (left plot on Fig.~\ref{fig:N10LT}) also violating the second law. The time averages (right plot on Fig.~\ref{fig:N10LT}) show that late-time plateaus of averaged spread complexities depend on $k$ and are higher for lower $k$ (here for $k=3$ and $k=4$ we see oscillations due to strong finite-size effects). 

Next, we move to the continuum model, choosing the initial state  $\ket{k}$ for $0\le k\le \infty$ gives the family of return amplitudes
\be
S(t)= \bra{k}e^{iHt}\ket{k}=\int^\pi_0 d\theta\phi_k(\theta)^2e^{iE(\theta)t}\,,
\ee
which, after using
\be
\sin^2((k+1)\theta)=\frac{1-\cos(2(k+1)\theta)}{2}\,,
\ee
can be written in terms of Bessel functions as
\be
S(t)=e^{iat}\left[J_{0}(2bt)+i^{2k}J_{2(k+1)}(2bt)\right]\,.
\ee
First, for $k=1$, we again find a split between odd and even Lanczos coefficients, similarly to the discrete model \eqref{OddEvenbDiscr}, but without finite-seize effects i.e., we have two families of infinite Lanczos coefficients
\be
a_n=a\,,\qquad b_{2n-1}=\sqrt{\frac{n+1}{n}}b\,,\qquad b_{2n}=\sqrt{\frac{n}{n+1}}b\,.\label{k1ContLancz}
\ee
Then, for general $k$, we find Lanczos coefficients $a_n=a$ and a clear pattern of Lanczos coefficients that can be written as
\be
b_{(k+1)(n-1)+1}=\sqrt{\frac{n+1}{n}}b\,,\qquad b_{(k+1)n}=\sqrt{\frac{n}{n+1}}b\,,\quad b_n=b \qquad \text{-- otherwise}\,.
\ee
Again, for all the choices $k\ge1$ of initial state we have
\be
b_1=\sqrt{2}b\,.
\ee
Then, we find a repeating gaps of $k-1$ coefficients $b_n=b$'s after $b_1$ and every other non-trivial pair. As we increase $k$, we also increase the number of constant $b_n=b$ in the sequence. For $k=1$ there is no gap and the coefficients simply reduce to even and odd $b_n$'s \eqref{k1ContLancz}.
\begin{figure}[b!]
\centering
\includegraphics[width=.6\textwidth]{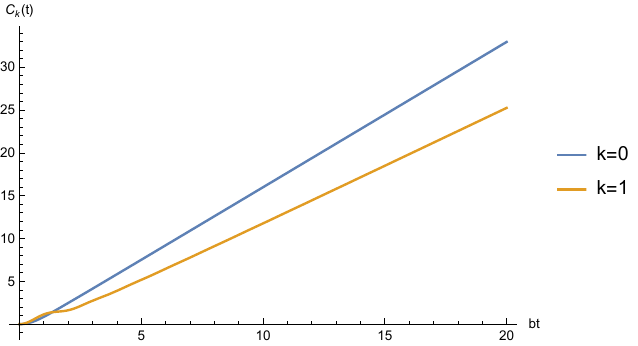}
\caption{Evolution of spread complexity for the initial state $\ket{0}$ (blue) and $\ket{1}$ (orange). }
\label{Spread01}
\end{figure}
It is clear from these expressions that for large $n$, Lanczos coefficients become constant $b_n\sim b$, similarly to the $k=0$ initial state. This is an interesting way in which the scrambling of information about the initial state (in the continuous model with infinite-dimensional Hilbert space) is manifested in Lanczos coefficients.
This behavior also predicts the linear growth of spread complexity at late times, irrespectively of the initial state $\ket{k}$. We will now confirm this after solving the Schr\"{o}dinger's equation below.

Staggering of Lanczos coefficients into even and odd families has been observed in examples of operator growth before \cite{Avdoshkin:2019trj,Camargo:2022rnt,Avdoshkin:2022xuw}. However, to our knowledge, solving the Schr\"{o}dinger equation in those examples was still too difficult and performed only numerically. Fortunately, in our model above, following the recursive procedure \eqref{eq:wf-rec}, we can find the wave functions for $k=1$ analytically. They also split into even end odd families given by
\be
\psi_{2n}(t)=(-1)^ne^{-iat}\left[J_{2n}(-2bt)-J_{2n+4}(-2bt)\right]\,,
\ee
for $n=0,1,2,...$ and
\be
\psi_{2n-1}(t)=\frac{i(-1)^{n+1}e^{-iat}}{\sqrt{n(n+1)}}\left[(n+1)J_{2n-1}(-2bt)+J_{2n+1}(-2bt)-nJ_{2n+3}(-2bt)\right]\,, 
\ee
with $n=1,2,...$. Unfortunately, a closed form for spread complexity is very tedious. Still, it is easy to plot  (Fig.~\ref{Spread01}) and compare with the $k=0$ results. 

For $\ket{1}$ we see a bit faster initial growth of complexity (consistent with bigger $b_1=\sqrt{2}b$ than $b_1=b$ for initial state $\ket{0}$) but is overtaken by spread complexity for $\ket{0}$ later. As predicted by constant Lanczos coefficients for large $n$, at late times, both quantities evolve linearly with time, but complexity of evolving $\ket{0}$ with a slightly steeper slope. This way, we can see that, although for large n all the Lanczos coefficients become constant (some notion of scrambling of information about the initial state), the slopes of spread complexity in this model differ allowing to distinguish between the initial states.

Analysis for higher $k$, even though more cumbersome, is also relatively straightforward and the interesting patterns of $b_n$'s find their counterparts in the solutions of the Schr\"{o}dinger's equation.

Firstly, we have wave functions with only two Bessel functions
\be
\psi_{(k+1)n}(t)=(i)^{(k+1)n}\left(J_{(k+1)n}(-2bt)+(-1)^kJ_{(k+1)(n+2)}(-2bt)\right)e^{-iat}\,,\, n=0,1,2...\,,
\ee
which for $k=1$ are simply the even ones. Then, they are separated by k wave functions with 3 Bessel functions in them (similarly as for $k=1$). For example, for $k=2$, using the notation $\psi_{k,n}=i^{-n}e^{iat}\psi_n(t)$, and $x=-2bt$, the first 16 solutions are expressed in terms of Bessel functions $J_n$ as
\bea
\begin{array}{c}
 \psi _{2,0}=J_0(x)+J_6(x)\,, \\
 \psi _{2,1}=\frac{2 J_1(x)-J_5(x)+J_7(x)}{\sqrt{2}}\,, \\
 \psi _{2,2}=\frac{2 J_2(x)+J_4(x)+J_8(x)}{\sqrt{2}}\,, \\
 \psi _{2,3}=J_3(x)+J_9(x)\,, \\
 \psi _{2,4}=\frac{3 J_4(x)-J_8(x)+2 J_{10}(x)}{\sqrt{6}}\,, \\
 \psi _{2,5}=\frac{3 J_5(x)+J_7(x)+2 J_{11}(x)}{\sqrt{6}}\,, \\
 \psi _{2,6}=J_6(x)+J_{12}(x)\,, \\
 \psi _{2,7}=\frac{4 J_7(x)-J_{11}(x)+3 J_{13}(x)}{2 \sqrt{3}}\,, \\
 \psi _{2,8}=\frac{4 J_8(x)+J_{10}(x)+3 J_{14}(x)}{2 \sqrt{3}}\,, \\
 \psi _{2,9}=J_9(x)+J_{15}(x)\,, \\
 \psi _{2,10}=\frac{5 J_{10}(x)-J_{14}(x)+4 J_{16}(x)}{2 \sqrt{5}}\,, \\
 \psi _{2,11}=\frac{5 J_{11}(x)+J_{13}(x)+4 J_{17}(x)}{2 \sqrt{5}}\,, \\
 \psi _{2,12}=J_{12}(x)+J_{18}(x)\,, \\
 \psi _{2,13}=\frac{6 J_{13}(x)-J_{17}(x)+5 J_{19}(x)}{\sqrt{30}}\,, \\
 \psi _{2,14}=\frac{6 J_{14}(x)+J_{16}(x)+5 J_{20}(x)}{\sqrt{30}}\,, \\
 \psi _{2,15}=J_{15}(x)+J_{21}(x)\,, \\
 \psi _{2,16}=\frac{7 J_{16}(x)-J_{20}(x)+6 J_{22}(x)}{\sqrt{42}}\,. \\
\end{array}\label{16Solk2}
\eea
\begin{figure}[t!]
\centering
\includegraphics[width=.6\textwidth]{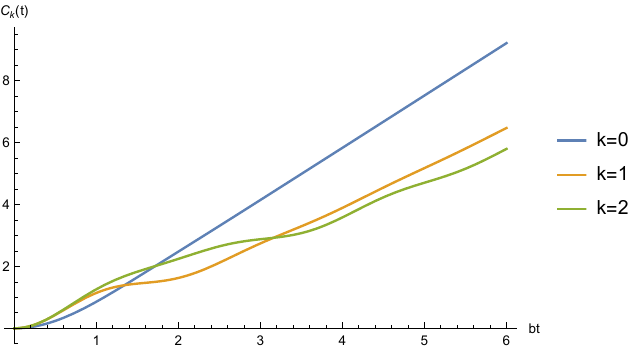}
\caption{Early time growth of spread complexity for $\ket{0}$, $\ket{1}$ and $\ket{2}$ as initial states. For $\ket{2}$ we used the initial 16 wave functions \eqref{16Solk2} that are sufficient to fix the early time growth. }
\label{Retamplk2}
\end{figure}
We can easily generate/find arbitrary higher-$n$ solutions in this pattern, and plot the corresponding spread complexity. For comparison, we plot spread complexities for $k=\{0,1,2\}$ on Fig.~ \eqref{Retamplk2}. 

Again, all higher-k states show initially faster growth than $\ket{0}$ but are later overtaken. At late times, all of them grow linearly with time but the lower the $k$, the steeper the slope of the spread complexity. Confirming that the evolution in this model remembers the information about initial state via the slope of linearly-growing spread complexity. Finally, for all values of $k$ (at least to the extend that we probed) spread complexity increases and the second law holds in the continuum.

\section{Energy gaps vs plateaus} \label{sce:gaps}
To probe the plateau in the spread complexity, one requires time scales $t\,|\Delta_{ij}| \gg 1$ for all energy gaps within the Krylov subspace. In particular, if the spectrum of the initial state $\ket{K_0}$ has different energy scales, the spread complexity will oscillate around the wrong value, till it will eventually saturate at the correct one.
 
To illustrate this fact, consider a Krylov subspace spanned by two sets of energy scales. First, a set of $E_k$ of dimension $d_\mt{K}$ with largest energy $E_*$ and energy gaps $\Delta_{jk} \sim \mathcal{O}(E_*)$. Then, a second set $E_\alpha$ of dimension $d_\alpha$ defined by $E_\alpha = E_*\left(1+ \delta E_\alpha\right)$ with $|\delta E_\alpha| \ll 1$. By construction, all energy gaps $\Delta_{\alpha*} = E_*\,\delta E_\alpha \ll E_\alpha \approx |\Delta_{jk}|$. If the initial state is typical, it follows that the plateau of the averaged spread complexity will be equal
\begin{equation}
  \overline{C_K(\infty)} = \frac{d_\mt{K} + d_\alpha -1}{2} \neq \frac{d_\mt{K} -1}{2}\,.
\end{equation}
However, the spread complexity will appear to oscillate around the fake value $\frac{d_\mt{K} -1}{2}$ for times scales $(E_*\,\delta E_\alpha)^{-1} \gg t \gg |\Delta_{jk}|$.

To illustrate this sensitivity to the energy gaps, it is sufficient to consider a simple example with $d_{\mt{K}}=3$. Fortunately, spread complexity and its time average for this case was already computed in a slightly different context in \cite{Caputa:2025ozd}, and can be generally written as
\be
\label{eq:TAd3}
\overline{C_K(t)}=
1-\left(\frac{M_{12}\sin(E_{12}t)}{9DE_{12}t}+\frac{M_{13}\sin(E_{13}t)}{9DE_{13}t}+\frac{M_{23}\sin(E_{23}t)}{9DE_{23}t}\right)\,,
\ee
where $E_{ij}=E_i-E_j$, $D=E^2_{12}+E^2_{13}+E^2_{23}$, $M_{12}=-E^2_{12}+5E^2_{13}+5E^2_{23}\,$, $M_{13}=5E^2_{12}-E^2_{13}+5E^2_{23}\,$ and  $M_{23}=5E^2_{12}+5E^2_{13}-E^2_{23}\,$.
\begin{figure}[b!]
\centering
\includegraphics[width=.49\textwidth]{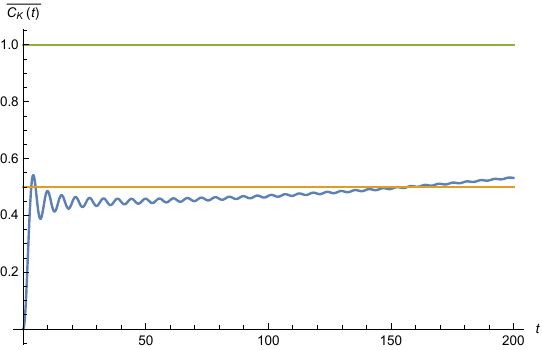}
\includegraphics[width=.49\textwidth]{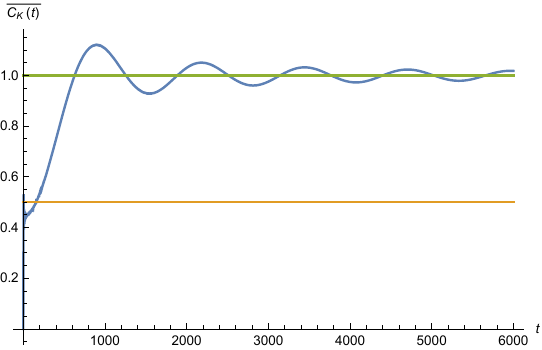}
\caption{Evolution of the time-average of spread complexity (blue) for initial typical pure state with $d_{\mt{K}}=3$ for early (left) and late (right) times. Plateau for $(d_{\mt{K}}-1)/2=1$ (green) and $1/2$ (for $d_{\mt{K}}=2$) (orange). 
Plot for $\bar{E}=5$ and $\{\delta E_1=-0.12,\delta E_2=0.1,\delta E_3=0.101\}$.}
\label{3states}
\end{figure}

In this setup,  we can consider a very tiny split between two of the three energy eigenvalues. For concreteness we take $\bar{E}=5$ and $\{\delta E_1=-0.12,\delta E_2=0.1,\delta E_3=0.101\}$ and plot the time-average of spread complexity on Fig.~\ref{3states}. The plots capture the physics we just described. First, at early times (left figure), the infinitesimal difference is negligible and the time-averaged complexity evolves as for the system with $d_{\mt{K}}=2$ oscillating around a plateau of value $1/2$. However, for much later times (right figure), complexity again starts increasing and reaches the target plateau of the $d_{\mt{K}}=3$ model at $1$.

In agreement with our general arguments, the timescales of the first and the second peak of the time-averaged spread complexity are determined by the differences between the infinitesimal fluctuations around the mean energy (and not $\bar{E}$ itself) 
\be
t\sim\frac{1}{|\delta E_1-\delta E_2|}\sim 4.5,\qquad t\sim\frac{1}{|\delta E_2-\delta E_3|}\sim 1000\,.
\ee
On the other hand, the time when averaged spread complexity starts to increase to the correct plateau depends on $\bar{E}$ as $t\sim (\bar{E}|\delta E_2-\delta E_3|)^{-1}\sim 200$.

\paragraph{Small amplitude perturbations.} One expects the same phenomenon should occur when we perturb a typical state in a Krylov subspace of dimension $d_{\mt{K}}$ to a much bigger Krylov subspace $d_\mt{K} + d_\alpha$ with $d_\alpha \gg d_\mt{K}$ adding a \textit{small initial amplitude} within this large subspace. More precisely, consider an initial Krylov vector
\begin{equation}
  |K_0\rangle = \sum_{k=1}^{d_\mt{K}} c_k |E_k\rangle \qquad \text{with} \qquad |c_k|^2 = \frac{1}{d_\mt{K}}\,,\label{TypPSK0}
\end{equation}
getting ``perturbed" to
\begin{equation}
  |K^\prime_0\rangle = \sum_{k=1}^{d_\mt{K}} c^\prime_k |E_k\rangle + \sum_{\alpha=1}^{d_\alpha} c_\alpha\,|E_\alpha\rangle\,, \qquad \text{with} \qquad |c^\prime_k|^2 = \frac{1-|\varepsilon|^2}{d_\mt{K}}\,, \quad |c_\alpha|^2 = \frac{|\varepsilon|^2}{d_\alpha}\,,
\label{eq:per-amp}
\end{equation}
with $E_\alpha = E_*\left(1+ \delta E_\alpha\right)$ and $|\delta E_\alpha| \ll 1$. Notice the initial amplitude $|c_\alpha|^2$ probing the new subspace is suppressed for two reasons : $|\varepsilon|^2 \ll 1$, as our small perturbative parameter and the perturbation being maximally coherent in the new subspace, leading to a further suppression by $d_\alpha^{-1}$.

A key difference between this perturbation and the typical states is that their energy and variances are, by construction, small perturbations of the original values. For example,
\begin{equation}
a^\prime_0 \approx (1-|\varepsilon|^2)\,a_0 + |\varepsilon|^2\,E_*\,, \qquad b^{\prime 2}_1 \approx b_1^2 + |\varepsilon|^2\,E_*^2\left(1-\frac{a_0}{E_*}\right)\,,
\end{equation}
where we ignored contributions from $E_*\,\delta E_\alpha$\footnote{This is indeed in contrast with the earlier discussion where $a^\prime_0 \sim E_*$ and $b^{\prime 2}_1 \sim \frac{d_\mt{K}}{d_\mt{T} + d_\alpha}\,E_*^2$. The change in variance would in fact allow to identify the correct Krylov subspace dimension by analyzing the $(b^\prime_1 t)^2$ short time behavior of the spread complexity.}. Thus, at short time scales, the spread complexity will indeed behave similarly to the unperturbed state. Unitarity guarantees this will not be the case for all time scales. In particular, given the perturbation probes a much larger Krylov subspace, one would anticipate a significant change in the plateau value if $d_\alpha$ scales with a negative power of the perturbative amplitude $|\varepsilon|^2$. Let us estimate if that this is indeed the case.

The exact value of the plateau for the perturbed state equals
\begin{equation}
  \overline{C^\prime_K(\infty)} = \sum_{n=0}^{d_\mt{K} + d_\alpha -1} n\,\left(\sum_k |c_k|^4\,P^{\prime 2}_n(E_k) + \sum_\alpha |c_\alpha|^4\,P^{\prime 2}_n(E_\alpha)\right)\,.
\end{equation}
The orthogonality relation \eqref{OrthoPnPmD} holds for the perturbed polynomials. Hence, the second sum over the perturbed Krylov subspace can be rewritten using the identity
\begin{equation}
  \sum_\alpha |c_\alpha|^2\,P^{\prime 2}_n(E_\alpha) = 1 - \sum_k |c_k|^2\,P^{\prime 2}_n(E_k) \,.
\end{equation}
Plugging this into the perturbed plateau value and using the specific amplitudes in \eqref{eq:per-amp}, we derive the \textit{exact} relation
\begin{equation}
   \overline{C^\prime_K(\infty)} = \sum_{n=0}^{d_\mt{K} + d_\alpha -1} n\,\left[ \frac{|\varepsilon|^2}{d_\alpha} + \left(1-\frac{|\varepsilon|^2}{d_\alpha}\right)  \frac{(1-|\varepsilon|^2)^2}{d_\mt{K}}  \sum_k \frac{1}{d_\mt{K}} P^{\prime 2}_n(E_k)\right]\,.
\end{equation}
The term proportional to $|\varepsilon|^2/d_\alpha$ is independent of $n$ leading to a contribution
\begin{equation}
  \frac{(d_\alpha + d_\mt{K})(d_\alpha + d_\mt{K} -1)}{2}\,\frac{|\varepsilon|^2}{d_\alpha} \approx \frac{1}{2} d_\alpha\,|\varepsilon|^2 \quad \text{for} \quad d_\alpha \gg d_\mt{K}\,.
\end{equation}
This is already of the expected form : for $d_\alpha \gtrsim |\varepsilon|^{-2}$, the effect on the plateau value will be non-negligible. One can estimate the remaining positive term as follows. First, by construction for $n \leq d_\mt{K}$, 
\begin{equation}
    \sum_k \frac{1}{d_\mt{K}} P^{\prime 2}_n(E_k) = 1 + \mathcal{O}(|\varepsilon|^2) \qquad \text{for} \quad n \leq d_\mt{K}\,,
 \end{equation}
 whereas for $n > d_\mt{K}$, the perturbed polynomial $P^\prime_n(E_k)$ is order $|\varepsilon|^2$. This does not mean it is small, but one can argue its contribution to the plateau must scale at least like $d_\alpha\,|\varepsilon|^2$.

\bibliography{RefsK}
\bibliographystyle{JHEP}

\end{document}